\documentclass[aps,prd,onecolumn,floatfix,nofootinbib,showpacs]{revtex4-1}

\pdfoutput=1
\usepackage{graphicx,color,float}
\usepackage{hyperref}
\usepackage{multirow}
\usepackage{verbatim}
\usepackage{longtable}
 \usepackage{amsmath}
\usepackage{amsfonts}
\usepackage{amssymb}
\usepackage{rotating}

\newcommand{\lsim}{\raisebox{-0.13cm}{~\shortstack{$<$ \\[-0.07cm] $\sim$}}~}
\newcommand{\gsim}{\raisebox{-0.13cm}{~\shortstack{$>$ \\[-0.07cm] $\sim$}}~}

\begin{document}

{\small
\begin{flushright}
IUEP-HEP-18-02
\end{flushright} }

\title{
Higgs-boson-pair production 
$H(\rightarrow b\overline{b})H(\rightarrow\gamma\gamma)$ from gluon fusion 
at the HL-LHC and HL-100 TeV hadron collider
}

\def\slash#1{#1\!\!/}

\renewcommand{\thefootnote}{\arabic{footnote}}

\author{
Jung Chang$^{1,2}$, Kingman Cheung$^{2,3,4}$, Jae Sik Lee$^{1,2,5}$, Chih-Ting Lu$^4$, 
and Jubin Park$^{5,1,2}$}
\affiliation{
$^1$ Department of Physics, Chonnam National University, \\
300 Yongbong-dong, Buk-gu, Gwangju, 500-757, Republic of Korea \\
$^2$ Physics Division, National Center for Theoretical Sciences,
Hsinchu, Taiwan \\
$^3$ Division of Quantum Phases and Devices, School of Physics, 
Konkuk University, Seoul 143-701, Republic of Korea \\
$^4$ Department of Physics, National Tsing Hua University,
Hsinchu 300, Taiwan \\
$^5$ Institute for Universe and Elementary Particles, Chonnam National University, \\
300 Yongbong-dong, Buk-gu, Gwangju, 500-757, Republic of Korea
}
\date{July 15, 2019}

\begin{abstract}
We perform the most up-to-date comprehensive signal-background
analysis for Higgs-pair production in $HH \to b\bar b \gamma\gamma$
channel at the HL-LHC and HL-100 TeV hadron collider, with the goal of
probing the self-coupling $\lambda_{3H}$ of the Higgs boson which is
normalized to its Standard Model value of $1$.
We simulate all the standard-model signal and background
processes and emphasize that the $ggH(\to\gamma\gamma)$ background
has been overlooked in previous studies.
We find that even for the most promising
channel $HH \to b\bar{b}\gamma\gamma$ at the HL-LHC
with a luminosity of 3000 fb$^{-1}$, the significance
is still not high 
enough to establish the Higgs self-coupling at the
standard model (SM) value. Instead, we can only constrain the self-coupling
to  $-1.0  < \lambda_{3H}  < 7.6 $
at $ 95\% $ confidence level
after considering the uncertainties associated with the top-Yukawa coupling
and the estimation of backgrounds.
Here we also extend the study to the HL-100 TeV hadron collider. 
With a luminosity of 3 ab$^{-1}$, we find there exists
a bulk region of $2.6  \lsim \lambda_{3H}  \lsim 4.8 $
in which one cannot pin down the trilinear coupling. Otherwise one can measure the coupling
with a high precision. At the SM value, for example,
we show that the coupling can be measured with about 20 \% accuracy.
While assuming 30 ab$^{-1}$, the bulk region reduces to 
$3.1  \lsim \lambda_{3H}  \lsim 4.3$ and the trilinear coupling can be measured with
about 7 \% accuracy at the SM value.
%
\end{abstract}

\maketitle

\section{Introduction}
Origin of mass is the most important question that one would ask for
our existence. This is related to the mechanism involved in
electroweak symmetry breaking (EWSB), which is believed to give
masses to gauge bosons and fermions. The simplest implementation in
our standard model (SM) is to introduce
a Higgs doublet field, whose non-vanishing vacuum
expectation value causes EWSB \cite{higgs}.
The by-product is a neutral scalar Higgs boson, which was  
eventually discovered in July 2012 \cite{discovery}. 
After accumulating enough data at the end of 8 TeV runs, the scalar boson
is best described by the SM Higgs boson \cite{higgcision}, in 
which the couplings to gauge bosons are confirmly established and those 
to fermions started to fall in the ball-park of the SM values.
However, the SM Higgs boson can hardly constitute a complete theory
because of, for example, the gauge hierarchy problem. 

The current measurements of the Higgs-boson properties mainly concern
the couplings of the Higgs boson to the SM particles.
There is no 
{\it a priori} reason why the EWSB sector simply contains only one 
Higgs doublet field. Indeed, many extensions of the EWSB sector consist of
more Higgs fields. 
Until now there is no information at all about the self-couplings of 
the Higgs boson, which depends on the dynamics of the EWSB sector.
The self-couplings of the Higgs boson are very different among the 
SM, two-Higgs doublet models (2HDM), and MSSM.
One of the probes of Higgs self-coupling is Higgs-boson-pair 
production at the LHC~\cite{hh-early,hh-mid,hh-later}. 
There have been a large number of works in literature on Higgs-pair production
in the SM \cite{hh-sm}, in model-independent formalism \cite{hh-in},
in models beyond the SM \cite{hh-bsm}, and in SUSY \cite{hh-susy}.

The predictions for various models are largely different such that the
production rates can give valuable information on the self-coupling $\lambda_{3H}$.  
In the SM, Higgs-pair production receives contributions from both
the triangle and box diagrams, which interfere with each other.
It is only the triangle diagram that 
involves the Higgs self-trilinear coupling $\lambda_{3H}$, yet
the top-Yukawa coupling appears in both triangle and box diagrams.
Therefore, we have to disentangle the triangle diagram from the box 
diagram in order to probe the Higgs trilinear coupling.
In Ref.~\cite{hh-ours}, we pointed out that 
the triangle diagram, with $s$-channel Higgs propagator,
is more important at low invariant-mass region than the box diagram.
Thus, the Higgs-boson pair from the triangle diagram tends to have lower
invariant mass, and therefore the opening angle in the decay products 
of each Higgs boson tends to be larger than that 
from the box diagram.
Indeed, the opening angle separations $\Delta R_{\gamma\gamma}$ and 
$\Delta R_{bb}$ between the decay products of the Higgs-boson
pair are very useful variables to disentangle the two sources.
However, in Ref.~\cite{hh-ours} we only assumed some level of signal
uncertainties to evaluate the sensitivity to the parameter space of
self-coupling $\lambda_{3H}$ and the top-Yukawa coupling $g^S_t$, without
calculating all the other SM backgrounds, e.g., jet-fake backgrounds,
single Higgs associated backgrounds, and non-resonant backgrounds.

In this work, we perform the most up-to-date comprehensive signal-background
analysis for Higgs-pair production 
through gluon fusion
and the $HH \to b\bar b \gamma\gamma$ decay channel. 
For other production and decay channels 
and some combined analyses, see Refs.~\cite{hh-others}.
We simulate the signal and all background processes
using simulation tools as sophisticated as what experimentalists use.
The signal subprocess is $ gg \to H H \to b\bar b \gamma\gamma$ with various
values for $\lambda_{3H}$. The background includes $t\bar t$, $t\bar t \gamma$, 
single Higgs associated backgrounds
(e.g. $ZH$, $t\bar t H$, $b \bar b H$, $ggH$  followed by $H \to
\gamma\gamma$), and non-resonant or jet-fake backgrounds (e.g. $b\bar b \gamma
\gamma $, $b\bar b j \gamma$, $b\bar b j j$, $ j j \gamma\gamma$, etc). 
We found a set of useful selection cuts to reduce the backgrounds.
We express the sensitivity that can be achieved in terms of
significance.  We find that even for the most promising
channel $HH \to b\bar{b}\gamma\gamma$ at the HL-LHC, the significance
is still not high 
enough to establish the Higgs self-coupling at the
SM value, though the self-coupling can be constrained 
to the range $ 0 < \lambda_{3H} < 7.1$ at $ 95\% $ confidence
level (CL) with an integrated luminosity of 3000 fb$^{-1}$.
Taking account of the uncertainties associated with the top-Yukawa coupling
and the estimation of backgrounds, we have found that the $ 95\% $ CL region
broadens into $-1.0  < \lambda_{3H}  < 7.6 $.
We also extend the analysis to the HL-100 TeV hadron collider.
With a luminosity of 3 ab$^{-1}$, we find a bulk region of 
$2.6  \lsim \lambda_{3H}  \lsim  4.8 $
in which one cannot pin down the trilinear coupling. Otherwise one can measure the coupling
with a high precision. At the SM value, for example,
we show that the coupling can be measured with about 20\% accuracy.
While assuming 30 ab$^{-1}$, the bulk region reduces to 
$3.1  \lsim \lambda_{3H}  \lsim 4.3$ and the trilinear coupling can be measured with
about 7 \% accuracy at the SM value.
This is the main result of this work.

This work has a number of improvements over our previous and other works
in literature, summarized as follows.
\begin{enumerate}
\item
We have included all the backgrounds, including $t\bar t$ related ones,
single Higgs associated production processes, 
non-resonant backgrounds, and
jet-fake backgrounds. 
Furthermore we would like to emphasize 
that we have implemented through detector simulations of 
all the backgrounds.

\item
While implementing all the relevant signal and background simulations, we find that
the $ggH(\to\gamma\gamma)$ background is possibly  very important and
has been overlooked in previous studies. Note that the similar
observation has been recently made by the authors of Ref.~\cite{Homiller:2018dgu}.

\item
For the signal, since the signal distributions  behave 
differently for different $\lambda_{3H}$, we evaluate the selection 
efficiency separately for each $\lambda_{3H}$
to properly cover the viable range of the non-standard values of
$\lambda_{3H}$.
%

%

\item At the HL-LHC, we firstly take into account the impact of
the uncertainty associated with the top-Yukawa coupling
on $95\%$ CL sensitivity. We find that, especially,
the lower boundary of the 95\% CL region of $\lambda_{3H}$ 
significantly varies upon the  expected
precision of the top-quark Yukawa coupling
in the HL-LHC era.

\item
Taking account of all the backgrounds known up to date and devising 
a {\it new} set of selection cuts,
we have most reliably estimated the potential reach of 
HL-100 TeV hadron collider for a broad range of $\lambda_{3H}$.

\item
At the HL-100 TeV collider,
we find there is a two-fold ambiguity in $\lambda_{3H}$ which 
could be lifted up by exploiting several kinematical 
distributions.
We also find that there exists a bulk region in which 
it would be difficult to establish the $\lambda_{3H}$
coupling even at the HL-100 TeV collider.

\end{enumerate}

The organization is as follows. 
In the next section, we briefly describe the effective Lagrangian for
Higgs-pair production. 
In Sec. III, we describe the signal and background processes and
simulation tools.
We also present the distributions, selection cuts, cut flows of signal and
backgrounds, and significance for the HL-LHC.
Section IV is dedicated to the case of HL-100 TeV hadron collider.
%
In Sec. V, we examine the impact of the NLO corrections considering 
full top-quark mass dependence, 
the effect of using a modern PDF set to include the LHC data on PDF, and
how the investigation of the uncertainties involved in the matching procedures 
affects the 95\% CL sensitivity region of $\lambda_{3H}$.
%
We discuss and conclude in Sec. VI.
We put some extra distributions and cut flow tables,
which can be ignored in the first reading,
into the appendices A and B. Appendix C, on the other hand,
gives the details for the procedures employed in the matching 
in calculating the cross sections of
the non-resonant backgrounds, as well as their uncertainties.

\section{Effective Lagrangian}

The contributing Feynman diagrams for Higgs-boson-pair production via 
gluon fusion include a triangle diagram with a Higgs-boson propagator 
and a box diagram with colored particles running in them. 
The relevant couplings involved 
are top-Yukawa and the Higgs trilinear self coupling, which
are given in this Lagrangian:
\begin{equation}
\label{lag}
-{\cal L}=\frac{1}{3!}\left(\frac{3M_H^2}{v}\right)\,\lambda_{3H}\,H^3
\ + \
 g^S_t \, \frac{m_t}{v}\,\bar{t} \,  t\,H
\end{equation}
In the SM, $\lambda_{3H}=g^S_t=1$.
The differential cross section for
the process $g(p_1)g(p_2) \to H(p_3)H(p_4)$ 
was obtained in Ref.~\cite{plehn} as 
\begin{equation}
\frac{d\hat\sigma(gg\to HH)}{d\hat{t}}=\frac{G_F^2 \alpha_s^2}{512(2\pi)^3}
\left[ \Big| \lambda_{3H} g^S_t D(\hat{s}) F_\triangle^S
+ (g^S_t)^2 F_\Box^{SS} \Big|^2 
+ \Big| (g^S_t)^2G_\Box^{SS} \Big|^2 \right] 
\end{equation}
where 
\begin{equation}
D(\hat{s})= \frac{3M_H^2}{\hat{s}-M_H^2+iM_H\Gamma_H}
\end{equation}
and
$\hat{s}=(p_1+p_2)^2$,
$\hat{t}=(p_1-p_3)^2$, and
$\hat{u}=(p_2-p_3)^2$ with $p_1+p_2=p_3+p_4$.
The loop functions $F_\triangle^S=F_\triangle$, 
$F_\Box^{SS}=F_\Box$, and $G_\Box^{SS}=G_\Box$ 
with $F_{\triangle\,,\Box}$ and $G_\Box$
given in Appendix A.1 of Ref.~\cite{plehn}.
In the heavy quark limit, one may have
\begin{eqnarray}
F_\triangle^S =+\frac{2}{3}+{\cal O}(\hat{s}/m_Q^2)\,, \ \ \
F_\Box^{SS}  = -\frac{2}{3}+{\cal O}(\hat{s}/m_Q^2)\,, \ \ \
G_\Box^{SS}  = {\cal O}(\hat{s}/m_Q^2)
\end{eqnarray}
leading to large cancellation between the triangle and box diagrams.

The production cross section normalized to the corresponding SM cross
section, with or without cuts, can be parameterized as follows:
\begin{eqnarray}
\frac{\sigma^{\rm LO}(gg\to HH)}{\sigma^{\rm LO}_{\rm SM}(gg\to HH)}
&=&
c_1(s) \, \lambda_{3H}^2 \, (g^S_t)^2 +
c_2(s) \, \lambda_{3H}\,(g^S_t)^3
+ c_3(s) \, (g^S_t)^4 
\label{cdef}
\end{eqnarray}
where the numerical coefficients $c_{1,2,3}(s)$ depend on $s$ and 
experimental selection cuts.  
Numerically, $c_1(s), c_2(s), c_3(s)$ are $0.263\,, - 1.310\,, 2.047$ 
at 14 TeV and $0.208\,, -1.108\,, 1.900$ at 100 TeV \cite{hh-ours}.
Upon our normalization, the ratio
should be equal to $1$ when $g_t^S=\lambda_{3H}=1$,
or $c_1(s)+c_2(s)+c_3(s)=1$. The coefficients $c_1(s)$ and $c_3(s)$ are for
the contributions from the triangle and box diagrams, respectively, and
the coefficient $c_2(s)$ for the interference between them.
Once we have the coefficients $c_i$
the cross sections can be easily obtained for 
any combinations of couplings. 

\begin{figure}[t!]
\centering
\includegraphics[width=3.2in,height=2.8in]{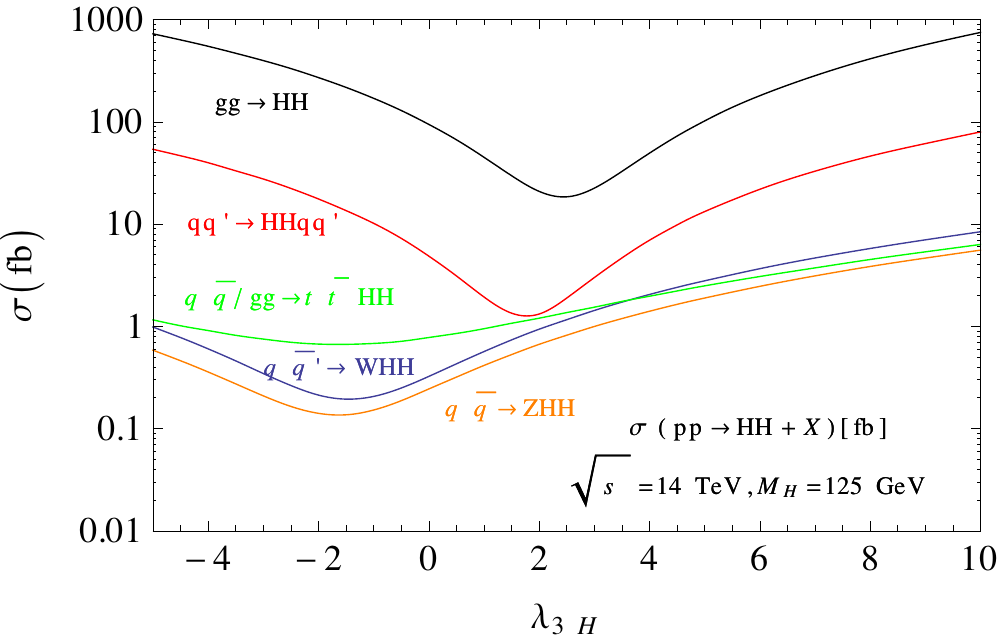}
\includegraphics[width=3.2in,height=2.8in]{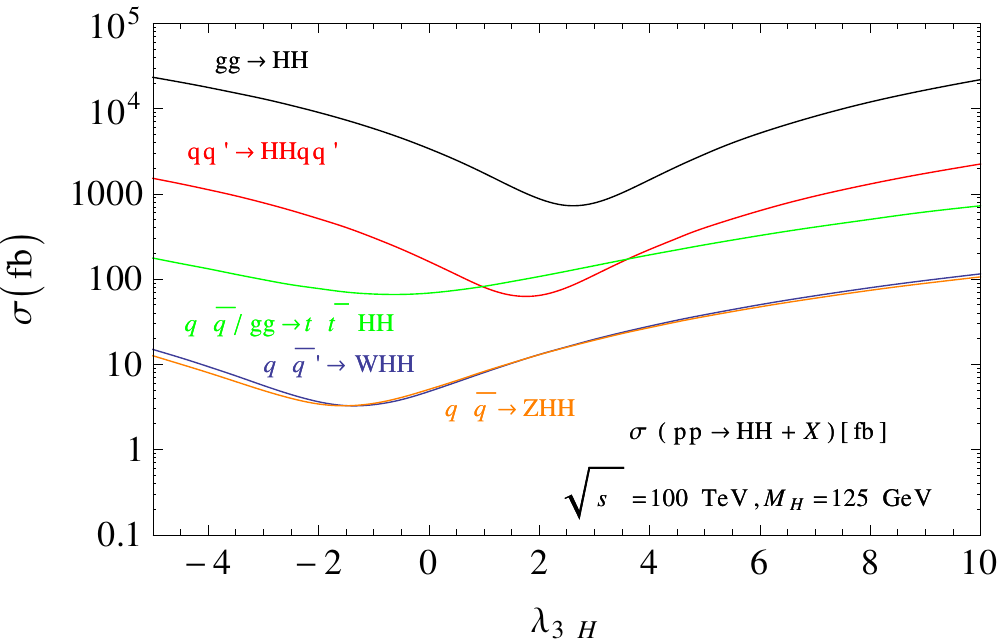}
\caption{\small \label{fig:total-cross}
Production cross sections for various channels for $HH$ production at
$\sqrt{s} = 14$ TeV (left) and $\sqrt{s} = 100$ TeV (right).
The {\tt NNPDF2.3LO} PDF set is used.  }
\end{figure}

\begin{figure}[th!]
\centering
\includegraphics[width=3.2in]{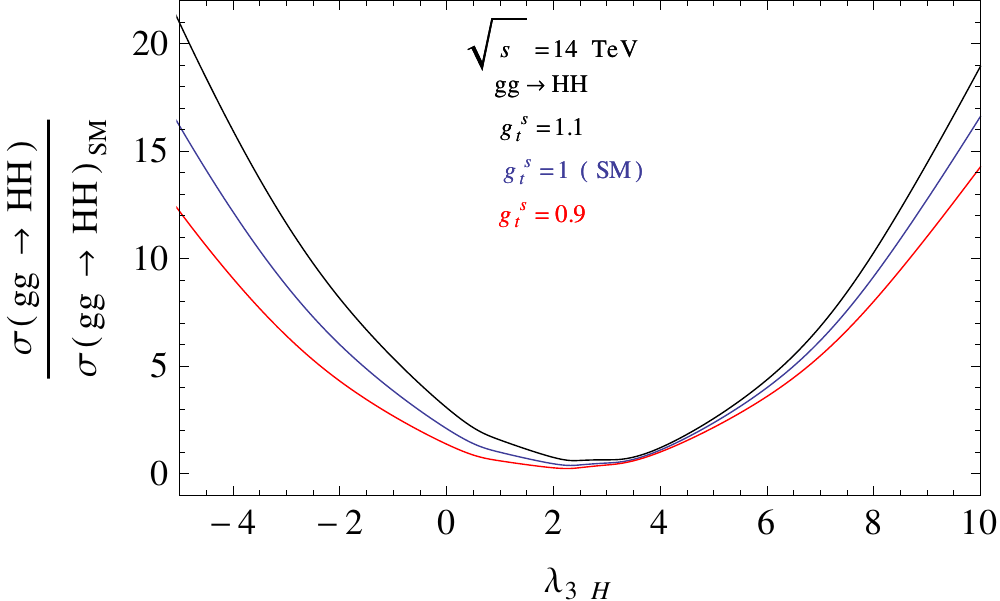}
\includegraphics[width=3.2in]{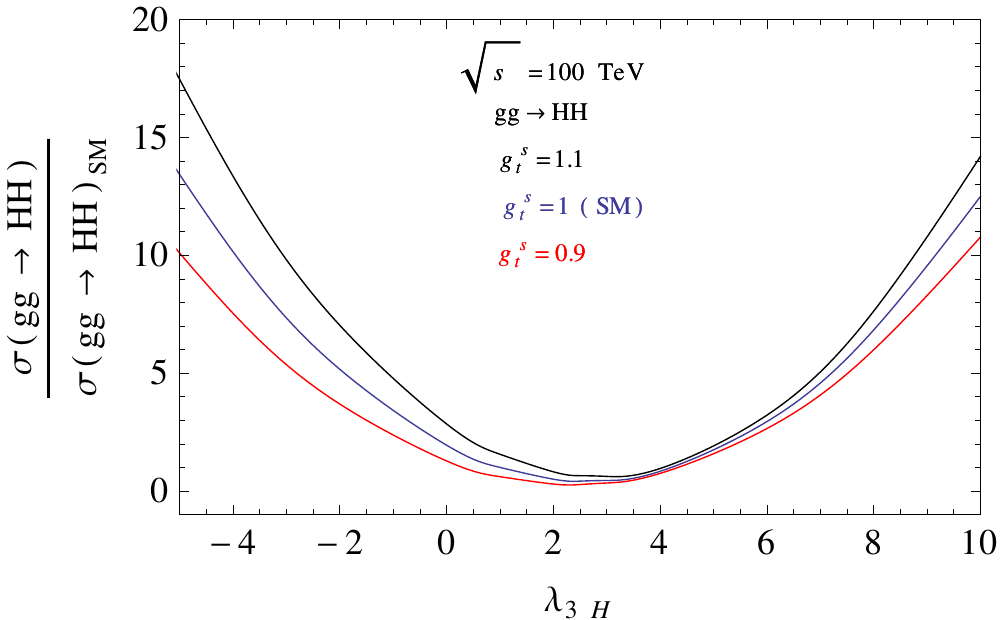}
\caption{\small \label{fig:ratio-cross}
Ratio of cross sections $\sigma(gg \to HH) / \sigma(gg\to HH)_{\rm SM}$
versus $\lambda_{3H}$ taking account of $10\%$  uncertainty of
the top-Yukawa coupling: $g^S_t=1.1$ (black),
$1$ (blue), and $0.9$ (red)
for $\sqrt{s} = 14$ TeV (left) and $\sqrt{s} = 100$ TeV (right).
}
\end{figure}

To get a feeling for the size of the cross sections that we are considering,
we show the total production cross sections for various $HH$ production channels
in Fig.~\ref{fig:total-cross}.
At 14 TeV, the SM cross sections 
$\sigma(gg\to HH)=45.05$ fb~\cite{deFlorian:2015moa}, 
$\sigma(qq'\to HHqq')=1.94$ fb~\cite{Frederix:2014hta}, 
$\sigma(q\bar q(')\to VHH=0.567(V=W^\pm)\,/0.415(V=Z)$ fb~\cite{Baglio:2012np}, 
and $\sigma(gg/q\bar q\to t\bar tHH)=0.949$ fb~\cite{Frederix:2014hta}
are calculated at NNLO+NNLL, NLO, NNLO, and NLO, respectively~\cite{deFlorian:2016spz}.
The 100 TeV cross sections
$\sigma(gg\to HH)=1749$ fb, 
$\sigma(qq'\to HHqq')=80.3$ fb, $\sigma(q\bar q(')\to VHH=8.00(V=W^\pm)\,/8.23(V=Z)$ fb, 
and $\sigma(gg/q\bar q\to t\bar tHH)=82.1$ fb
are calculated at the same orders as at 14 TeV~\cite{100tevcx,Contino:2016spe}.
From Fig.~\ref{fig:total-cross},
it is clear that the gluon fusion into $HH$
gives the largest cross sections independently of $\lambda_{3H}$
with its minimum occurring at $\lambda_{3H}\simeq 2.5$.
From now on we shall focus on the gluon fusion mechanism.
We show the ratio of the cross sections for the $gg \to HH$ process as a function of 
$\lambda_{3H}$ in Fig.~\ref{fig:ratio-cross}, 
in which we also indicate the effects of allowing the top-Yukawa coupling to have
$\pm 10\%$ uncertainty or $\delta g^S_t = \pm 0.1$.
At the HL-LHC, the  expected
precision of measurement of the top-quark Yukawa coupling is $10\%$~\cite{Vos:2017ses}.
Currently, without knowing the absolute value of the top-quark Yukawa coupling 
no better than $10\%$ precision,
we also consider the $\delta g_t^S=\pm 10\%$ effect at 100 TeV
though the expected uncertainty is $1\%$ at the 100-TeV $pp$ colliders.
%

\section{Simulations, Event Selections, and Analysis at the 14 TeV HL-LHC}
Our goal is to disentangle the effects of trilinear Higgs coupling, which is
present in the triangle diagram, in Higgs-pair production.  We focus
on the decay channel $ H H \to b\bar b \gamma\gamma$, in which the
final state consists of a pair of $b$ quarks and a pair of photons
reconstructed at the invariant mass around the Higgs-boson mass
($M_H \simeq 125$ GeV). We shall vary the value for the trilinear
coupling $\lambda_{3H}$ between $-5$ and $10$  to visualize the effects of
$\lambda_{3H}$.
The backgrounds then include
\begin{itemize}
\item
 single-Higgs associated production, such as $ggH,\, t\bar t H,\, ZH,\,
b\bar b H$ followed by $H \to \gamma\gamma$,
\item
 non-resonant backgrounds and jet-fake backgrounds, such as
$b\bar b\gamma\gamma$, $c\bar c \gamma\gamma$, $jj \gamma\gamma$,
$b \bar b j\gamma$, $c\bar c j \gamma$, $b\bar b jj$, and $Z\gamma\gamma \to
b\bar b \gamma\gamma$,
\item
 $t\bar t(\geq 1$ lepton) and $t\bar t\gamma (\geq 1$ lepton) backgrounds .
\end{itemize}
All the signal and backgrounds are summarized in Table~\ref{tab:ParticleList},
together with the information of the corresponding
event generator, the cross section times
the branching ratio ($\sigma\cdot BR$), the order in QCD
for the calculation of $\sigma\cdot BR$, and the Parton Distribution Function (PDF) used.

\begin{table}[t!]
\caption{Monte Carlo samples used in Higgs-pair production
  analysis $H(\rightarrow b\bar{b})H(\rightarrow \gamma\gamma)$, and
  the corresponding codes for the matrix-element generation, parton
  showering, and hadronization. The third (fourth) column shows their cross
  section times branching ratio (the order in perturbative QCD of the
  cross section calculation applied), and the final column shows their
  PDF set used in the simulation.
For the generation of
non-resonant  and $t\bar t\gamma$ backgrounds,
some pre-selection cuts  are applied at the parton level
in order to remove the divergence associated with the photons or jets,
see Eq.~(\ref{eq:presel}).
Note that, except the $ggH(\to\gamma\gamma)$ and
$t\bar t$ backgrounds which are generated at NLO, all the signal and 
backgrounds are generated at LO and normalized to the 
cross sections computed at the accuracy denoted in `Order in QCD'.
}\vspace{3mm}
\label{tab:ParticleList}
\centering
\begin{ruledtabular}
\begin{tabular}{  c  c  c  c  c  c }
\multicolumn{6}{c}{Signal} \\
\hline
\multicolumn{2}{c}{Signal process} & Generator/Parton Shower &
$\sigma \cdot BR$ [fb] & Order  & PDF used  \\
&&&& in QCD &   \\
\hline
\multicolumn{2}{c}{$gg \to HH \to b\bar b \gamma\gamma$
\cite{deFlorian:2016spz} } &
$\mathtt{MG5\_aMC@NLO}$/$\mathtt{PYTHIA8}$ & 0.119
 & NNLO & NNPDF2.3LO \\
 &&& & \!\!\!\!\!$+$NNLL  &  \\
\hline
\hline
\multicolumn{6}{c}{Backgrounds} \\
\hline
Background(BG)  & Process  & Generator/Parton Shower & $\sigma\cdot BR$~[fb] &  Order  &
PDF used\\
&&&& in QCD &   \\
\hline
\multirow{4}{*}{}
& $ggH(\rightarrow \gamma\gamma)$  &  $\mathtt{POWHEG-BOX}$/$\mathtt{PYTHIA6}$
  & $1.20 \times 10^2$ & NNNLO & $\mathtt{CT10}$ \\ \cline{2-5}
Single-Higgs  & $t \bar{t} H(\rightarrow \gamma\gamma)$ & $\mathtt{PYTHIA8}$/$\mathtt{PYTHIA8}$  & 1.37 & NLO & \\ \cline{2-5}
 associated BG \cite{deFlorian:2016spz}                 &  $ZH(\rightarrow \gamma\gamma)$   &  $\mathtt{PYTHIA8}$/$\mathtt{PYTHIA8}$   & 2.24 & NLO &\\ \cline{2-5}
                  & $b\bar{b}H(\rightarrow \gamma\gamma)$ &  $\mathtt{PYTHIA8}$/$\mathtt{PYTHIA8}$ & 1.26 & NLO  &\\ \hline
\multirow{7}{*}{Non-resonant BG} & $b\bar{b} \gamma\gamma$ & $\mathtt{MG5\_aMC@NLO}$/$\mathtt{PYTHIA8}$ & $1.40 \times 10^2$  &  LO & $\mathtt{CTEQ6L1}$ \\ \cline{2-5}
                  &  $c\bar{c} \gamma\gamma$ & $\mathtt{MG5\_aMC@NLO}$/$\mathtt{PYTHIA8}$ & $1.14 \times 10^3$ & LO &  \\ \cline{2-5}
                  &  $jj\gamma\gamma$ & $\mathtt{MG5\_aMC@NLO}$/$\mathtt{PYTHIA8}$ & $1.62 \times 10^4$ & LO &  \\ \cline{2-5}
                  &  $b\bar{b}j\gamma$ & $\mathtt{MG5\_aMC@NLO}$/$\mathtt{PYTHIA8}$ & $3.67 \times 10^5$ & LO &  \\ \cline{2-5}
                  &  $c\bar{c}j\gamma$ & $\mathtt{MG5\_aMC@NLO}$/$\mathtt{PYTHIA8}$ & $1.05 \times 10^6$ & LO &  \\ \cline{2-5}
                  &  $b\bar{b}jj$  & $\mathtt{MG5\_aMC@NLO}$/$\mathtt{PYTHIA8}$ &$4.34 \times 10^8$ & LO &  \\ \cline{2-5}
                  & $Z(\rightarrow b\bar{b})\gamma\gamma$ & $\mathtt{MG5\_aMC@NLO}$/$\mathtt{PYTHIA8}$ & $5.17$ & LO & \\ \hline
\multirow{2}{*}{$t\bar{t}$ and $t\bar{t}\gamma$ BG} & $t\bar{t}$
\cite{Czakon:2011xx}
 & $\mathtt{POWHEG-BOX}$/$\mathtt{PYTHIA8}$  &
$5.30 \times 10^5$ & NNLO &  $\mathtt{CT10}$  \\ 
 &  &  & & \!\!\!\!\!$+$NNLL &    \\ \cline{2-6}
    ($\geq 1$ lepton)    & $t\bar{t}\gamma$ \cite{Melnikov:2011ta}
 & $\mathtt{MG5\_aMC@NLO}$/$\mathtt{PYTHIA8}$  &
$1.60 \times 10^3$ & NLO & $\mathtt{CTEQ6L1}$
\end{tabular}
\end{ruledtabular}
\end{table}

\subsection{Parton-level event generations and detector simulations}

Parton-level events for the backgrounds $\big(
b\bar{b}\gamma\gamma$, $c\bar{c}\gamma\gamma$, $jj\gamma\gamma$,
$b\bar{b}j\gamma$, $c\bar{c}j\gamma$, $b\bar{b}jj$, $t\bar{t}\gamma$,
and $Z(\rightarrow b\bar{b})\gamma\gamma \big)$ and for the signal
$\big($with $-5\leq \lambda_{3H} \leq 10\big)$ are
generated with {\fontfamily{qcr}\selectfont \textbf
{MadGraph5{\_}aMC@NLO}}\,({\fontfamily{qcr}\selectfont \textbf
{MG5{\_}aMC@NLO}}) \cite{Alwall:2014hca}. 
Backgrounds for gluon fusion and top-quark pair are generated with
{\fontfamily{qcr}\selectfont \textbf {POWHEG BOX}}
\cite{Nason:2004rx}.  
The single-Higgs associated backgrounds for 
$t \bar{t}H(\rightarrow \gamma\gamma), ZH(\rightarrow \gamma\gamma), 
b\bar{b}H(\rightarrow
\gamma\gamma)$ are generated with $\textbf{Pythia8}$
\cite{Sjostrand:2014zea}.
Here we would like to provide more detailed information
on the parton-level generation of signal and background events.
The signal cross sections are calculated with the adjustable Higgs 
self-coupling in UFO format \cite{Degrande:2011ua} and events are 
generated in the loop induced mode \cite{Hirschi:2015iia}. 
The {\fontfamily{qcr}\selectfont \textbf {MadSpin}} code
\cite{Artoisenet:2012st} is then employed to let the Higgs-boson pair 
decay into $b\bar{b} \gamma \gamma$.
Further on the parton-level generation of
non-resonant  and $t\bar t\gamma$ backgrounds,
the following pre-selection cuts at parton level are imposed
in order to avoid any divergence in the parton-level calculations
\cite{atlas_hh17}:
\begin{eqnarray}
\label{eq:presel}
P_{T_j} > 20\  \text{ GeV},\ P_{T_b} > 20\  \text{ GeV},\ P_{T_\gamma} > 25\  \text{ GeV},\ P_{T_l} > 10\  \text{ GeV}, \nonumber\\
 |\eta_j|<5,\ |\eta_\gamma|<2.7,\ |\eta_l|<2.5,\ \Delta R_{jj,ll,\gamma\gamma,\gamma j,jl, \gamma l} > 0.4, \nonumber\\
M_{jj} > 25\ \text{GeV},\ M_{bb} > 45\ \text{GeV},\ 60<M_{\gamma\gamma} < 200\ \text{GeV}. 
\end{eqnarray}

For parton showering, hadronization, and decays of unstable particles, 
{\fontfamily{qcr}\selectfont  \textbf {Pythia8}}\cite{Sjostrand:2014zea}
is used both for signal and backgrounds.
Finally, fast detector simulation and analysis at the HL-LHC are 
performed using {\fontfamily{qcr}\selectfont  \textbf{Delphes3}} 
\cite{deFavereau:2013fsa} with the ATLAS template.
In the template,
we use the expected performance for photon efficiency, 
photon fake rates,
$b$-jet tagging efficiency, and $b$-jet fake rates 
obtained with a mean pile-up $\langle\mu\rangle=200$ 
(see Refs.~\cite{atlas_hh17,atlas_perform}). 
For the photon efficiency, we use the 
$P_T$-dependent formula
$$\epsilon_\gamma =  0.888*\tanh(0.01275*P_{T_\gamma}/{\rm GeV})\,,$$
which we obtain by fitting to the ATLAS simulation results.
At $P_{T_\gamma} \sim 50\ \text{GeV}$, $\epsilon_\gamma\sim 50\%$
as in Ref.~\cite{atlas_hh17}
and it approaches $\epsilon_\gamma \sim 85\%$ in the saturation
region of the curve, at $P_{T_\gamma} \sim 150\ \text{GeV}$
to be specific, being consistent with ATLAS simulation~\cite{atlas_perform}.
The photon fake rates are taken
from Ref. \cite{atlas_hh17}: $P_{j\rightarrow \gamma}=5\times
10^{-4}$ and $P_{e\rightarrow \gamma}=2\%\,(5\%)$ in 
the barrel (endcap) region.
The $b$-jet tagging efficiency $\epsilon_b$
depends on $P_T$ and $\eta$ of $b$ jet 
and we have fully considered its $P_T$ and $\eta$ dependence,
see Fig.7(b) of Ref.~\cite{atlas_perform}.
The charm-jet fake rate $P_{c \to b}$
depends on $\epsilon_b$ and, accordingly, on
$P_T$ and $\eta$ of $c$ jet. For the 
multi-variate MV1 $b$-tagging algorithm
taken in our analysis,
$P_{c \to b}\sim 1/5$  when $\epsilon_b=0.7$
and it approaches $1$ as $\epsilon_b\to 1$~\cite{atlas_2015}.
In our simulation, the $P_T$ and $\eta$ dependence
of $P_{c \to b}$ is also considered.
For the light-jet fake rate, we are taking $P_{j \to b} = 1/1300$~\cite{atlas_hh17}.
Incidentally, we have also considered the energy loss due to the 
$b$ momentum reconstruction from the $b$-tagged jet
and set the jet-energy scale using the scaling 
formula~\cite{deFavereau:2013fsa} 
$$\sqrt{ \frac{(3.0 - 0.2|\eta_b|)^2 }{ P_{T_b}/{\rm GeV}} +1.27 }$$
where the factor 1.27 is tuned to get a correct peak position at $M_H$
in the invariant mass distribution of a $b$-quark pair in the signal process.

In this study, we do not include the pile-up effects into our
simulation. There are a couple of reasons for this.
First, it is expected that the pile-up
effects can be dealt with by the upgraded event trigger in future,
and its overall effect could be negligible in the channel of our interests
\footnote{It is shown that the rejection factor for pile-up jets
could be 1350 with a mean pile-up $\langle\mu\rangle=200$~\cite{atlas_hh17}.  
According to ATLAS simulation,
only $0\%\,(1.28\%)$  and $0.54\%\,(4.03\%)$ of
jets identified as (sub)leading  $b$-jets and 
reconstructed (sub)leading photons, respectively, 
originate from pile-up jets. }.
More importantly, by imposing a narrow $M_{\gamma\gamma}$ invariant mass window cut
in event selection, we could eventually obtain similar results 
independently of including the pile-up effects.
This is because pile-up causes the stronger impact 
on photons than on $b$-jets and the soft fake photons from pile-up jets 
make the width of $M_{\gamma\gamma}$ peak wider.
Incidentally, we also have checked that the simulation results using
the ATLAS $b$-tagging efficiency in the presence of pile-up 
are similar to those
obtained by using the $b$-tagging efficiency
in the absence of pile-up (the MV1 algorithm).


\subsection{Signal Event Samples}

The dominant mechanism for Higgs-pair production is the gluon fusion
process at the hadron colliders. Other processes are more than an
order of magnitude smaller. Thus, only the gluon fusion production
mode is used for the signal process $HH \rightarrow
b\bar{b}\gamma\gamma$. These samples are generated with
$\mathtt{MADGRAPH5\_aMC@NLO}$ at LO
\footnote{We use $m_t=172$ GeV.}. 
They are showered by $\mathtt{PYTHIA8}$ to model
the parton showering and hadronization. Note that the A14 tune and the
$\mathtt{NNPDF2.3LO}$ PDF~\cite{Ball:2014uwa}
set are used together. 

The signal event samples are generated with various self-coupling strengths 
in order to show their characteristics: $-5\leq \lambda_{3H}\leq 10$
with $\lambda_{3H} = 1$ being corresponding to the
SM Higgs self-coupling strength.
And, the expected signal yields are normalized to the 
cross section computed at next-to-next-to-leading-order (NNLO) accuracy 
including next-to-next-to-leading-log (NNLL) gluon resummation~\cite{deFlorian:2016spz}
\footnote{
For the signal event normalization, we take the cross section
computed in the infinite top quark mass approximation~\cite{deFlorian:2016spz}.
}.
In Table \ref{tab:x-cross}, we show the production cross section times 
the branching ratio at
the 14 TeV LHC for six selected values of $\lambda_{3H} = -4, 0,1,
2, 6, 10$.
To obtain the production cross section $\sigma$ 
for the non-SM values of $\lambda_{3H}\neq 1$, we 
have used\footnote{See also Fig.~\ref{fig:ratio-cross}.}
\begin{equation}
\sigma=
\frac{\sigma^{\rm LO}}{\sigma^{\rm LO}_{\rm SM}}\times
\sigma^{\rm NNLO+NNLL}_{\rm SM}\,.
\end{equation}

\begin{table}[t!]
\caption{\small 
The production cross section times the branching ratio 
$\sigma \cdot BR(HH \to b\bar b \gamma\gamma)$ at the 14 TeV LHC.
}\vspace{3mm}
\label{tab:x-cross}
\centering
\begin{ruledtabular}
\begin{tabular}{c|cccccc}
  $\lambda_{3H}$& -4 & 0 & 1 & 2 & 6 & 10 \\
\hline
$\sigma\cdot BR(HH \to b\bar b \gamma\gamma)$ [fb] & 1.45 & 0.25 & 0.12 & 0.06 
& 0.48 & 1.97 \\
\end{tabular}
\end{ruledtabular}
\end{table}

\subsection{Background Samples}

The backgrounds mainly come from the processes with multiple jets and
photons. They can mimic the signal-like two photons and two $b$-jets in
the final state. These backgrounds can be categorized into 
\begin{itemize}
\item Single-Higgs associated backgrounds:
$ggH(\gamma\gamma)$, $t \bar{t} H(\gamma\gamma)$,
$ZH(\gamma\gamma)$ and $b\bar{b}H(\gamma\gamma)$,
\item Non-resonant (continuum) backgrounds:
$b\bar{b} \gamma\gamma$, $c\bar{c} \gamma\gamma$,
$jj\gamma\gamma$, $b\bar{b}j\gamma$, $c\bar{c}j\gamma$, $b\bar{b}jj$
and $Z(b\bar{b})\gamma\gamma$ events with an additional jet, 
\item $t\bar{t}$ and $t\bar{t}\gamma$ backgrounds 
in which at least one of the top quarks decays leptonically.
\end{itemize}
The information is summarized in Table~\ref{tab:ParticleList}.

\subsubsection{Single-Higgs associated backgrounds}
The gluon-fusion process $ggH(\gamma\gamma)$ is generated using
{{\fontfamily{qcr}\selectfont \textbf{POWHEG-BOX}} \cite{Nason:2004rx}
and then the background yield is
normalized using the cross section at next-to-next-to-next-leading order 
(NNNLO) in QCD \cite{deFlorian:2016spz}.
The samples for $t \bar{t} H(\gamma\gamma)$, $ZH(\gamma\gamma)$ and
$b\bar{b}H(\gamma\gamma)$ are generated using $\mathtt{PYTHIA8}$
and they are normalized to the cross sections calculated at
NLO in QCD \cite{deFlorian:2016spz}.

\subsubsection{Non-resonant backgrounds}
The non-resonant or continuum background (BG) processes included for the analysis
are $b\bar{b} \gamma\gamma$, $c\bar{c} \gamma\gamma$, $jj\gamma\gamma$,
$b\bar{b}j\gamma$, $c\bar{c}j\gamma$, $b\bar{b}jj$ and
$Z(b\bar{b})\gamma\gamma$. 
They are all generated with
$\mathtt{MADGRAPH5\_aMC@NLO}$ and interfaced with $\mathtt{PYTHIA8}$
for showering and hadronization, and the $\mathtt{CTEQ6L1}$ 
PDF~\cite{Pumplin:2002vw} set is taken.
Note that these samples are generated inclusively with an
additional hard jet to capture the bulk of the NLO corrections.
We then avoid the double counting problems
in our non-resonant background samples by applying the pre-selection
cuts listed in Eq.~(\ref{eq:presel}).
We have found that the resulting cross sections
for the non-resonant backgrounds, presented in Table I, agree with
those presented in Ref.~\cite{atlas_hh17}
within errors of less than 5\%.

Among them, as will be shown, the $b\bar{b} \gamma\gamma$ and
$b \bar b j \gamma$ samples give the dominant BG yields. 
In the latter, $j$ is faking $\gamma$.
The sub-dominant BGs come from the $c\bar{c} \gamma\gamma$,
$c\bar{c}j\gamma$, and $b\bar{b}jj$ processes with
$c$ faking $b$ and/or $j$ faking $\gamma$.
And the next sub-leading BG is from the $jj\gamma\gamma$ sample.
Here, one should be cautious about the $jj\gamma\gamma$ process 
because it receives contributions not only from the light hard quarks and gluons
but also from hard charm quarks. Schematically, one may write
\footnote{For our  $jj\gamma\gamma$ analysis,
first we have removed $c$ jet from a set of an additional hard jet.}
\begin{eqnarray}
\label{eq:jjaa}
jj\gamma\gamma &\simeq & \sum_{j_h^l,j_h,\cal S}
[1\oplus j_h^l]\otimes [j_hj_h\gamma\gamma] \otimes [1\oplus {\cal S}]
\nonumber \\[2mm]
&\simeq & \sum_{\{j_h^l\},\cal S}
\left\{
[1\oplus j_h^l]\otimes [c_h\bar c_h\gamma\gamma] \otimes [1\oplus {\cal S}]\right\}
\oplus\left\{
[1\oplus j_h^l]\otimes [j_h^lj_h^l\gamma\gamma] \otimes [1\oplus {\cal S}]\right\}\,.
\end{eqnarray}
In the first line, $j_h^l$ 
\footnote{
Here, $j_h^l$ denotes a light hard jet originating from
light $u$, $d$, and $s$ quarks and gluons.
Do not confuse it with $j_h$ which is for a hard jet 
originating not only from the light quarks and gluons
but also from $c$ quarks.}
in the first bracket denotes the additional light hard jet
and ${\cal S}$ in the last bracket is for jets generated 
during the showering process or
${\cal S}=j_s^l,j_s^lj_s^l, c_s\bar c_s, b_s\bar b_s$, etc with the 
subscript $s$ standing for showering jets.
In the second line, we use
$j_hj_h\simeq c_h\bar c_h\oplus j_h^l j_h^l$ with  the 
subscript $h$ standing for jets from hard scatterings.
We definitely see that the first part of Eq.~(\ref{eq:jjaa}) constitutes 
a part of the $c\bar c\gamma\gamma$ sample
and should be removed from the $jj\gamma\gamma$ sample
to avoid a double counting.
After removing it,
we find that the process with ${\cal S}=c_s\bar c_s$ dominates
the $jj\gamma\gamma$ BG with $c_s$ faking $b$.
Note that charm quarks should be treated separately from
the light quarks since
the $c$-quark fake rate $P_{c \rightarrow b}$ is much
larger than the light-jet fake rate of $P_{j \rightarrow b}=1/1300$.
Incidentally, we recall that $P_{j \rightarrow \gamma}=5\times10^{-4}$.
Finally, the $Z(b\bar{b})\gamma\gamma$ sample has the least
contribution to the non-resonant backgrounds.
In Table \ref{tab:FakeRate}, we are summarizing the main
fake processes and rates in each sample of backgrounds.

\begin{table}
\caption{
The main fake processes and the corresponding rates in each sample of
non-resonant and $t\bar t(\gamma)$ backgrounds.
We recall that $P_{j\to\gamma}=5\times10^{-4}$ and
$P_{e\rightarrow \gamma}=2\%$/$5\%$ in the barrel/endcap
calorimeter region.
For $c_s$ quarks produced during showering in the $jj\gamma\gamma$ sample,
we use $P_{c_s\to b}=1/8$ as in Ref.~\cite{atlas_hh17}.
Otherwise
the $P_T$ and $\eta$ dependence of $P_{c\to b}$ is fully considered
as explained in the text.
}\vspace{3mm}
\label{tab:FakeRate}
\begin{tabular}{ | c | c | c | c | }
\hline
Background(BG)  & Process & Fake Process & Fake rate \\
\hline
\multirow{7}{*}{Non-resonant} & $b\bar{b} \gamma\gamma$ & N/A & N/A    \\
                  &  $c\bar{c} \gamma\gamma$ & $c \rightarrow b$, $\bar{c} \rightarrow \bar{b}$ & $(P_{c\to b})^2$
\\
                  &  $jj\gamma\gamma$ & $c_s \rightarrow b$, $\bar{c_s} \rightarrow \bar{b}$ & $(P_{c_s\to b})^2$
\\
                  &  $b\bar{b}j\gamma$ & $j \rightarrow \gamma$ & $5 \times 10^{-4}$     \\
      BG           &  $c\bar{c}j\gamma$ & $c\rightarrow b$, $\bar{c} \rightarrow \bar{b}$,
$j \rightarrow \gamma$ & $(P_{c\to b})^2 \cdot (5\times 10^{-4})$    \\
                  &  $b\bar{b}jj$  & $j \rightarrow \gamma$, $j \rightarrow \gamma$ & $(5\times 10^{-4})^2$    \\
                  & $Z(\rightarrow b\bar{b})\gamma\gamma$ & N/A & N/A  \\ \hline
\multirow{2}{*}{$t\bar{t}$}
               & Leptonic decay &  $e \rightarrow \gamma$, $e \rightarrow \gamma$ &
$(0.02)^2$/$0.02\cdot 0.05$/$(0.05)^2$  \\
               & Semi-leptonic decay & $e \rightarrow \gamma$, $j \rightarrow \gamma$ &
$(0.02)\cdot 5\times 10^{-4}$/$(0.05)\cdot 5\times 10^{-4}$  \\
\hline
\multirow{2}{*} {$t\bar{t}\gamma$}
                &  Leptonic decay &  $e \rightarrow \gamma$ & $0.02$/$0.05$  \\
                & Semi-leptonic    &  $e \rightarrow \gamma$ &  $0.02$/$0.05$  \\ \hline
\end{tabular}
\end{table}

\subsubsection{$t\bar{t}$ and $t\bar{t}\gamma$ backgrounds}

The $t\bar{t}$ background is 
generated  at NLO in QCD using
{\fontfamily{qcr}\selectfont \textbf{POWHEG-BOX}},
and interfaced to $\mathtt{PYTHIA8}$ for parton showering and
hadronization. And for the PDF set, $\mathtt{CT10}$~\cite{Lai:2010vv} is taken.
Since it mimics the signal with an electron in the final state
faking a photon, we have required the final state should include at least 
$1$ lepton
\footnote{Here a lepton means $e$, $\mu$, or $\tau$.}.  
And the BG yield is normalized using the cross section calculated 
with $\mathtt{Top++2.0}$ program at NNLO in QCD 
which also includes soft-gluon resummation to NNLL~\cite{Czakon:2011xx}. 
Here we are taking $m_t=172.5$ GeV. 

A background with a similar size
comes from the $t\bar{t}$ production with one photon in
the final state.  The $t\bar{t}\gamma$ sample is generated at LO
in QCD with $\mathtt{MADGRAPH5\_aMC@NLO}$ and interfaced with
$\mathtt{PYTHIA8}$ for showering and hadronization. 
For $t\bar{t}\gamma$, we are taking the $\mathtt{CTEQ6L1}$ PDF set 
and the BG yield is normalized using the cross section
calculated in NLO in QCD~\cite{Melnikov:2011ta}.
Also, as in $t\bar t$, we require the final state to contain at least
$1$ lepton.
In Table \ref{tab:FakeRate}, we are summarizing the main
fake processes and rates also for 
the $t\bar{t}$ and $t\bar{t}\gamma$ backgrounds.

%
\begin{table}[t!]
\caption{Sequence of event selection criteria at the HL-LHC applied
in this analysis.}
\label{tab:selection}
\begin{center}
  \begin{tabular}{|c|l| }
  \hline
    Sequence &~ Event Selection Criteria at the HL-LHC\\
    \hline
    \hline
    1 &~ Di-photon trigger condition,
$\geq $ 2 isolated photons with $P_T > 25$ GeV, $|\eta| < 2.5$
\\
    \hline
    2 &~ $\geq $ 2 isolated photons with $P_T > 30$ GeV,
$|\eta| < 1.37$ or $1.52 < |\eta| <2.37$, $\Delta R_{j\gamma} > 0.4$ \\
    \hline
    3 &~ $\geq$ 2 jets identified as b-jets with leading(sub-leading) $P_T > 40(30)$ GeV, $|\eta|<2.4$ \\
    \hline
    4 &~ Events are required to contain $\le 5$ jets with
 $P_T >30$ GeV within $|\eta|<2.5$ \\
\hline
    5 &~ No isolated leptons with $P_T > 25$ GeV, $|\eta| <2.5$ \\
    \hline
    6 &~ $0.4 < \Delta R_{b \bar{b}} < 2.0$, $0.4 < \Delta R_{\gamma \gamma} < 2.0$ \\
    \hline
    7 &~ $122 < M_{\gamma\gamma}/{\rm GeV} < 128$ and $100 < M_{b \bar{b}}/{\rm GeV} < 150$ \\
    \hline
    8 &~ $P_T^{\gamma\gamma} > 80$ GeV, $P_T^{b  \bar{b}} > 80$ GeV \\
    \hline
    \end{tabular}
\end{center}
\end{table}
\begin{figure}[th!]
\centering
\includegraphics[width=3.2in]{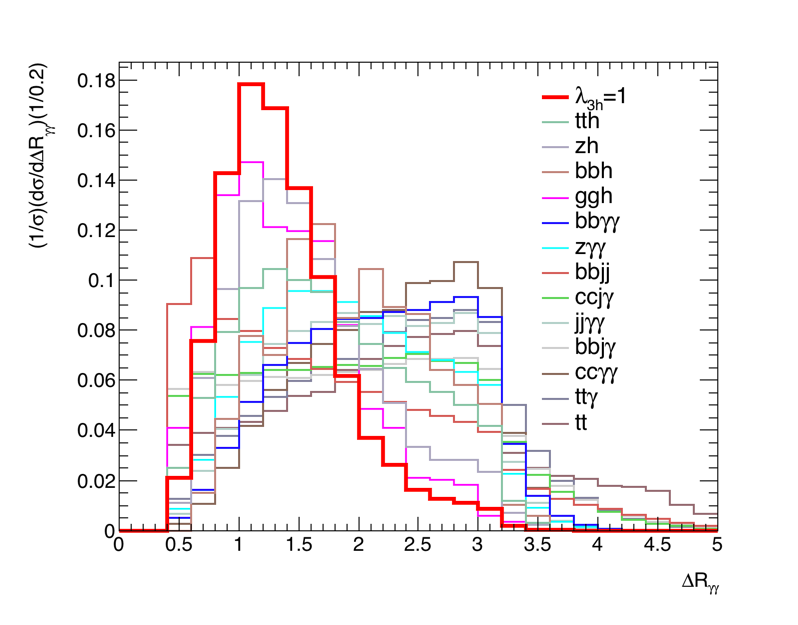}
\includegraphics[width=3.2in]{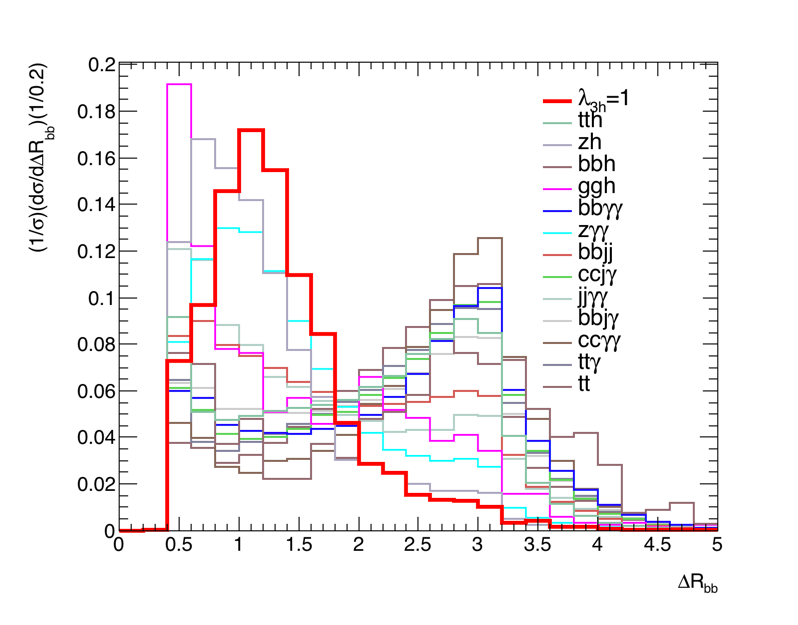}
\caption{\small \label{fig:distr1}
The $\Delta R_{\gamma\gamma}$ and $\Delta R_{b\bar b}$ distributions for the 
signal with $\lambda_{3H}=1$ and all the other backgrounds.
}
\end{figure}

\begin{figure}[ht!]
\centering
\includegraphics[width=3.2in]{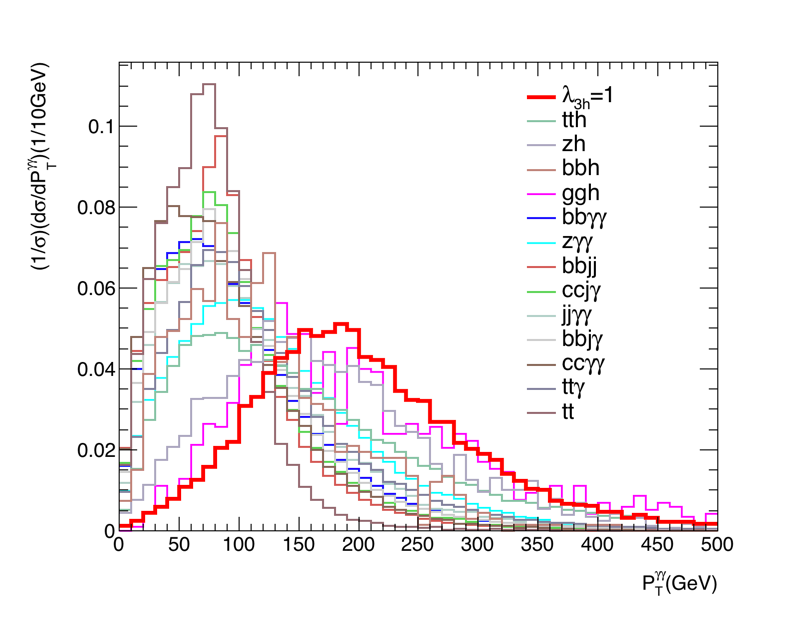}
\includegraphics[width=3.2in]{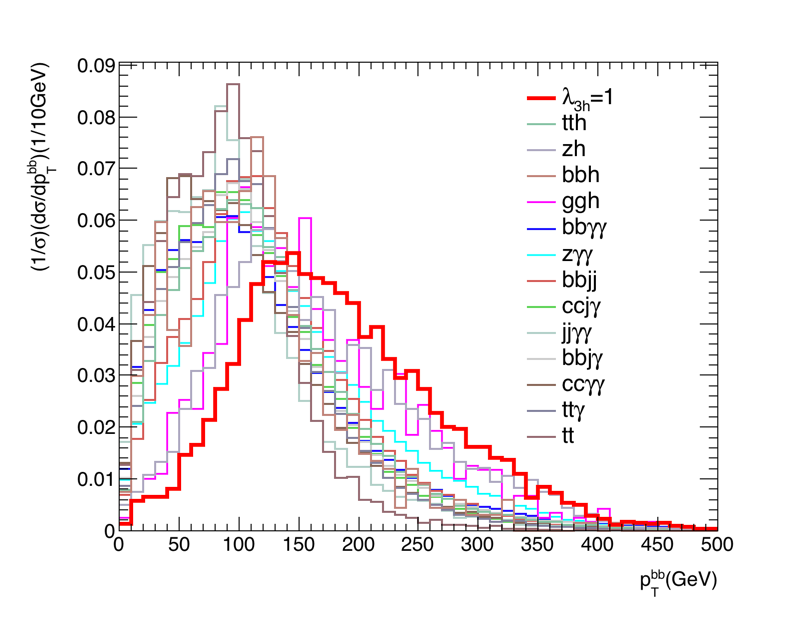}
\caption{\small \label{fig:distr2}
The transverse momentum distributions $P_T^{\gamma\gamma}$ and 
$P_T^{b\bar b}$ for the signal with $\lambda_{3H}=1$ and all the 
other backgrounds.
}
\end{figure}

\subsection{Event Selections}

A sequence of event selections is applied to the signal and background
samples. It is clearly listed in Table~\ref{tab:selection}.
We follow closely the steps reported in an ATLAS conference report 
\cite{atlas_hh17}.
The goal is to obtain a pair of isolated photons and a pair of 
isolated $b$ quarks.  Both pairs are reconstructed near the Higgs-boson mass.
In particular, the cuts $\Delta R_{\gamma\gamma} < 2.0$ and 
$\Delta R_{bb} < 2.0$ are imposed so as to 
suppress the main backgrounds which are more populated in the
regions of  $\Delta R_{\gamma\gamma\,, bb} > 2.0$, see Fig.~\ref{fig:distr1}
\footnote{For larger values of $|\lambda_{3H}|$, the cuts of
$\Delta R_{\gamma\gamma\,,bb} < 2.0$ remove more signal events
compared to the SM case, see the upper left frame of Fig.~\ref{appfig:distr1}
in Appendix A. 
This leads to the smaller efficiencies as shown in Table~\ref{tab:14TeVsig}
when $\lambda_{3H}=-4$, $6$, and $10$. 
We find that the different choices of $\Delta R_{\gamma\gamma\,,bb}$ cuts
hardly improve the signal significance and employ the same cuts
taken by the ATLAS group~\cite{atlas_hh17}.
}.
We show the angular separation between photons and 
that between $b$ jets
for all the backgrounds and the signal with $\lambda_{3H}=1$
in the left and right frame of Fig.~\ref{fig:distr1}, respectively.  
It is clear that the majority of the signal 
and a very few backgrounds lie in the region $\Delta R_{\gamma\gamma} < 2$
and $\Delta R_{bb} < 2$.
In Fig.~\ref{fig:distr2}, we show the transverse momentum distributions 
$P_T^{\gamma\gamma}$ and $P_T^{b\bar b}$ for the signal with $\lambda_{3H}=1$ 
and all the backgrounds. We observe the signal tends to have
larger transverse momentum. 
Distributions of $\Delta R_{\gamma\gamma}$
and $P_T^{\gamma\gamma}$ with other values of $\lambda_{3H}$ can be 
found in Appendix \ref{app-hl} where we also show  the
$\Delta R_{\gamma j}$ and $M_{\gamma\gamma b b}$ distributions.
The details of cuts are summarized in Table~\ref{tab:selection}.

All events passing the above selection criteria are classified into two
categories, depending on the pseudorapidities of the photons.
If both photons appear in the barrel region ($|\eta| < 1.37$)
the event is labeled as ``barrel-barrel'', otherwise it is labeled as
``other''.

\begin{sidewaystable}
\caption {Efficiencies ($\%$) and event yields ($\#$):
the signal cut flows for
Higgs-pair production at LHC 14 TeV with an integrated luminosity of
3000 fb$^{-1}$ for $\lambda_{3H}$=-4,0,1,2,6,10.
}\vspace{3mm}
\label{tab:14TeVsig} 
\centering
\begin{ruledtabular}
  \begin{tabular}{ l | c c| c c | c  c | c  c | c c |c  c  }
 $\lambda_{3H}$ & \multicolumn{2}{c|}{$-4$} & \multicolumn{2}{c|}{0}  &
  \multicolumn{2}{c|}{1}  & \multicolumn{2}{c|}{2}  &
  \multicolumn{2}{c|}{6}  &
   \multicolumn{2}{c}{10}  \\ \hline
 Cross section (fb) & \multicolumn{2}{c|}{1.45} & 
                      \multicolumn{2}{c|}{0.25} & 
                      \multicolumn{2}{c|}{0.12} & 
                      \multicolumn{2}{c|}{0.06} & 
                      \multicolumn{2}{c|}{0.48} & 
                      \multicolumn{2}{c}{1.97} \\ \hline
  Cuts & Eff.$\%$ & No.$\#$  & $\%$ & No.$\#$  & $\%$ 
       &            No.$\#$  & $\%$ & $\#$& $\%$ & $\#$& $\%$ & $\#$ \\  \hline
1. diphoton trigger & 23.15 & 1007 & 25.63 & 192 &  27.47 & 99 & 28.94 & 52
	& 20.50 & 295 &	21.01 &	1242 \\ \hline
2. $\ge 2$ isolated photons & 20.79 & 904 & 23.33 & 175 & 25.21	& 91 & 26.73	
& 48 & 17.82 & 257 & 18.38 & 1086 \\ \hline
3-1.  jet candidates & 14.58 & 634 &17.10 & 128 & 19.07 & 69 & 
20.85 & 38 & 11.62 & 167 & 12.14 & 717 \\ \hline
3-2 $\ge 2$ two b-jet & 4.61 & 200 & 5.65 & 42 & 6.46 & 23 & 7.26 & 13	&
3.34 & 48 &	3.55 & 210 \\ \hline
4.  no. of jets $\leq 5$ & 4.47 & 194 & 5.43	& 41 &
6.23 & 22 & 6.97 & 13 & 3.26 &
47 & 3.45 &	204	\\ \hline
5.  lepton veto & 4.41 & 192 & 5.36	& 40 & 
	6.15 & 22 & 6.88 & 12
 	& 3.22 & 46 & 3.41 & 202 \\ \hline
6. $\Delta R_{\gamma\gamma,bb}$ cut & 2.72 & 118 & 4.00	& 30
 & 4.98	& 18 & 5.87	& 11 & 1.19 & 
17 & 1.49 &	88 \\ \hline
7-1.  Higgs mass window $M_{\gamma\gamma}$  & 2.65 & 115 & 3.88	& 29 &
	4.82 & 17 & 5.64 &
 10	& 1.15 & 17	& 1.46 & 86	\\ \hline
7-2.  Higgs mass window $M_{bb}$ & 1.80 & 78 & 2.62	& 20 &
 3.20 &	12	& 3.82	&
	7 & 0.78 & 11 &	0.99 & 59 \\ \hline
8.  $p_{T_{\gamma\gamma}}, 
     p_{T_{bb}}$ cuts & 1.77 & 77 & 2.60 & 20 &
	3.17 & 11 &	3.79 & 7 &
	0.77 & 11 &	0.97 &	57	\\ \hline
 \hline
   other/barrel ratio & \multicolumn{2}{c|}{35.27 \%} & 
  \multicolumn{2}{c|}{36.08\%} &
\multicolumn{2}{c|}{33.89\%} &\multicolumn{2}{c|}{32.66\%} 
  &\multicolumn{2}{c|}{39.47\%}
  & \multicolumn{2}{c}{37.01\%} 
  \end{tabular}
\end{ruledtabular}
\end{sidewaystable}
\subsection{Cut Flows and  Efficiencies}
We follow closely the steps used in the ATLAS
conference note \cite{atlas_hh17}.
We compare the cut flow of our current analysis with ATLAS results
for the $\lambda_{3H}=1$ case, and they agree with each other within
about 5--15\%.
We show in Table~\ref{tab:14TeVsig} the efficiencies and event yields
for Higgs-pair production
in the channel $HH \to b\bar b \gamma \gamma$
at the HL-LHC  with an integrated luminosity of 3000 fb$^{-1}$
for various values of
$\lambda_{3H}=-4,0,1,2,6,10$.
In the last row, ``other/barrel ratio'' is the ratio of events
for the two photon candidates falling in the ``other'' region 
to those
in the ``barrel-barrel'' region, after applying all the event selection cuts.
The overall other/barrel ratios are all similar.

The overall signal efficiency has its peak value of 3.79 \%
at $\lambda_{3H} =2$ and it decreases when
$\lambda_{3H}$ deviates from $2$.
We observe that the overall efficiency drops
quickly when $\lambda_{3H}$ moves to a larger value
and becomes smaller than 1 \% when $\lambda_{3H}\gsim 4$.
While when $\lambda_{3H}$ becomes smaller and starts to take on
negative values, it decreases to 3.17 \% at the SM value of $\lambda_{3H}=1$
and reaches 1.77 \% at $\lambda_{3H}=-4$.
This is because of the strong destructive (constructive) interference between 
the
triangle and box diagrams for the positive (negative) values of $\lambda_{3H}$
and the enhancement of kinematical features of the triangle diagram for
$|\lambda_{3H}|>1$.
Thus, these two effects are combined to give strong dependence of
the $\Delta R_{\gamma\gamma\,,bb}$ distributions on $\lambda_{3H}$,
and therefore leading to the strong dependence of the signal efficiency on
$\lambda_{3H}$.
On the other hand, the number of signal events, which is given by the 
product of the
cross section, signal efficiency, and luminosity of 3000 fb$^{-1}$,
is only 7 at $\lambda_{3H}=2$ but it becomes 11
at the SM value of $\lambda_{3H}=1$.
Note that one may have the same number of signal events 
also at $\lambda_{3H}=6$.

The cut flow tables of all the backgrounds in terms of
efficiencies at the HL-LHC are presented in Appendix \ref{cutflow}.

\begin{table}[ht!]
\caption {HL-LHC yields: Expected number of signal
and background events
at the HL-LHC assuming $3000~ \mathrm{fb}^{-1}$.
We separate the backgrounds into three categories (See text).
The significance for $\lambda_{3H}=1$ (SM) is also shown,
see Eq.~(\ref{eq:Z}).
The combined significance is given by the
square root of the sum of the squares of the
``barrel-barrel'' and ``other'' significances.
} 
\label{tab:Result-HL_LHC}
\begin{center}
\begin{tabular}{|l|r|r|r|r||r|}
\hline
Expected yields $(3000~ \mathrm{fb}^{-1})$ &Total & Barrel-barrel & Other & Ratio
(O/B) & $\#$ of Gen. \\
Samples &  &  & (End-cap) & & Events \\
\hline
$H(b\,\bar{b})\,H(\gamma\,\gamma)$, $\lambda_{3H} = -4$ & $77.14 \pm 0.94$ &
$57.03 \pm 0.75$ &  $20.11 \pm 0.34$ & $0.35\pm 0.01$ & $3\times 10^5$\\
$H(b\,\bar{b})\,H(\gamma\,\gamma)$, $\lambda_{3H} = 0$ &  $19.50 \pm 0.20$ &
$14.33 \pm 0.16$ & $5.17 \pm 0.13$ & $0.36 \pm 0.01$ & $3\times 10^5$\\
$H(b\,\bar{b})\,H(\gamma\,\gamma)$, $\lambda_{3H} = 1$ & $11.42 \pm 0.082$ & $8.53
\pm 0.092$ & $2.89 \pm 0.048$ & $0.34 \pm 0.01$ & $3\times 10^5$\\
$H(b\,\bar{b})\,H(\gamma\,\gamma)$, $\lambda_{3H} = 2$ &
$6.82 \pm 0.05$ & $5.14 \pm 0.04$ & $1.68 \pm 0.03$ & $0.33 \pm 0.01$ & $3\times
10^5$ \\
$H(b\,\bar{b})\,H(\gamma\,\gamma)$, $\lambda_{3H} = 6$ & $11.03 \pm 0.21$ & $7.91
\pm 0.23$ & $3.12 \pm 0.10$ & $0.39 \pm 0.02$ & $3\times 10^5$ \\
$H(b\,\bar{b})\,H(\gamma\,\gamma)$, $\lambda_{3H} = 10$ & $57.46 \pm 1.01$ &
$41.94 \pm 0.60$ & $15.52 \pm 0.62$ & $0.37 \pm 0.02$ & $3\times 10^5$ \\
\hline
$gg\,H(\gamma\,\gamma)$ & $6.60 \pm 0.69$ & $4.50 \pm 0.71$ & $2.10 \pm 0.30$ &
$0.47 \pm 0.10$ & $6 \times 10^6$ \\
$t\,\bar{t}\,H(\gamma\,\gamma)$ & $13.21 \pm 0.23$ & $9.82 \pm 0.19$ & $3.39 \pm
0.17$ & $0.35 \pm 0.02$ & $10^6$ \\
$Z\,H(\gamma\,\gamma)$ & $3.62 \pm 0.16$ & $2.44 \pm 0.16$ & $1.18 \pm 0.08$ &
$0.48 \pm 0.05$ & $10^6$ \\
$b\,\bar{b}\,H(\gamma\,\gamma)$ & $0.15 \pm 0.024$ & $0.11 \pm 0.027$ & $0.04 \pm
0.014$ & $0.40 \pm 0.16$ & $10^6$ \\
\hline
$b\,\bar{b}\,\gamma\,\gamma$ & $18.86 \pm 0.9$ & $11.15 \pm 0.7$ & $7.71 \pm 0.5$
& $0.69 \pm 0.06$ & $1.1 \times 10^7$ \\
$c\,\bar{c}\,\gamma\,\gamma$ & $7.53 \pm 1.06$  & $4.79 \pm 1.10$  & $2.74 \pm
0.81$  & $0.57 \pm 0.21$ & $10^7$ \\
$j\,j\,\gamma\,\gamma$ & $3.34 \pm 0.46$ & $1.59 \pm 0.31$ & $1.75 \pm 0.32$  &
$1.10 \pm 0.29$ & $10^7$ \\
$b\,\bar{b}\,j\,\gamma$ & $18.77 \pm 1.00$ & $10.40 \pm 0.83$ & $8.37 \pm 0.63$ &
$0.80 \pm 0.09$ & $10^7$ \\
$c\,\bar{c}\,j\,\gamma$ & $5.52 \pm 1.4$ & $3.94 \pm 1.0$  & $1.58 \pm 0.6$  &
$0.40 \pm 0.18$ & $10^7$ \\
$b\,\bar{b}\,j\,j$ & $5.54 \pm 0.5$ & $3.81 \pm 0.3$ & $1.73 \pm 0.2$  & $0.45 \pm
0.06$ & $5 \times 10^6$ \\
$Z(b\,\bar{b})\,\gamma\gamma$ & $0.90 \pm 0.03$ & $0.54 \pm 0.02$ & $0.36 \pm
0.02$  & $0.67 \pm 0.04$ & $10^7$ \\
\hline
$t\,\bar{t}$~($\geq$ 1 leptons) & $4.98 \pm 0.23$ & $3.04 \pm 0.12$ & $1.94 \pm
0.21$ & $0.64 \pm 0.07$ & $10^7$ \\
$t\,\bar{t}\,\gamma$~($\geq$ 1 leptons) & $3.61 \pm 0.21$ & $2.29 \pm 0.15$ &
$1.32 \pm 0.15$ & $0.58 \pm 0.08$ & $10^7$ \\
\hline
Total Background & $92.63 \pm 2.5$ & $58.42 \pm 2.0$ & $34.21 \pm 1.4$  &  $0.59
\pm 0.03$  \\
\cline{1-5}
\cline{1-5}
\multicolumn{1}{|l|}{Significance $Z$} & \multicolumn{1}{r|}{$1.163$} &
\multicolumn{1}{r|}{$1.090$} & \multicolumn{1}{r|}{$0.487$}   \\ \cline{1-4}
\multicolumn{2}{|l|}{Combined significance}   & \multicolumn{2}{c|}{$1.194$ }
 \\ \cline{1-4}
\end{tabular}
\end{center}
\end{table}
\subsection{Analysis and Results}

Here we show the main results of our analysis in
Table~\ref{tab:Result-HL_LHC}
-- the resultant signal rates for various $\lambda_{3H}$ against
all the backgrounds.
The last column is for the number of generated events in each sample.
The statistical uncertainties originating from the limited number of
generated events are estimated by 
dividing each of the background and signal samples into roughly 
$O(10)$ subsamples. The fluctuations among the subsamples are
then taken as the uncertainty of the sample.
We have made detailed comparisons with the results from ATLAS~\cite{atlas_hh17}.
In
general we agree, except for $ggH$ and $t\bar t$.
In the $ggH$ sample, we figure out that about half of the disagreement
is caused by the differences in $b$-tagging algorithm and detector simulations.
While, for the $t\bar t$ sample, our estimation is made based on
the \textbf{Delphes3} algorithm for electron reconstruction and
identification which is about 20 times more
efficient than that taken by ATLAS.

More precisely, in the $ggH$ sample,
our number of the $ggH$ background is $6.60$ which is $2.4$ times larger than
the ATLAS number of $2.74$~\cite{atlas_hh17}. By noting that
the $ggH$ sample is dominated by $b$ quarks from showering processes,
we find a part of the difference can be attributed to
different $b$-tagging algorithm taken in our work:
ours is from \cite{atlas_perform} while, in the ATLAS paper, the
algorithm from \cite{atlas_2014} is used. If we use the same algorithm
taken in the ATLAS paper, we find the number of background reduces to about
$5$ which is still a bit above the ATLAS number $2.74$.
We also find another reason in the detector simulations but, again, it is not
enough to fully explain the difference.
Indeed, the similar observation has been recently made
by the authors of Ref.~\cite{Homiller:2018dgu}.
When they used the same selection cuts as ATLAS, the result was also larger than
the ATLAS result, but consistent with ours.
 
We note that the kinematic distributions for the signal with different
$\lambda_{3H}$ would not be very different, as seen by the
ratio $(O/B)$ in the last of Table~\ref{tab:Result-HL_LHC}, which
are more or less the same for different $\lambda_{3H}$.
On the other hand, the ratio $(O/B)$ for the backgrounds, on average,
is larger than the signal, which means the backgrounds are 
in general more forward.
We further note that the combined significance obtained by splitting events
into two categories of barrel-barrel and other is improved by 
3 \% over the total one when $\lambda_{3H}=1$.
For our analysis, we use the combined significance.

The most dominant one in the single-Higgs associated backgrounds is
$t\bar t H$ followed by $ggH$.
The single-Higgs associated processes
contribute about 23 events to the total background.
Meanwhile the dominant ones in non-resonant backgrounds are
$b\bar b \gamma\gamma$ and $b\bar b j \gamma$ with each of it
contributing 19 events to the total background.
A similar size of background comes from combined $c\bar c \gamma\gamma 
\oplus c\bar cj\gamma \oplus jj\gamma\gamma$, in which
either hard or showering $c$ quarks are faking $b$ jets basically.
Among the non-resonant backgrounds, $b\bar b jj$ contributes the least.  
Including $t\bar t$ and $t\bar t\gamma$ in which one or two electrons are faking
photons, we note that more than
one half of the total background is due to fakes.

\begin{figure}[t!]
\centering
\includegraphics[width=3.2in]{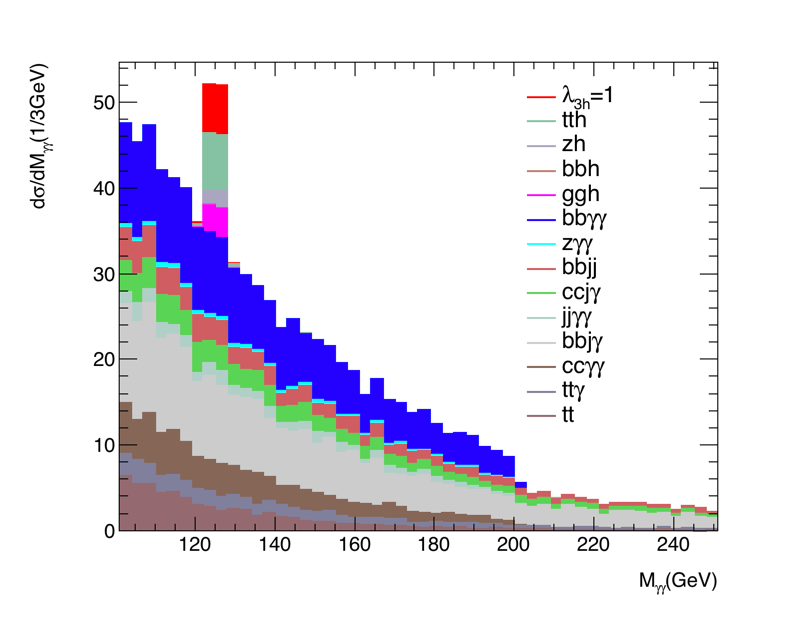}
\includegraphics[width=3.2in]{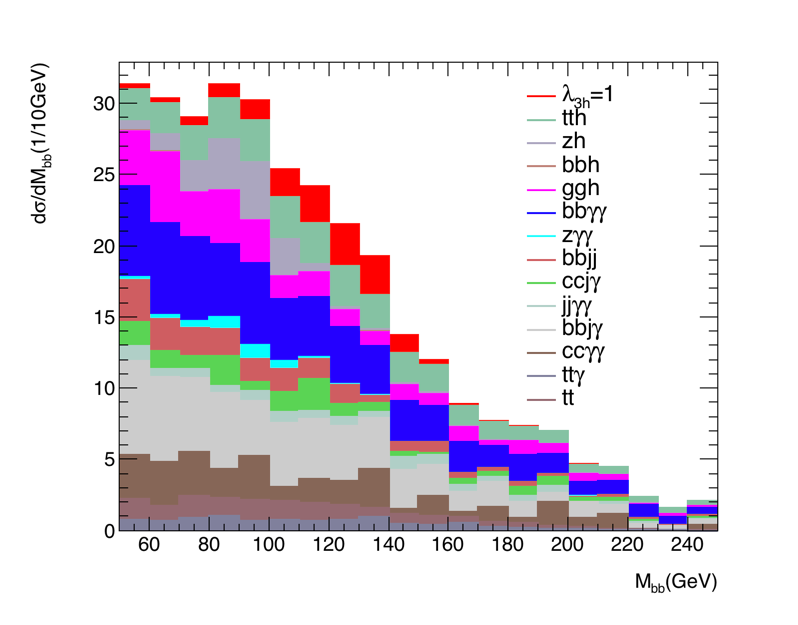}
\caption{\small \label{fig:inv-mass}
The $M_{\gamma\gamma}$ (upper) and $M_{b\bar b}$ (lower) distributions 
for the signal on top of the backgrounds at the HL-LHC.
}
\end{figure}
In Fig.~\ref{fig:inv-mass}, we show the resultant invariant-mass distributions
of the two photon (upper) and two $b$ (lower) candidates for the signal
on top of all the backgrounds.
We have applied all the selection cuts except for 
the cut on $M_{\gamma\gamma}$ ($M_{b\bar b}$) in 
the upper (lower) frame.
The photon peak $M_{\gamma\gamma} \sim 125$ GeV is very clear while
that of $M_{b\bar b} \sim 125$ GeV is rather broad, due to the $b$-jet
resolution.

%
\begin{figure}[t!]
\centering
\includegraphics[width=4.5in]{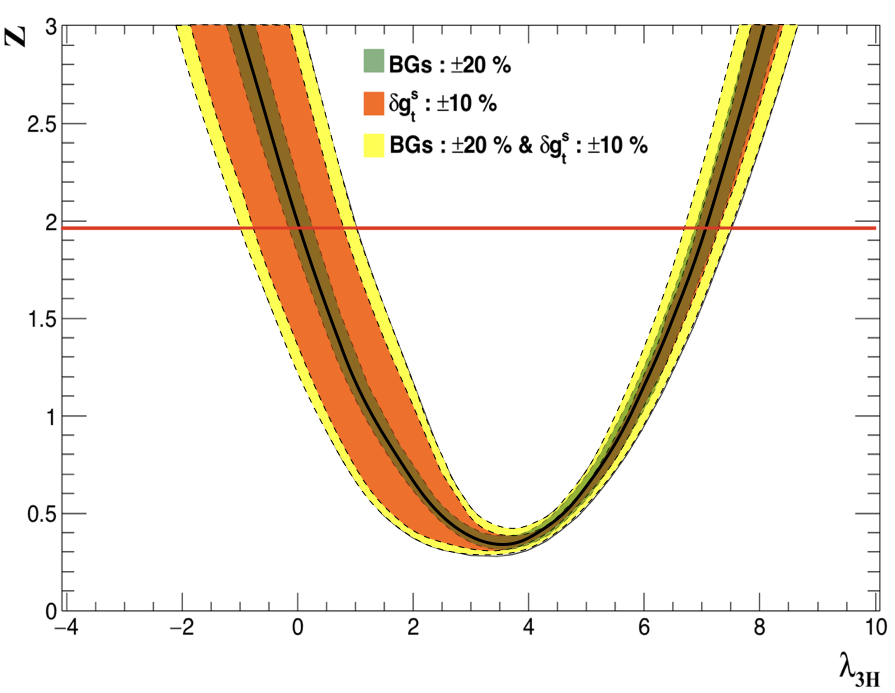}
\caption{\small \label{fig:signif}
{\bf HL-LHC}:
Significance of the signal over the background versus
$\lambda_{3H}$. 
The orange and green bands represents the impact of
the uncertainties associated with the top-Yukawa coupling
and the estimation of backgrounds, respectively, and
the yellow one the impact of both of the uncertainties.
The black solid line is for the case when 
$g_t^S=1$ and $b=92.63$, see Table~\ref{tab:Result-HL_LHC}.
}
\end{figure}
In Fig.~\ref{fig:signif}, we show the significance defined by
\begin{equation}
\label{eq:Z}
 Z = \sqrt{ 2 \cdot \left[ \left( (s+b) \cdot \ln( 1+ s/b) - s \right)
  \right ] }
\end{equation}
where $s$ and $b$ represent the numbers of signal and background events, respectively.
The central curve is for the case when the top-Yukawa coupling takes
on the SM value of $g_t^S=1$ and $b=92.63$, see Table~\ref{tab:Result-HL_LHC}.
The orange and green bands have been obtained by
varying the top-Yukawa coupling by 10 \% 
\footnote{In our work, we also take account of the effect of the 10 \% uncertainty of the
top-Yukawa coupling on the $ggH$ and $t\overline{t}H $ backgrounds
while neglecting its effect on the Higgs decay mode into two photons
since it is dominated by the $W$ loops.  Incidentally, we have taken the SM values for
the Higgs couplings to $b$ quarks and $W$ bosons for $H\rightarrow\gamma\gamma$.}
($|\delta g_t^S| \leq 0.1$)
and the total background yield by 20 \% ($|\delta b/b| \leq  0.2$), respectively.
The yellow band has been obtained by considering 
both of the uncertainties simultaneously.
The uncertainty associated with the estimation of backgrounds
may arise from pile-up, the photon and $b$-tagging efficiencies,
several fake rates, the choices of renormalization and factorizations scales and PDF, etc.
We note that the $\delta g_t^S$ effect becomes larger when 
$\lambda_{3H}$ decreases from $3.5$.
For $\lambda_{3H}\gsim 3.5$, 
the $\delta b$ effect could be comparably important.
Given all the uncertainties can be minimized and the top-Yukawa at
the SM value, the 95\% CL sensitivity region for $\lambda_{3H}$ is
$0 < \lambda_{3H} < 7.1$.
However, given the worst uncertainties with $\delta g_t^S = \pm 0.1$ and
$\delta b/b = \pm 0.2$, the sensitivity range widens to 
$ -1.0 < \lambda_{3H} < 7.6$.
We note that the lower boundary of
the 95\% CL region of $\lambda_{3H}$ is  
sensitive to the top-Yukawa $g_t^S$ while the impact of the uncertainty
associated with the estimation of backgrounds turns out minor
upon the 20 \% variation over the total background.

\begin{figure}[th!]
\centering
\includegraphics[width=4.5in]{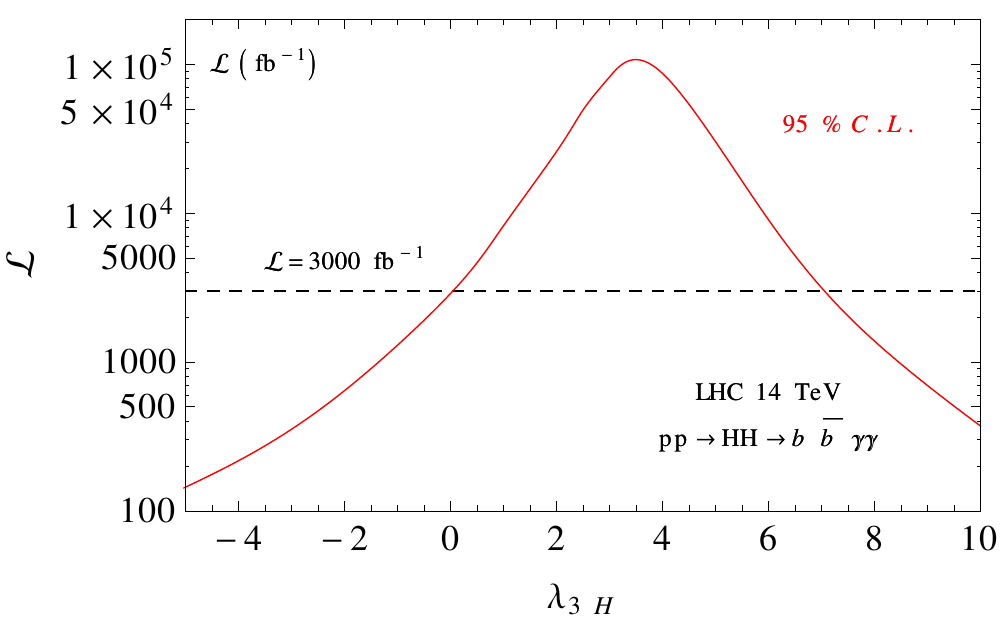}
\caption{\small \label{fig:lumin}
{\bf HL-LHC}:
Required luminosity for 95\% CL sensitivity at the 14 TeV HL-LHC
versus $\lambda_{3H}$. Here we assume
that the top-Yukawa coupling takes the SM value.
}
\end{figure}
Finally, we show in Fig.~\ref{fig:lumin} the luminosity required to 
achieve 95\% CL sensitivity versus $\lambda_{3H}$. 
We observe that the SM value of $\lambda_{3H}=1$ can
only be established with about 8.5 ab$^{-1}$ luminosity.
Note that the required luminosity peaks at $\lambda_{3H} \simeq 3.5$
while the $gg\to HH$ production takes its smallest value at 
$\lambda_{3H} \simeq 2.5$, see Fig.~\ref{fig:total-cross}.
This is because of 
the strong dependence of the signal efficiency on $\lambda_{3H}$
induced by the substantial interference between the triangle and box diagrams 
together with, especially for $|\lambda_{3H}|>1$, 
the enhancement of kinematical features of the triangle diagram 
or the smaller Higgs-pair invariant mass of $M_{\gamma\gamma b b}$,
the wider angular separations of $\Delta R_{\gamma\gamma\,,bb}$,
and the smaller transverse momenta of $P_T^{\gamma\gamma\,,bb}$.

\newpage

\section{Simulations, Event Selections, and Analysis at the
HL-100 TeV Collider}

In this section, through the $HH \to b\bar b \gamma\gamma$ channel,
we estimate how well one can measure the $\lambda_{3H}$ coupling
at a 100 TeV hadron collider assuming a nominal luminosity of 3 ab$^{-1}$
or at the HL-100 TeV hadron collider.
We basically follow the procedures that we took in the last 
section for the 14 TeV HL-LHC case, though some selection cuts may be changed because of 
the much higher center-of-mass energy. 
We have taken a crude estimate projected from the current LHC detectors
for the $P_T$ and $\eta$ coverage for jets, leptons, and photons
without any specific detector designs available for the 100 TeV 
hadron collider.

\subsection{Parton-level event generations and detector simulations}

\begin{table}[th!]
\centering
\caption{\small
The same as in Table \ref{tab:ParticleList} but for a 100 TeV 
hadron collider.  In the row for
$b\bar b H(\to\gamma\gamma)$, 5FS stands for the 5-flavor scheme.
Note that, except the $ggH(\to\gamma\gamma)$ 
background which is generated at NLO, all the signal and
backgrounds are generated at LO and normalized to the
cross sections computed at the accuracy denoted in `Order in QCD'.
}
\label{tab:ParticleList_100TeV}
\begin{ruledtabular}
\begin{tabular}{  c  c  c  c  c  c }
\multicolumn{6}{c}{Signal} \\
\hline
\multicolumn{2}{c}{Signal process} & Generator/Parton Shower &
$\sigma \cdot BR$ [fb] & Order  & PDF used  \\
&&&& in QCD &   \\
\hline
\multicolumn{2}{c}{$gg \to HH \to b\bar b \gamma\gamma$
\cite{Contino:2016spe} } &
$\mathtt{MG5\_aMC@NLO}$/$\mathtt{PYTHIA8}$ & 4.62
 & NNLO & NNPDF2.3LO \\
 &&& & \!\!\!\!\!$+$NNLL  &  \\
\hline
\hline
\multicolumn{6}{c}{Backgrounds} \\
\hline
Background(BG)  & Process  & Generator/Parton Shower & $\sigma\cdot BR$~[fb] &  Order  &
PDF used\\
&&&& in QCD&   \\
\hline
\multirow{4}{*}{}
& $ggH(\rightarrow \gamma\gamma)$ \cite{Contino:2016spe} &  $\mathtt{POWHEG-BOX}$/$\mathtt{PYTHIA8}$
  & $1.82 \times 10^{3}$ & $\mathrm{NNNLO}$ & $\mathtt{CT10}$ \\ \cline{2-5}
Single-Higgs  & $t \bar{t} H(\rightarrow \gamma\gamma)$ \cite{Contino:2016spe} &
$\mathtt{PYTHIA8}$/$\mathtt{PYTHIA8}$  & $7.29\times 10^1$ & NLO & \\ \cline{2-5}
 associated BG             &  $ZH(\rightarrow \gamma\gamma)$  \cite{Contino:2016spe} &
$\mathtt{PYTHIA8}$/$\mathtt{PYTHIA8}$   & $2.54\times 10^1$ & NNLO &\\ \cline{2-5}
                  & $b\bar{b}H(\rightarrow \gamma\gamma)$ \cite{bbH_twiki} &
$\mathtt{PYTHIA8}$/$\mathtt{PYTHIA8}$ & $1.96\times 10^1$ & NNLO(5FS)  &\\ \hline
\multirow{7}{*}{Non-resonant BG} & $b\bar{b} \gamma\gamma$ & $\mathtt{MG5\_aMC@NLO}$/$\mathtt{PYTHIA8}$ & $4.93 \times 10^3$  &  LO & $\mathtt{CTEQ6L1}$ \\ \cline{2-5}
                  &  $c\bar{c} \gamma\gamma$ & $\mathtt{MG5\_aMC@NLO}$/$\mathtt{PYTHIA8}$ & $4.54 \times 10^4$ & LO &  \\ \cline{2-5}
                  &  $jj\gamma\gamma$ & $\mathtt{MG5\_aMC@NLO}$/$\mathtt{PYTHIA8}$ & $5.38 \times 10^5$ & LO &  \\ \cline{2-5}
                  &  $b\bar{b}j\gamma$ & $\mathtt{MG5\_aMC@NLO}$/$\mathtt{PYTHIA8}$ & $1.44 \times 10^7$ & LO &  \\ \cline{2-5}
                  &  $c\bar{c}j\gamma$ & $\mathtt{MG5\_aMC@NLO}$/$\mathtt{PYTHIA8}$ & $4.20 \times 10^7$ & LO &  \\ \cline{2-5}
                  &  $b\bar{b}jj$  & $\mathtt{MG5\_aMC@NLO}$/$\mathtt{PYTHIA8}$ &$1.60 \times 10^{10}$ & LO &  \\ \cline{2-5}
                  & $Z(\rightarrow b\bar{b})\gamma\gamma$ &
$\mathtt{MG5\_aMC@NLO}$/$\mathtt{PYTHIA8}$ & $9.53\times 10^1$ & LO & \\ \hline
\multirow{2}{*}{$t\bar{t}$ and $t\bar{t}\gamma$ BG~\cite{Contino:2016spe}} 
& $t\bar{t}$ & $\mathtt{MG5\_aMC@NLO}$/$\mathtt{PYTHIA8}$  &
$1.76 \times 10^7$ & NLO &  $\mathtt{CT10}$  \\ 
\cline{2-6}
  ($\geq 1$ lepton)   & $t\bar{t}\gamma$
 & $\mathtt{MG5\_aMC@NLO}$/$\mathtt{PYTHIA8}$  &
$4.18 \times 10^4$ & NLO & $\mathtt{CTEQ6L1}$
\end{tabular}
\end{ruledtabular}
\end{table}

The same signal and backgrounds are considered as in the 14 TeV case.
The Monte Carlo generators, the cross sections, and the orders of QCD calculation
are shown in Table~\ref{tab:ParticleList_100TeV}.
Note that, for some backgrounds, the orders in QCD 
are different compared to the 14 TeV case.
Otherwise, the calculational methods taken for the signal and background samples
are essentially the same as 
those what we employed for the HL-LHC.

On the other hand, pre-selection cuts, detector energy resolutions, and
tagging efficiencies and fake rates may undergo significant changes
because of different designs and projected performance of the detectors 
in the future. 
Below, we describe in detail what we use in our analysis.
\begin{itemize}
\item Pre-selection cuts, which are imposed in order to avoid 
any divergence in the parton- level calculations, are modified 
as follows to match the wider $\eta$ coverage of 
future particle detectors:
\begin{eqnarray}
P_{T_j} > 20\  \text{ GeV},\ P_{T_b} > 20\  \text{ GeV},\ P_{T_\gamma} > 25\  \text{ GeV},\ P_{T_l} > 10\  \text{ GeV}, \nonumber\\
 |\eta_j|<6,\ |\eta_\gamma|<6,\ |\eta_l|<6,\ \Delta R_{jj,ll,\gamma\gamma,\gamma j,jl, \gamma l} > 0.4, \nonumber\\
M_{jj} > 25\ \text{GeV},\ M_{bb} > 45\ \text{GeV},\ 60<M_{\gamma\gamma} < 200\ \text{GeV}.  \nonumber
\end{eqnarray}
\item Fast detector simulation and analysis at the HL-100 TeV hadron collider 
are
performed using {\fontfamily{qcr}\selectfont \textbf{Delphes3}}
\cite{deFavereau:2013fsa} with the FCChh template. 
For the energy resolution of the detector, we have chosen 
the ``Medium" detector performance for 
ECAL and HCAL~\cite{Contino:2016spe}
\footnote{In Ref.~\cite{Contino:2016spe}, three scenarios of
ECAL and HCAL performance are considered: ``Low'', ``Medium'', and ``High''.}
because we could get the best significance for this choice.
In the ``Medium" performance scenario, the ECAL energy resolution is given by
$$
\left.\Delta E/E\right|_{\rm ECAL} = \sqrt{0.01^2+0.1^2\, {\rm GeV}/E}
$$
and the HCAL energy resolution by
$$
\left.\Delta E/E\right|_{\rm HCAL} = \left\{
\begin{array}{lll}
\sqrt{0.03^2+0.5^2\, {\rm GeV}/E} & {\rm for} & |\eta|\leq 4 \,, \\[2mm]
\sqrt{0.05^2+1.0^2\, {\rm GeV}/E} & {\rm for} & 4< |\eta|\leq 6 \,. \end{array} \right.
$$
Further we set the magnetic field 6 T and the jet energy scale of $1.135$ is taken
to get the 
correct peak position at $M_H$ in the invariant mass distribution of the 
$b$-quark pair in the signal process.

\item For the $b$-jet tagging efficiency and related jet fake rates,
we are taking  $\epsilon_b=75$ \%, $P_{c\to b}=10$ \%, 
and $P_{j\to b}=1$ \%~\cite{Contino:2016spe}.

\item For the photon efficiency and jet fake rate, we are taking:
$\epsilon_\gamma=95\, \%\, (|\eta_\gamma|\leq 1.5)$, 
$90\, \%\ (1.5<|\eta_\gamma|\leq 4)$, 
$80\, \%\ (4<|\eta_\gamma|\leq 6)$, and
$P_{j\to \gamma}=1.35\times 10^{-3}$~\cite{Contino:2016spe}.
For the $e \rightarrow \gamma$ fake rate,
with a separation between the barrel and endcap regions at $|\eta| = 2$,
we take $P_{e\rightarrow \gamma } = 2\,\%\,(5\,\%)$ in the  
barrel (endcap) region as a reference~\cite{atlas_hh17} .

\end{itemize}

\subsection{Signal Event Samples}

The signal event samples are generated in exactly the same way as in 
the HL-LHC case.
We show the production cross section times the branching ratio at the 100
TeV $pp$ collider for six selected values of $\lambda_{3H} =
-4,0,1,2,6,10$ in Table~\ref{tab:x-cross-100}.

\begin{table}[ht!]
\caption{\small 
Production cross section times the branching ratio 
$\sigma \cdot BR(HH \to b\bar b \gamma\gamma)$ at the 100 TeV $pp$ collider.
}\vspace{3mm}
\label{tab:x-cross-100}
\centering
\begin{ruledtabular}
\begin{tabular}{c|cccccc}
  $\lambda_{3H}$& -4 & 0 & 1 & 2 & 6 & 10 \\
\hline
$\sigma\cdot BR(HH \to b\bar b \gamma\gamma)$ [fb] & 
46.97 & 8.99 & 4.62 & 2.32 & 13.61 & 57.78  \\
\end{tabular}
\end{ruledtabular}
\end{table}

\subsection{Background Samples}
As in the HL-LHC case, we categorize the backgrounds into
single-Higgs associated backgrounds,
non-resonant backgrounds, and
$t\bar t$ and $t\bar t\gamma$ backgrounds. 
The information is summarized in Table~ \ref{tab:ParticleList_100TeV}.
Note that the $t\bar t$ sample is generated with 
$\mathtt{MADGRAPH5\_aMC@NLO}$, and
for showering, hadronization and decays of unstable particles
only $\mathtt{PYTHIA8}$ is used
\footnote{Note $\mathtt{PYTHIA6}$ is used
for the $ggH(\to\gamma\gamma)$ process at the HL-LHC.}.
Otherwise, the descriptions of the backgrounds are the same
as in the HL-LHC case.

The cross sections increase as we move from 14 TeV to 100 TeV. 
The signal cross section increases by a factor of about $40$.
The cross section for the single-Higgs associated backgrounds
increases by a factor of about $15$ except $t\bar t H(\to\gamma\gamma)$:
the increment factor for the $t\bar t H(\to\gamma\gamma)$ process 
is about $50$.
The cross section for the $Z(\to b\bar b)\gamma\gamma$ process
increases by a factor of about
$20$ while the increment factor of the other non-resonant backgrounds is about $40$.
The cross sections for the $t\bar t$ related backgrounds
increase by about $30$ times.
As we will show, the non-resonant backgrounds constitutes
more than 75 \% of the total backgrounds. Roughly, 
the cross sections for the signal and dominant background processes
increase by a factor of about $40$.
Finally, in Table~\ref{tab:FakeRate_100TeV},
we summarize the faking rates of non-resonant and $t\bar t$-related
backgrounds which we use for the HL-100 TeV collider.
\begin{table}[ht]
\caption{\small 
The main fake processes and the corresponding faking rates in each sample of
non-resonant and $t\bar t(\gamma)$ backgrounds.
We recall that $P_{j\to\gamma}=1.35\times10^{-3}$, 
$P_{c\to b}=P_{c_s\to b}=0.1$ \cite{Contino:2016spe} and
$P_{e\rightarrow \gamma}=2\%$/$5\%$ in the barrel/endcap calorimeter region.
}\vspace{3mm}
\label{tab:FakeRate_100TeV}
\begin{tabular}{ | c | c | c | c | }
\hline
Background(BG)  & Process & Fake Process & Fake rate \\
\hline
\multirow{7}{*}{Non-resonant} & $b\bar{b} \gamma\gamma$ & N/A & N/A    \\
                  &  $c\bar{c} \gamma\gamma$ & $c \rightarrow b$, $\bar{c} \rightarrow \bar{b}$ & $(0.1)^2$
\\
                  &  $jj\gamma\gamma$ & $c_s \rightarrow b$, $\bar{c}_s \rightarrow \bar{b}$ & $(0.1)^2$
\\
                  &  $b\bar{b}j\gamma$ & $j \rightarrow \gamma$ & $1.35 \times 10^{-3}$     \\
      BG           &  $c\bar{c}j\gamma$ & $c\rightarrow b$, $\bar{c} \rightarrow \bar{b}$,
$j \rightarrow \gamma$ & $(0.1)^2 \cdot (1.35 \times 10^{-3})$    \\
                  &  $b\bar{b}jj$  & $j \rightarrow \gamma$, $j \rightarrow \gamma$ & $(1.35\times 10^{-3})^2$    \\
                  & $Z(\rightarrow b\bar{b})\gamma\gamma$ & N/A & N/A  \\ \hline
\multirow{2}{*}{$t\bar{t}$}
               & Leptonic decay &  $e \rightarrow \gamma$, $e \rightarrow \gamma$ &
$(0.02)^2$/$0.02\cdot 0.05$/$(0.05)^2$  \\
               & Semi-leptonic decay & $e \rightarrow \gamma$, $j \rightarrow \gamma$ &
$(0.02)\cdot 1.35\times 10^{-3}$/$(0.05)\cdot 1.35\times 10^{-3}$  \\
\hline
\multirow{2}{*} {$t\bar{t}\gamma$}
                &  Leptonic decay &  $e \rightarrow \gamma$ & $0.02$/$0.05$  \\
                & Semi-leptonic    &  $e \rightarrow \gamma$ &  $0.02$/$0.05$  \\ \hline
\end{tabular}
\end{table}

%
\begin{table}[t!]
\caption{\small 
Sequence of event selection criteria at the HL-100 TeV hadron collider 
applied in this analysis.}
\label{tab:selection100}
\begin{center}
  \begin{tabular}{|c|l| }
  \hline
    Sequence &~ Event Selection Criteria at the HL-100 TeV hadron collider\\
    \hline
    \hline
    1 &~ Di-photon trigger condition,
$\geq $ 2 isolated photons with $P_T > 30$ GeV, $|\eta| < 5$
\\
    \hline
    2 &~ $\geq $ 2 isolated photons with $P_T > 40$ GeV,
$|\eta| < 3$, $\Delta R_{j\gamma} > 0.4$ \\
    \hline
    3 &~ $\geq$ 2 jets identified as b-jets with leading(sub-leading) $P_T > 50(40)$ GeV, $|\eta|<3$ \\
    \hline
    4 &~ Events are required to contain $\le 5$ jets with
 $P_T >40$ GeV within $|\eta|<5$ \\
\hline
    5 &~ No isolated leptons with $P_T > 40$ GeV, $|\eta| <3$ \\
    \hline
    6 &~ $0.4 < \Delta R_{b \bar{b}} < 3.0$, $0.4 < \Delta R_{\gamma \gamma} < 3.0$ \\
    \hline
    7 &~ $122.5 < M_{\gamma\gamma}/{\rm GeV} < 127.5$ and $90 < M_{b \bar{b}}/{\rm GeV} < 150$ \\
    \hline
    8 &~ $P_T^{\gamma\gamma} > 100$ GeV, $P_T^{b  \bar{b}} > 100$ GeV \\
    \hline
    \end{tabular}
\end{center}
\end{table}
\subsection{Event Selections}
A sequence of event selections is applied to the signal and background
samples, see Table~\ref{tab:selection100}.  
We basically follow our HL-LHC analysis but using more relaxed $\Delta R$
condition to inclusively cover the broad range of $\lambda_{3H}$ 
still allowed after the HL-LHC era.
Also considered are the wider $|\eta|$ coverage at 100 TeV
and the more energetic jets and photons.

The distributions in $\Delta R_{\gamma\gamma}$, $\Delta R_{bb}$, 
$P_T^{\gamma\gamma}$, $P_T^{bb}$, $\Delta R_{\gamma b}$, and $M_{\gamma\gamma bb}$
are very similar to the case of HL-LHC. We collect some of them in appendix
\ref{app-hl} in order not to interrupt smooth reading of the main text.

\begin{sidewaystable}
\caption {The same as in Table \ref{tab:14TeVsig}  but
at the 100 TeV hadron collider with an integrated luminosity of
3 ab$^{-1}$. 
}\vspace{3mm}
\label{tab:100-signal} 
\centering
\begin{ruledtabular}
  \begin{tabular}{ l | c c| c c | c  c | c  c | c c |c  c  }
 $\lambda_{3H}$ & \multicolumn{2}{c|}{$-4$} & \multicolumn{2}{c|}{0}  &
  \multicolumn{2}{c|}{1}  & \multicolumn{2}{c|}{2}  &
  \multicolumn{2}{c|}{6}  &
   \multicolumn{2}{c}{10}  \\ \hline
 Cross section (fb) & \multicolumn{2}{c|}{46.97} & 
                      \multicolumn{2}{c|}{8.99} & 
                      \multicolumn{2}{c|}{4.62} & 
                      \multicolumn{2}{c|}{2.32} & 
                      \multicolumn{2}{c|}{13.61} & 
                      \multicolumn{2}{c}{57.78} \\ \hline
  Cuts & Eff.$\%$ & No.$\#$  & $\%$ & No.$\#$  & $\%$ 
       &            No.$\#$  & $\%$ & $\#$& $\%$ & $\#$& $\%$ & $\#$ \\  \hline
1. diphoton trigger & 56.06 & 78988 & 57.78 & 15582 &  58.99 & 8176 & 60.00 & 4176
	& 53.44 & 21818 &	53.82 &	93293 \\ \hline
2. $\ge 2$ isolated photons & 36.31 & 51158 & 39.21 & 10575 & 41.29	& 5722 & 43.40	
& 3021 & 32.39 & 13225 & 32.94 & 57105 \\ \hline
3-1.  jet candidates & 29.07 & 40965 & 32.77 & 8838 & 35.36 & 4901 & 
37.94 & 2641 & 23.87 & 9746 & 24.74 & 42881 \\ \hline
3-2 $\ge 2$ two b-jet & 9.57 & 13492 & 11.41 & 3076 & 12.75 & 1767 & 14.18 & 987	&
7.31 & 2986 &	7.65 & 13252 \\ \hline
4.  no. of jets $\leq 5$ & 9.03 & 12724 & 10.60	& 2860 &
11.79 & 1634 & 13.04 & 907 & 6.99 &
2856 & 7.29 & 12638	\\ \hline
5.  lepton veto & 9.03 & 12724 & 10.60	& 2860 & 
	11.79 & 1634 & 13.04 & 907
 	& 6.99 & 2856 & 7.29 & 12637 \\ \hline
6. $\Delta R_{\gamma\gamma,bb}$ cut & 8.32 & 11730 & 10.08	& 2718
 & 11.34	& 1572 & 12.57	& 875 & 5.92 & 
2419 & 6.39 &	11023 \\ \hline
7-1.  Higgs mass window $M_{\gamma\gamma}$  & 7.78 & 10968 & 9.35	& 2523 &
	10.51 & 1456 & 11.57 &
 805	& 5.55 & 2268 & 5.97 & 10341	\\ \hline
7-2.  Higgs mass window $M_{bb}$ & 6.14 & 8650 & 7.32	& 1974 &
 8.23 &	1140	& 9.08	&
	632 & 4.48 & 1830 &	4.77 & 8264 \\ \hline
8.  $p_{T_{\gamma\gamma}}, 
     p_{T_{bb}}$ cuts & 3.98 & 5604 & 5.61 & 1514 &
	6.79 & 941 &	8.01 & 557 &
	1.84 & 753 &	2.21 &	3838	\\ \hline
 \hline
   other/barrel ratio & \multicolumn{2}{c|}{31.64\%} & 
  \multicolumn{2}{c|}{30.14\%} &
\multicolumn{2}{c|}{30.05\%} &\multicolumn{2}{c|}{29.18\%} 
  &\multicolumn{2}{c|}{33.03\%}
  & \multicolumn{2}{c}{31.26\%} 
  \end{tabular}
\end{ruledtabular}
\end{sidewaystable}
\subsection{Cut Flows and Efficiencies}
We closely follow the procedures that we employed for the HL-LHC. 
We show in Table~\ref{tab:100-signal} the efficiencies and event yields 
for Higgs-pair production in the channel $HH \to b\bar b \gamma\gamma$
with $\lambda_{3H} = -4, 0, 1, 2, 6, 10$ and an integrated luminosity of 
3000 fb$^{-1}$ at the 100 TeV collider.

The overall signal efficiency has its peak value of 8.01 \%
at $\lambda_{3H} =2$ and its behavior is similar to that at 14 TeV with
$\sim 2$ \% when $\lambda_{3H}\gsim 4$, 
6.79 \% at the SM value of $\lambda_{3H}=1$, and
3.98 \% at $\lambda_{3H}=-4$.
On the other hand, the number of signal event
is 557 at $\lambda_{3H}=2$ and it becomes 941 at the SM value of 
$\lambda_{3H}=1$.
Note that one may have a similar number of signal events at $\lambda_{3H}=6$.

The cut flow table of all the backgrounds in terms of efficiencies 
at the HL-100 TeV hadron collider
is presented in Appendix \ref{cutflow}.

\subsection{Analysis and Results}

\begin{table}
\caption {The same as in Table
\ref{tab:Result-HL_LHC}  but at the HL-100 TeV hadron collider with an integrated
luminosity of 3 ab$^{-1}$. 
} \vspace{0mm}
\label{tab:Result-100TeV_I}
\begin{center}
\begin{tabular}{|l|r|r|r|r||r|}
\hline
Expected yields $(3000~ \mathrm{fb}^{-1})$ &Total & Barrel-barrel & Other & Ratio
(O/B) & $\#$ of Gen. \\
Samples &  &  & (End-cap) &   &   Events \\
\hline
$H(b\,\bar{b})\,H(\gamma\,\gamma)$, $\lambda_{3H} = -4$ &  $5604.46 \pm 63.36$ &
$4257.36 \pm 57.90$  & $1347.10 \pm 23.22$  & $0.32 \pm 0.007$ & $3 \times 10^5$
\\
$H(b\,\bar{b})\,H(\gamma\,\gamma)$, $\lambda_{3H} = 0$ & $1513.56 \pm 14.81$ &
$1163.04 \pm 14.09$ & $350.52\pm 3.57$ & $0.30 \pm 0.005$ & $3 \times 10^5$  \\
$H(b\,\bar{b})\,H(\gamma\,\gamma)$, $\lambda_{3H} = 1$ & $941.37 \pm 7.65$ &
$723.86 \pm 6.64$ & $217.51 \pm 3.66$ & $0.30 \pm 0.006$  & $3 \times 10^5$\\
$H(b\,\bar{b})\,H(\gamma\,\gamma)$, $\lambda_{3H} = 2$ & $557.36 \pm 1.93$
 & $431.45 \pm 1.87$ & $125.91 \pm 1.21$  & $0.29 \pm 0.003$ & $3 \times 10^5$\\
$H(b\,\bar{b})\,H(\gamma\,\gamma)$, $\lambda_{3H} = 6$ & $753.18 \pm 6.02$ &
$566.18 \pm 5.59$ & $187.00 \pm 5.33 $ & $0.33 \pm 0.010$ & $3 \times 10^5$\\
$H(b\,\bar{b})\,H(\gamma\,\gamma)$, $\lambda_{3H} = 10$ & $3838.33 \pm 36.82$  &
$2924.25 \pm 32.11$  & $914.08 \pm 28.01$  & $0.31 \pm 0.010$ & $3 \times 10^5$ \\
\hline
$gg\,H(\gamma\,\gamma)$ & $890.47 \pm 72.91$ & $742.97 \pm 58.43$ & $147.50 \pm
20.51$ & $0.20 \pm 0.03$ & $10^6$ \\
$t\,\bar{t}\,H(\gamma\,\gamma)$ & $868.73 \pm 8.53$ & $659.33 \pm 12.94$ & $209.40
\pm 7.04$ & $0.32 \pm 0.01$ & $9.63 \times 10^5$ \\
$Z\,H(\gamma\,\gamma)$ & $168.86 \pm 5.91$ & $122.91 \pm 4.68$ & $45.95 \pm 1.69$
&  $0.37 \pm 0.02$ & $10^6$ \\
$b\,\bar{b}\,H(\gamma\,\gamma)$ & $9.82 \pm 0.59$ & $7.00 \pm 0.58$ & $2.82 \pm
0.25 $ & $0.40 \pm 0.05$ & $10^6$ \\
\hline
$b\,\bar{b}\,\gamma\,\gamma$ & $770.42 \pm 23.48$ & $514.96 \pm 20.81$ & $255.46
\pm 15.10$ & $0.50 \pm 0.04$ & $1.1 \times 10^7$\\
$c\,\bar{c}\,\gamma\,\gamma$ & $222.88 \pm 40.55$  & $111.44 \pm 32.55$  & $111.44
\pm 26.92$  & $1.00 \pm 0.38 $ & $1.1 \times 10^7$ \\
$j\,j\,\gamma\,\gamma$ & $32.28 \pm 3.23$ & $20.98 \pm 3.99$ & $11.30 \pm 2.34$  &
$0.54 \pm 0.15$ & $10^7$  \\
$b\,\bar{b}\,j\,\gamma$ & $1829.13 \pm 75.08$ & $1288.34 \pm 45.27$ & $540.79 \pm
49.79$ & $0.42 \pm 0.04$  & $1.1 \times 10^7$ \\
$c\,\bar{c}\,j\,\gamma$ & $293.81 \pm 40.11$ & $216.49 \pm 36.71$  & $77.32 \pm
32.97$  & $0.36 \pm 0.16$ & $1.1 \times 10^7$  \\
$b\,\bar{b}\,j\,j$ & $3569.73 \pm 209.93$ & $2294.83 \pm 207.69$ & $1274.90 \pm
189.68$ & $0.56 \pm 0.10$ & $3.43 \times 10^6$  \\
$Z(b\,\bar{b})\,\gamma\gamma$ & $54.87 \pm 3.79$ & $35.72 \pm 3.36$ & $19.15 \pm
2.02$  & $0.54 \pm 0.08$ & $10^6$\\
\hline
$t\,\bar{t}$~($\geq$ 1 leptons) & $59.32 \pm 7.40$ & $38.32 \pm 5.79$ & $21.00 \pm
5.61$ & $0.55 \pm 0.17$ & $1.1 \times 10^7$ \\
$t\,\bar{t}\,\gamma$~($\geq$ 1 leptons) & $105.68 \pm 8.22$ & $62.53 \pm 5.07$ &
$43.15 \pm 7.95$ & $0.69 \pm 0.14$ & $10^6$ \\
\hline
Total Background & $8876.00 \pm 243.07$ & $6115.82 \pm 227.41$ & $2760.18 \pm
202.67$  & $0.45 \pm 0.04$  \\
\cline{1-5}
\cline{1-5}
\multicolumn{1}{|l|}{Significance $Z$} & \multicolumn{1}{r|}{$9.823$} &
\multicolumn{1}{r|}{$9.082$} & \multicolumn{1}{r|}{$4.087$ }  \\ \cline{1-4}
\multicolumn{2}{|l|}{Combined significance}   & \multicolumn{2}{c|}{$9.959$ }   \\
\cline{1-4}
\end{tabular}
\end{center}
\end{table}

Here we show the main results of the analysis for the 100 TeV hadron collider,
see Table~\ref{tab:Result-100TeV_I}.
Among the single-Higgs associated backgrounds, the major ones come from
$ggH$ and $t\bar t H$, comprising about 20 \% of the total background.
Meanwhile the dominant ones in non-resonant backgrounds are
$b\bar b jj$ followed $b\bar b j \gamma$ which make up about 60 \%
of the total background.
Including other backgrounds, we note that 70 \% 
of the total background is due to fakes.
Being contrary to the HL-LHC case, the combined
significance achieved is much higher:
$Z=9.981$ at the SM value of $\lambda_{3H}=1$, 
which is mainly because of much higher signal event rates though
the signal to background 
ratios are similar at HL-LHC and HL-100 TeV collider.

\begin{figure}[t!]
\centering
\includegraphics[width=3.2in]{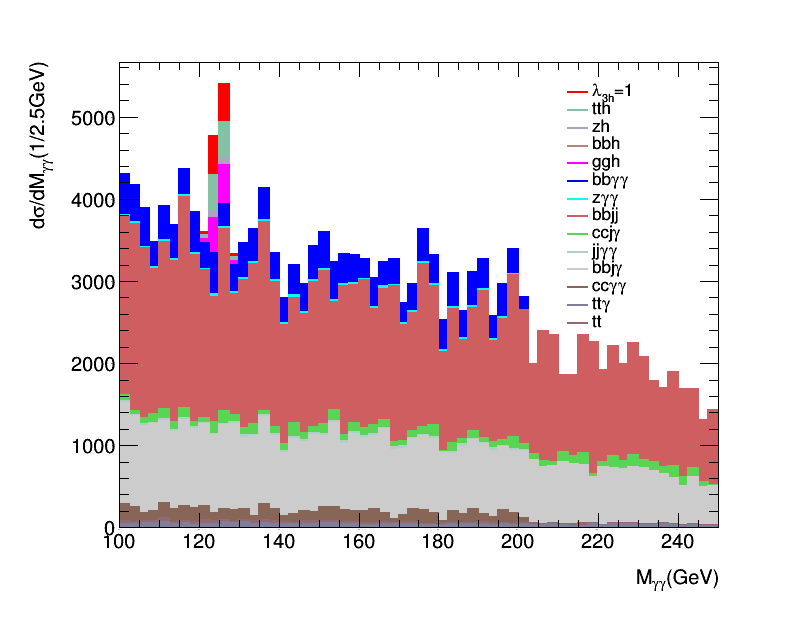}
\includegraphics[width=3.2in]{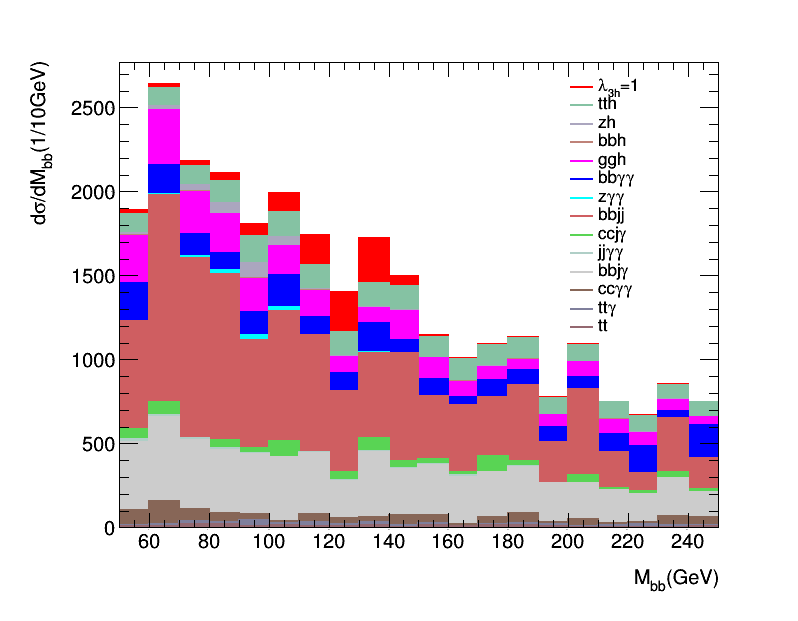}
\caption{\small \label{fig:inv-mass100}
The $M_{\gamma\gamma}$ (upper) and $M_{b\bar b}$ (lower)
distributions for the signal on top of all the backgrounds 
at the HL-100 TeV hadron collider.  
}
\end{figure}
In Fig.~\ref{fig:inv-mass100}, we show the resultant invariant-mass 
distributions
of the two photon (upper) and two $b$ (lower) candidates for the signal
on top of all the backgrounds at the HL-100 TeV collider,
as similar to HL-LHC in Fig.~\ref{fig:inv-mass}.
We observe the similar behavior as in the HL-LHC case.

\begin{figure}[t!]
\centering
\includegraphics[width=3.2in]{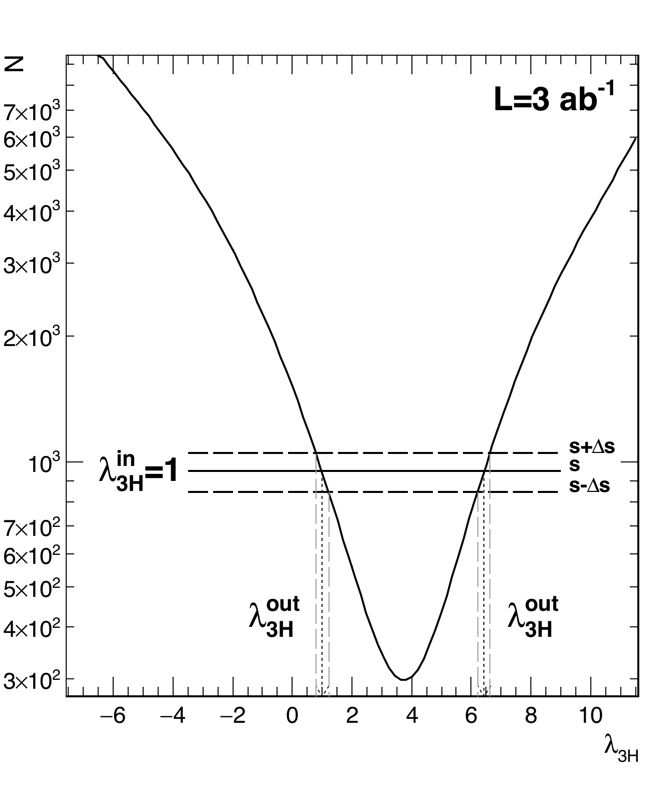}
\includegraphics[width=3.2in]{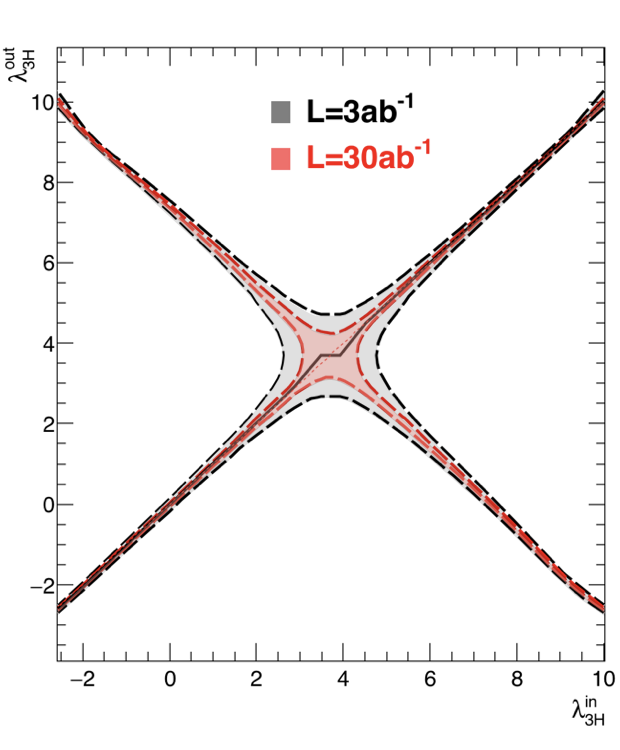}
\caption{\small \label{fig:l3h_100}
{\bf HL-100 TeV}:
(Left) The number of signal events $N$ versus $\lambda_{3H}$
with 3 ab$^{-1}$.
The horizontal solid line is for the number of signal events $s$
when $\lambda_{3H}^{\rm in}=1$
and the dashed lines for
$s\pm\Delta s$ with the statistical error of $\Delta s = \sqrt{s+b}$.
(Right) The 1-$\sigma$ error regions versus
the input values of $\lambda_{3H}^{\rm in}$ assuming
3 ab$^{-1}$ (black) and 30 ab$^{-1}$ (red).
}
\end{figure}
%
\begin{figure}[t!]
\centering
\includegraphics[width=4.5in]{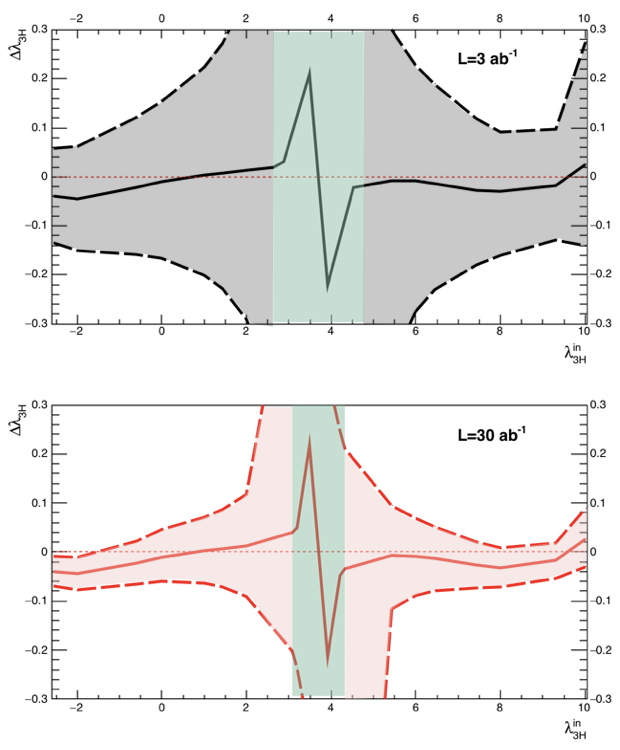}
\caption{\small \label{fig:dlam_100}
{\bf HL-100 TeV}: $\Delta \lambda_{3H} = \lambda_{3H}^{\rm out}-\lambda_{3H}^{\rm in}$
versus $\lambda_{3H}^{\rm in}$
along the $\lambda_{3H}^{\rm out}=\lambda_{3H}^{\rm in}$ line
with 3 ab$^{-1}$ (upper) and 30 ab$^{-1}$ (lower).
The lines are the same as in the right frame of Fig.~\ref{fig:l3h_100}.
We consider $|\Delta\lambda_{3H}|\leq 0.3$ to find the regions in which
one can pin down the $\lambda_{3H}$ coupling with an absolute error
smaller than $0.3$.
}
\end{figure}
Since the achieved significance is high enough, we try to estimate 
how well one can measure the $\lambda_{3H}$ coupling
at the HL-100 TeV hadron collider.
In the left frame of Fig.~\ref{fig:l3h_100}, 
we show the number of signal events $N$ as a function of $\lambda_{3H}$.
To obtain the curve we assume the luminosity of 3 ab$^{-1}$
and take into account the $\lambda_{3H}$-dependent overall 
signal efficiencies, see Table~\ref{tab:100-signal}.
One may find the values of 
$N$ for some representative choices of $\lambda_{3H}$
in Table~\ref{tab:Result-100TeV_I}.
On the other hand,
the solid horizontal line shows the number of signal events $s$, as an example,
when the input value of $\lambda_{3H}$ or $\lambda_{3H}^{\rm in}$ takes the SM value of 1.
The dotted lines delimit
the 1-$\sigma$ region considering the statistical error of
$\Delta s = \sqrt{s+b}$ with $b=9147.63$.
For this purpose, we generate another pseudo dataset for the signal.
By locating the points where the $N$ curve and the horizontal lines
meet, one can obtain the two center values of output $\lambda_{3H}$
and the corresponding two regions of 1-$\sigma$ error. 
Note that, usually, there is a two-fold 
ambiguity in this approach.
By repeating this procedure for different input values of $\lambda_{3H}$,
we can obtain the center output $\lambda_{3H}$ values
together with the regions of 1-$\sigma$ error, 
as shown in the right frame of Fig.~\ref{fig:l3h_100}.

The black-shaded region (delimited by the black dashed lines)
in the right frame of Fig.~\ref{fig:l3h_100}
shows the 1-$\sigma$ errors versus
the input values of $\lambda_{3H}^{\rm in}$ with the luminosity of 3 ab$^{-1}$.
Incidentally, the black solid line shows 
the center values of output $\lambda_{3H}$ values or $\lambda_{3H}^{\rm out}$
along the $\lambda_{3H}^{\rm out}=\lambda_{3H}^{\rm in}$ line
denoted by the thin dotted line.
We note that there exists
a bulk region of $2.6  \lsim \lambda_{3H}  \lsim 4.8 $
in which one cannot pin down the $\lambda_{3H}$ coupling. 
We find that
the bulk region reduces to $3.1  \lsim \lambda_{3H}  \lsim 4.3$
assuming the luminosity of 30 ab$^{-1}$ as shown by 
the red-shaded region (delimited by the red dashed lines)
in the same frame of Fig.~\ref{fig:l3h_100}.

Even though it would be difficult to pin down the 
$\lambda_{3H}$ coupling in the bulk region, yet one goes
a bit away from it and is able to measure the coupling
with a high precision as indicated by the narrowness of 
the 1-$\sigma$ error regions.
And, the two-fold ambiguity can be lifted up by exploiting the kinematical 
differences found in the distributions of 
$\Delta R_{\gamma\gamma}$, $P_T^{\gamma\gamma}$, $M_{\gamma\gamma b b}$ when
$\lambda_{3H}$ takes on different values:
see Fig.~\ref{appfig:100}.
Keeping these all in mind, in Fig.~\ref{fig:dlam_100},
we show the regions in which one can determine
the $\lambda_{3H}$ coupling within an absolute error
of $0.3$ (either upper or lower error)
along the $\lambda_{3H}^{\rm out}=\lambda_{3H}^{\rm in}$ line
assuming 3 ab$^{-1}$ (upper panel) and 30 ab$^{-1}$ (lower panel).
The green-shaded regions around $\lambda_{3H}=3.5$ denote the bulk regions.
We observe that, when
$\lambda_{3H}\lsim 1.6\,(2.4)$ or $\lambda_{3H}\gsim 5.9\,(5.3)$,
one can pin down the $\lambda_{3H}$ coupling with 
an absolute error smaller than $0.3$ assuming $3\,(30)$ ab$^{-1}$.
At the SM value of $\lambda_{3H}=1$, specifically, 
we observe that the coupling can be measured 
with about $20\,(7)$ \% accuracy assuming 
the integrated luminosity of $3\,(30)$ ab$^{-1}$.
Our results are about 2 times better than those reported 
in Ref.~\cite{Barr:2014sga}
and comparable with those in Ref.~\cite{Kling:2016lay}
taking account of the more sophisticated and comprehensive treatment
of the background processes taken in this work.

Before moving to the next Section, 
we would like to comment that the bulk region 
can be shifted by adopting a different set of selection cuts and it may help
if it turns out that $\lambda_{3H}$ falls into the bulk region in future.

\section{Further Improvements Envisaged}
In our analysis, we are taking the SM cross sections of
$\sigma(gg\to HH)=45.05$ fb and
$\sigma(gg\to HH)=1749$ fb at 14 TeV and 100 TeV, respectively,
which are calculated at NNLO accuracy including
NNLL gluon resummation
in the infinite top quark mass approximation.
We have taken these values of cross sections to confirm, especially, 
the ATLAS results~\cite{atlas_hh17}.
Recently, the NLO corrections considering full top-quark mass dependence
have been available~\cite{Borowka:2016ehy,Borowka:2016ypz}. The calculation reveals
that
the full top-quark mass dependence is vital to get reliable predictions for
Higgs boson pair production. Precisely, 
the total cross section is reduced by 14 \% at 14 TeV compared to that obtained by
the Born improved Higgs Effective Field Theory (HEFT) in which
the infinite top mass approximation is taken. At 100 TeV,
the larger reduction of 24 \% is found.

\begin{figure}[t!]
\centering 
\includegraphics[width=3.2in,height=2.7in]{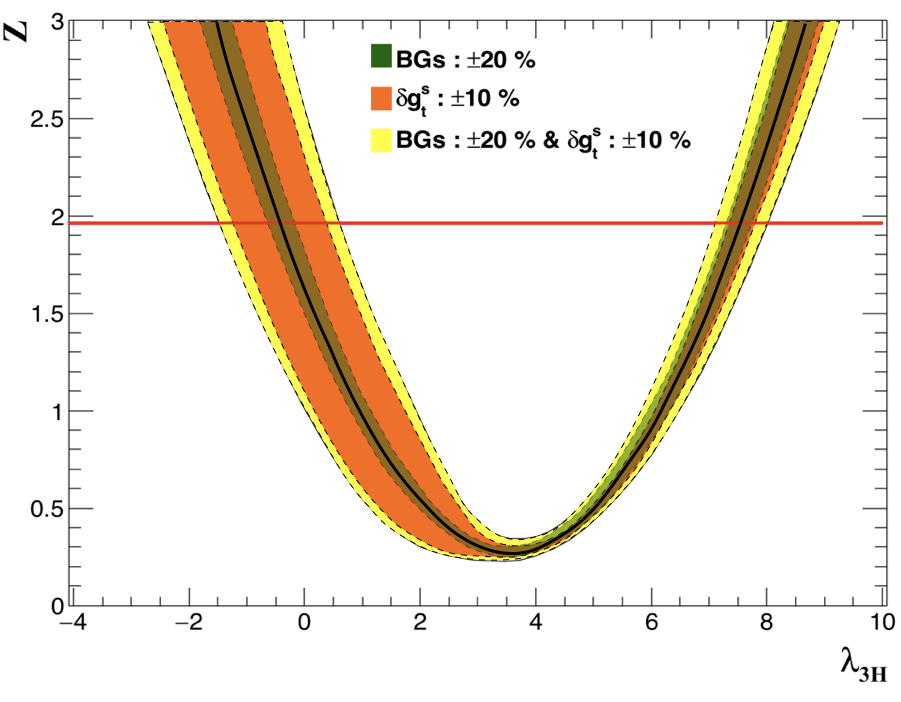}
\includegraphics[width=3.2in,height=2.7in]{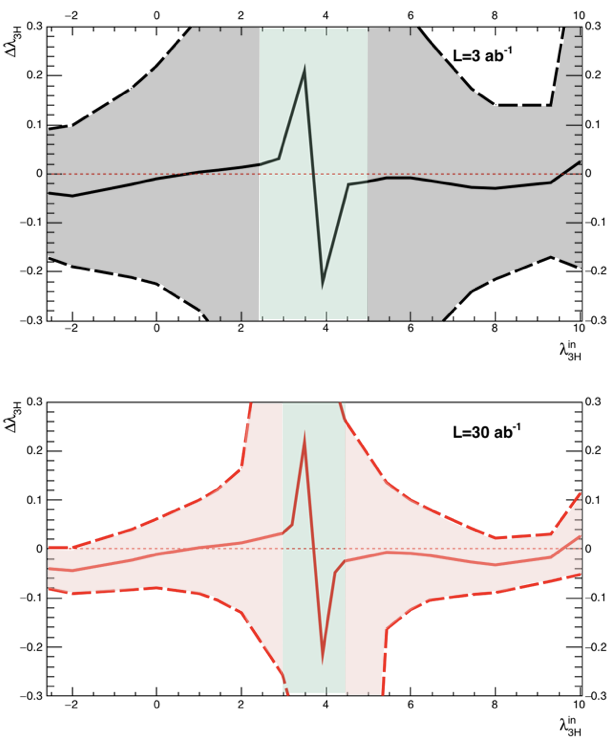}
\caption{\small \label{fig:new}
(Left)  {\bf HL-LHC}:
The same as in Fig.~\ref{fig:signif} but taking the NNLO cross section
$\sigma(gg\to HH)=36.69$ fb in the FT approximation.
(Right) {\bf HL-100 TeV}:
The same as in Fig.~\ref{fig:dlam_100} but taking the NNLO cross section
$\sigma(gg\to HH)=1224$ fb in the FT approximation.
}
\end{figure}
At the moment,  as suggested in Ref.~\cite{Grazzini:2018bsd},
the best way to incorporate the finite top-quark mass effects 
at NNLO
might be by adopting the 
FT approximation~\cite{Frederix:2014hta,Maltoni:2014eza}
in which the full top-quark mass dependence is considered only in 
the real radiation while the HEFT is taken in the virtual part.
At NNLO in the FT approximation,
$\sigma(gg\to HH)=36.69$ fb and
$\sigma(gg\to HH)=1224$ fb at 14 TeV and 100 TeV, respectively~\cite{Grazzini:2018bsd}. 
We observe that
20 (30) \%  reduction  at 14 (100) TeV
compared to the cross sections used in Sections III and IV.
To see the impact  of the reduced cross sections on our main results, 
in Fig.~\ref{fig:new},
we show the signal significance over the background versus
$\lambda_{3H}$ at the HL-LHC  (left) and 
the regions in which one can determine the $\lambda_{3H}$ coupling with an absolute error of
$0.3$ at the HL-100 TeV collider (right).
At 14 TeV with 3000 fb$^{-1}$,
the trilinear coupling is constrained to be 
$ -1.5 < \lambda_{3H} < 8.1$ at 95\% CL taking account of 
the uncertainties associated with the top-Yukawa coupling
and the estimation of backgrounds.
Taking the central line, the 95\% CL sensitivity region for $\lambda_{3H}$ is
$-0.4 < \lambda_{3H} < 7.5$ which becomes broader by the amount of $\pm 0.4$
compared to the results presented in Section III
\footnote{Recall that the corresponding region is
$0 < \lambda_{3H} < 7.1$ if the NNLO+NNLL cross section of 45.05 fb is taken.}.
At 100 TeV,
we find a little bit broader bulk regions of
$2.4  \lsim \lambda_{3H}  \lsim 5.0 $ and $3.0  \lsim \lambda_{3H}  \lsim 4.4 $
with 3 ab$^{-1}$ and 30 ab$^{-1}$, respectively, compared to the results
presented in Section IV
\footnote{Recall that, when the NNLO+NNLL cross section of 1749 fb is taken at 100 TeV,
the bulk regions are $2.6\,(3.1) < \lambda_{3H} < 4.8\,(4.3)$  
and $\lambda_{3H}$ can be measured with an accuracy of 20 (7) \% at its SM value
with 3 ab$^{-1}$ (30 ab$^{-1}$).}.
And,  $\lambda_{3H}$ can be measured with an accuracy of 30 (10) \% with 
an integrated luminosity
of 3 (30) ab$^{-1}$ when it takes on its SM value of $1$.
We observe that the effects of the reduced cross sections
are less significant in the case with 30 ab$^{-1}$ at 100 TeV
in which the number of signal events is comparable to or larger than
that of backgrounds.

The QCD corrections also affect
the ratio $\sigma(gg\to HH)/\sigma(gg\to HH)_{\rm SM}$ which is used to obtain the cross
sections for non-SM values of $\lambda_{3H}$. The QCD corrections depend on $\lambda_{3H}$
and become larger when $\lambda_{3H}$  deviates from the SM value $1$ due to
the nontrivial interference between the triangle and box diagrams~\cite{Borowka:2016ypz}.
We observe that the ratio increases by  about
10 (35) \% at $\lambda_{3H}=-1\,(5)$, see Fig.~\ref{fig:kfactor}.
It is clear that the QCD corrections are less significant than
the uncertainties associated with the top-Yukawa coupling,
see Fig.~\ref{fig:ratio-cross}.
In this respect,
we have not taken account of the $\lambda_{3H}$-dependent QCD corrections
on the ratio $\sigma(gg\to HH)/\sigma(gg\to HH)_{\rm SM}$ in this work
\footnote{
Taking account of the $\lambda_{3H}$-dependent QCD corrections,
at 14 TeV,
we observe that the central 95\% CL sensitivity region reduces from
$-0.4 < \lambda_{3H} < 7.5$  to $-0.4 < \lambda_{3H} < 6.9$ since
the QCD corrections enhance the signal cross section for 
$\lambda_{3H} \lsim 1$ and
$\lambda_{3H} \gsim 2.5$.}.
On the other hand,
when $|\lambda_{3H}|$ is significantly larger than $1$,
vertex corrections proportional to $\lambda_{3H}^3$ 
appear at the amplitude level. This may bring sizeable distortion
to $\sigma(gg\to HH)/\sigma(gg\to HH)_{\rm SM}$. In this case, it might be practical 
to consider
$\lambda_{3H}$ as an effective parameter, not as a fundamental one.
\begin{figure}[t!]
\centering
\includegraphics[width=3.2in]{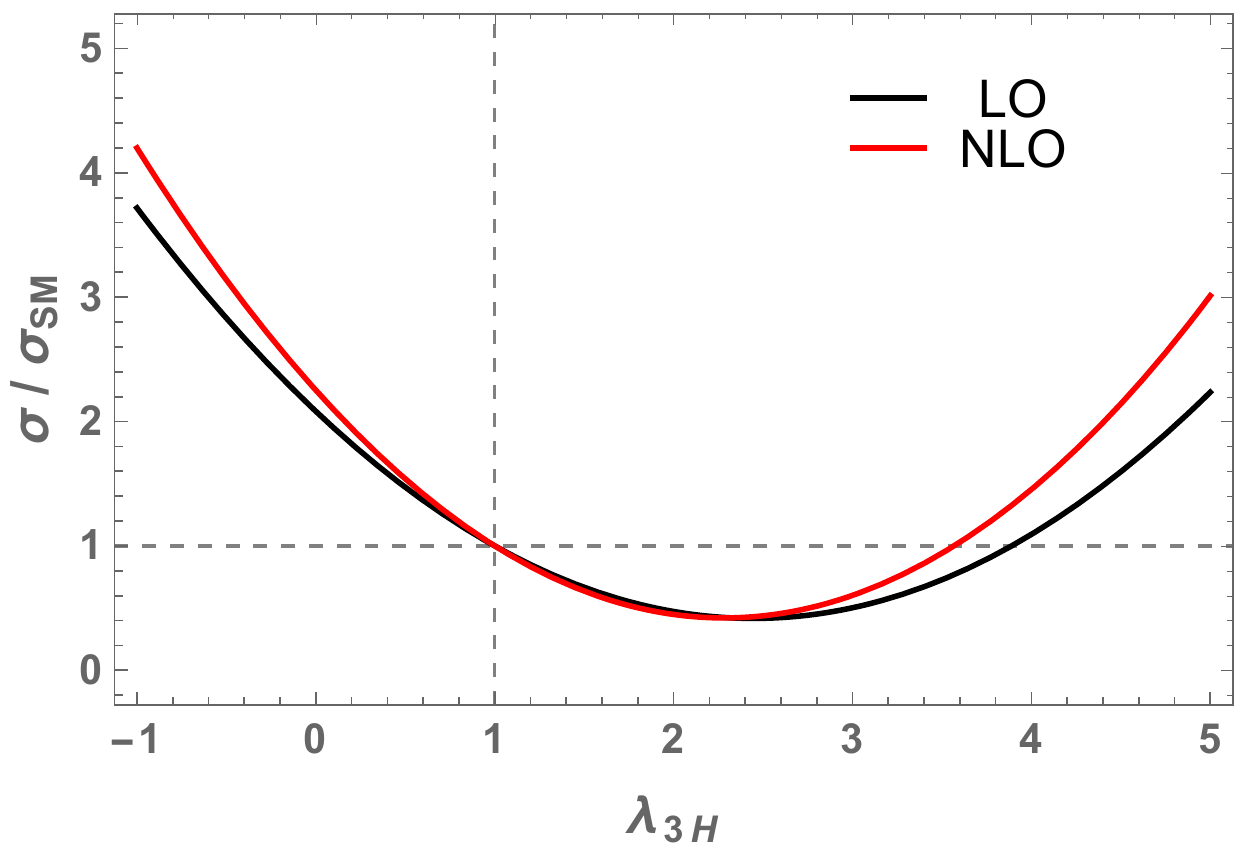} 
\includegraphics[width=3.2in]{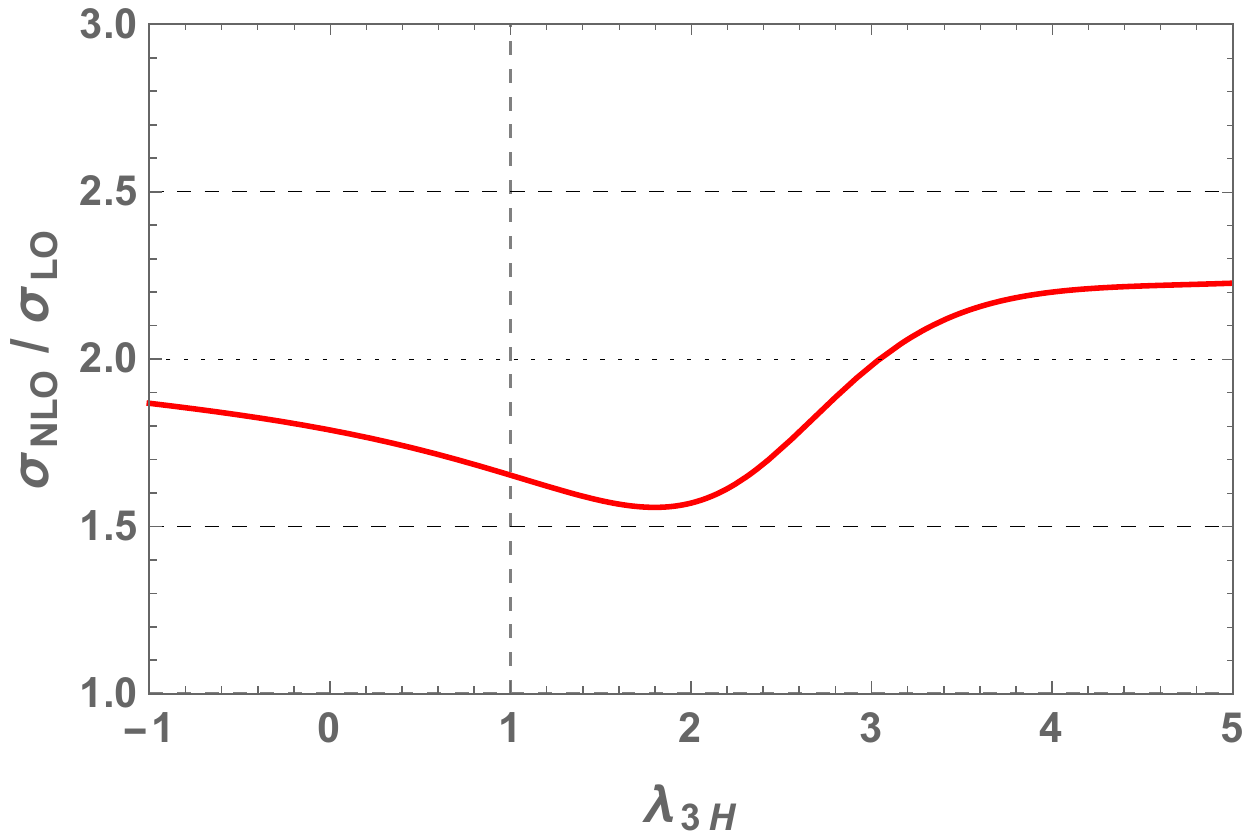}
\caption{\small \label{fig:kfactor}
(Left) The ratio $\sigma(gg\to HH)/\sigma(gg\to HH)_{\rm SM}$ versus $\lambda_{3H}$
at LO (black) and NLO (red) at 14 TeV. We have taken the NLO cross sections
considering full top-quark mass dependence.
(Right) The ratio $\sigma^{\rm NLO}(gg\to HH)/\sigma^{\rm LO}(gg\to HH)$ 
versus
$\lambda_{3H}$ at 14 TeV.
We refer to Ref.~\cite{Borowka:2016ypz} for absolute cross sections as functions of
$\lambda_{3H}$.
}
\end{figure}

Note that the $P_T^{\gamma\gamma,bb}$ and $M_{\gamma\gamma b b}$ 
distributions are affected by the QCD corrections 
at NLO and NNLO
as shown in, for example, Refs.~\cite{Borowka:2016ypz,Grazzini:2018bsd}.
For more precise predictions at the HL-LHC and HL-100 TeV collider and to
lift up the two-fold ambiguity in $\lambda_{3H}$ especially,
one may need to incorporate them in the future.

The PDF set of {\tt CTEQ6L1} taken to calculate the non-resonant backgrounds
does not include the use of data from LHC experiments.
To study the impact of the LHC data on PDF, instead of {\tt CTEQ6L1},
we take the PDF set of {\tt CT14LO} \cite{Dulat:2015mca} and 
re-simulate all the non-resonant backgrounds at 14 TeV.
Taking the example of $b\bar b\gamma\gamma$ background,
which is one of the two most severe non-resonant backgrounds,
we obtain the overall efficiency of $4.34\times 10^{-3}$ by generating $10^7$ events.
This is very similar to the efficiency of
$4.49\times 10^{-3}$ obtained using {\tt CTEQ6L1}, see Table~\ref{apptab:14bkgs}. 
Actually, we observe that
the two efficiencies in each step of cut flow coincide within less than 10\%
and there are no
significant differences in kinematic distributions caused by {\tt CT14LO}.
Meanwhile, the real effect of {\tt CT14LO} is the reduction of the 
cross sections for the non-resonant backgrounds. 
For $b\bar b\gamma\gamma$, as an example, it reduces to 112 fb
\footnote{For other backgrounds at 14 TeV, see $\sigma_{\rm Eq.\,(6)}$
presented in Table.~\ref{table:updatedXsection2}.}.
Compared to the cross section of 140 fb obtained
using {\tt CTEQ6L1}, the cross section reduces by 20\%.

Furthermore, 
the pre-selection cuts listed in Eq.~(\ref{eq:presel}) 
may not be enough to avoid the double counting
problems in the non-resonant background samples.
To address this point, we implement MLM matching~\cite{Mangano:2006rw,Alwall:2007fs}.
We observe that there are no
significant differences in kinematic distributions due to MLM matching.
For details of the matching precesses and the calculation of the merged cross
sections, we refer to Appendix~\ref{app:C}.
Taking account of the NNLO cross section $\sigma(gg\to HH)=36.69$ fb 
in the FT approximation and the $\lambda_{3H}$-dependent QCD corrections,
we obtain the central 95\% CL sensitivity region of $-0.4 < \lambda_{3H} < 6.9$
at 14 TeV, see the black dash-dotted line in Fig.~\ref{fig:Zcomp}. 
Incorporating the impact of {\tt CT14LO} and the reduction of the 
non-resonant background cross sections 
by MLM matching, the region reduces to $0.1 < \lambda_{3H} < 6.6$, 
see the blue dashed line in Fig.~\ref{fig:Zcomp}.

Last but not least,
we also take into account the contribution from the Higgs production
accompanied by a hard $b\bar b$ pair via gluon-fusion at 14 TeV.
For this purpose, we calculate
the $gg\to H b\bar b$ process, which is supposed to be the
leading hard process for the contribution \cite{Homiller:2018dgu}.
Adopting the cuts suggested in Ref~\cite{Homiller:2018dgu} and
using 
{\tt MG5\_aMC@NLO} and {\tt NNPDF2.3LO},
we obtain $\sigma(gg\to H b\bar b)\simeq 4.8$ fb at 14 TeV
\footnote{This is about 4 times smaller than the corresponding cross section
of $\sim 22$ fb at 27 TeV~\cite{Homiller:2018dgu}.}.
Then we find a selection efficiency of $2.7$\% for the process
$gg\to H(\to\gamma\gamma)b\bar b$,
which leads to $0.9$ event at 14 TeV with 3 ab$^{-1}$
after all the selection cuts are applied.
Therefore, the total number of
the $ggH(\to\gamma\gamma)$ background may increase into
$6.6+0.9=7.5$ after including the hard process.
We conclude that about 10\% of
the background might come from the hard $b\bar b$ pair
production at 14 TeV.

\section{Conclusions}

One of the major goals of the HL-LHC and HL-100 TeV hadron collider
is to unfold the mystery of the EWSB mechanism, which is related
to the origin of mass. We have investigated the trilinear self-coupling of 
the Higgs boson in Higgs-pair production using the most promising 
channel $pp \to HH \to \gamma \gamma b\bar b$ with a fully comprehensive 
signal-background analysis.  It turns out that various fake backgrounds,
including $c \to b$, $j\to \gamma$, $e\to\gamma$, are among the most
dominant backgrounds that have to be discriminated against the signal.

The high-luminosity option of the LHC (HL-LHC) with an integrated
luminosity of 3000 fb$^{-1}$ can only constrain the trilinear coupling
by $-1.0 < \lambda_{3H} < 7.6$ at 95\% CL after taking into account
the uncertainties associated with the top-Yukawa coupling and estimation
of total background. This is unfortunate if the trilinear  coupling
takes on the SM value, it cannot be confirmed at the HL-LHC due to very
small event rates.
On the other hand, a much larger signal event rate at 
the HL-100 hadron collider enables one to pin down the value of $\lambda_{3H}$
with an absolute error smaller than $0.3$, except for a near-bulk region
$1.6 < \lambda_{3H} < 5.9$ ($2.4 < \lambda_{3H} < 5.3$), with an integrated
luminosity of 3 ab$^{-1}$ (30 ab$^{-1}$).  If $\lambda_{3H}$ takes on the
SM value, it can be measured with an accuracy of 20 (7) \% with luminosity
of 3 (30) ab$^{-1}$.

Before closing we would like to offer a few more comments.
\begin{enumerate}
\item
Variations of cross sections with $\lambda_{3H}$ for different production
channels differ from one another. Indeed, if $\lambda_{3H}$ falls at the
minimum of $\sigma(gg \to HH)$, one can use, for example, $q\bar q^{(')}
\to W/Z + HH$ to probe the trilinear coupling. See Fig.~\ref{fig:total-cross}.

\item 
We do not investigate the vector-boson fusion mechanism in this work. Though
its cross section is at least one order magnitude smaller than gluon
fusion, it has an additional handle to discriminate against backgrounds due
to two very energetic and forward jets in the final state. 

\item 
Currently, the reconstruction of the $b$-quark momentum is far from ideal
as can be shown from the invariant mass $M_{b\bar b}$ spectrum. We expect
that the $b$-jet tagging and $b$-jet reconstruction can be substantially improved
with Deep Learning techniques in future, such that the invariant mass
cut on $M_{b\bar b}$ can be much more effective.


\item 
In many other Higgs-sector extensions of the SM, there usually exist
heavy neutral scalar bosons, which can be produced via gluon fusion and 
decays into Higgs-boson pair. Our approach of signal-background 
analysis can be adopted to analyze such kinds of models.
Although specialized cuts tailored for particular models may generate 
higher significance, our approach can be applied in general.

\item
Adopting the most recent NNLO calculations in the FT approximation,
the inclusive cross section is reduced by
20 \%  at 14 TeV
compared to the NNLO+NNLL cross section and, accordingly, 
the 95 \% sensitivity range of $\lambda_{3H}$ broadens by about 10 \%.
On the other hand, the inclusive cross section is reduced by 30 \%
at 100 TeV which results in about 20 \% increment of bulk regions.
And the accuracy at $\lambda_{3H}=1$ worsens to 30 (10) \%
with 3 (30) ab$^{-1}$.

\item
When we compare our HL-100 TeV results to those of Ref.~\cite{Contino:2016spe},
we found that
their results have higher significance. This is because we have considered
more backgrounds in our analysis such as the category of single-Higgs
backgrounds and $bbjj$.

\item
We observe that the non-resonant backgrounds could be significantly reduced
by reflecting the impact of the LHC data on PDF and considering MLM matching.

\end{enumerate}

%

\section*{Acknowledgment}
We thank Tie-Jiun Hou for helpful comments on PDFs.
We also thank Olivier Mattelaer and Stefan Prestel
for helpful comments on MLM matching and DJR distribution
in {\bf MadGraph5{\_}aMC@NLO} with {\tt PYTHIA8}.
This work was supported by the National Research Foundation of Korea
(NRF) grant No. NRF- 2016R1E1A1A01943297. 
K.C. was supported by the
MoST of Taiwan under grant number MOST-105-2112-M-007-028-MY3.
J.P. was supported by the NRF grant No. NRF-2018R1D1A1B07051126.
%

$$ $$
\newpage

\appendix

\section{Kinematical distributions for the signal and backgrounds at the HL-LHC
and HL-100 TeV hadron collider}
\label{app-hl}

In Fig.~\ref{appfig:distr1},
we show the $\Delta R_{\gamma\gamma}$, $P_T^{\gamma\gamma}$,
$\Delta R_{\gamma b}$, and $M_{\gamma\gamma b b}$
distributions for the signal 
taking $\lambda_{3H} = -4,0,1,2,6$, and $10$ at the HL-LHC.
We observe the $M_{\gamma\gamma b b}$ distribution
becomes narrower and softer for the larger values of $|\lambda_{3H}|$
due to the $s$-channel Higgs propagator.

In the left frame of Fig.~\ref{appfig:distr2}, 
we show the angular separation between one of the
photons and one of the $b$ quarks at the HL-LHC for
the SM signal ($\lambda_{3H}=1$) and all the backgrounds
considered in this work.
The signal tends to have relatively larger
$\Delta R_{\gamma b}$ implying that $\gamma$ and 
$b$ originated from the signal are
more or less back-to-back.
The right frame of Fig.~\ref{appfig:distr2} is for
the invariant mass distributions $M_{\gamma\gamma bb}$.

\begin{figure}[th!]
\centering
\includegraphics[width=3.2in]{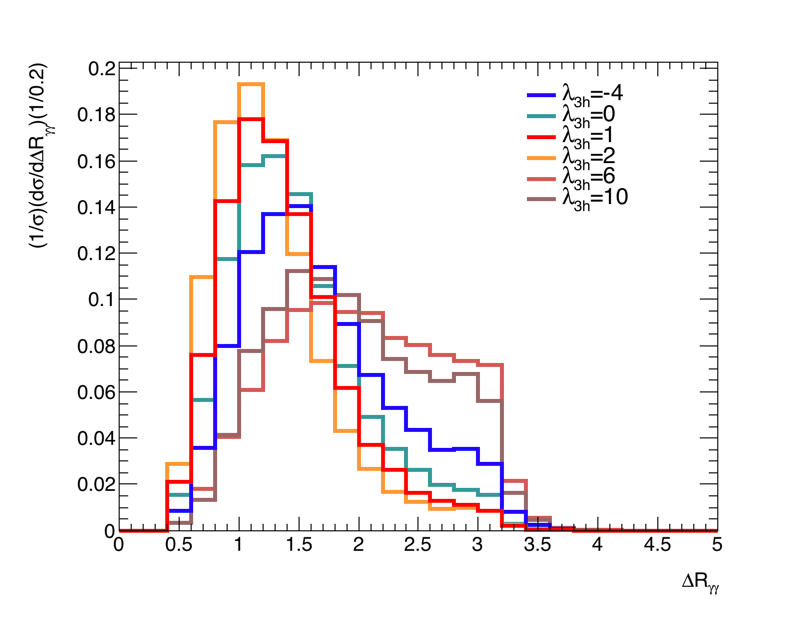}
\includegraphics[width=3.2in]{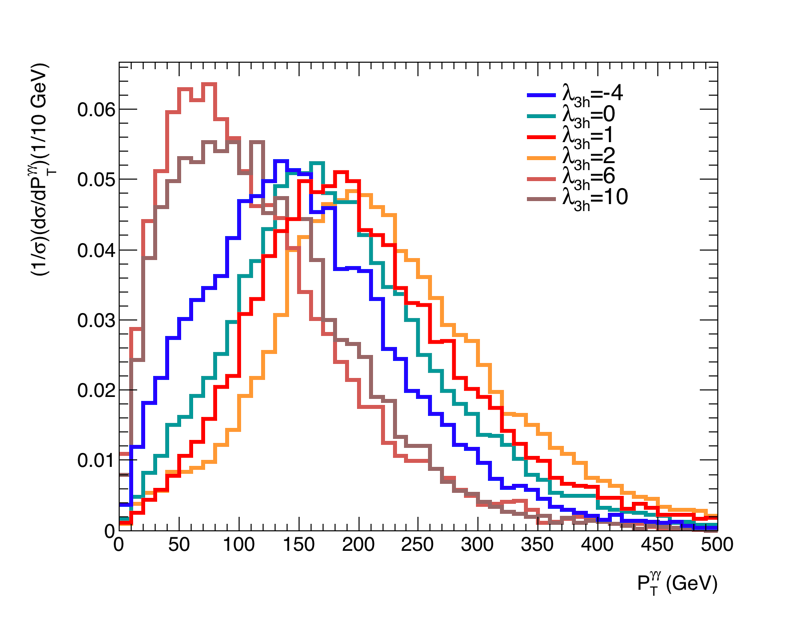}
\includegraphics[width=3.2in]{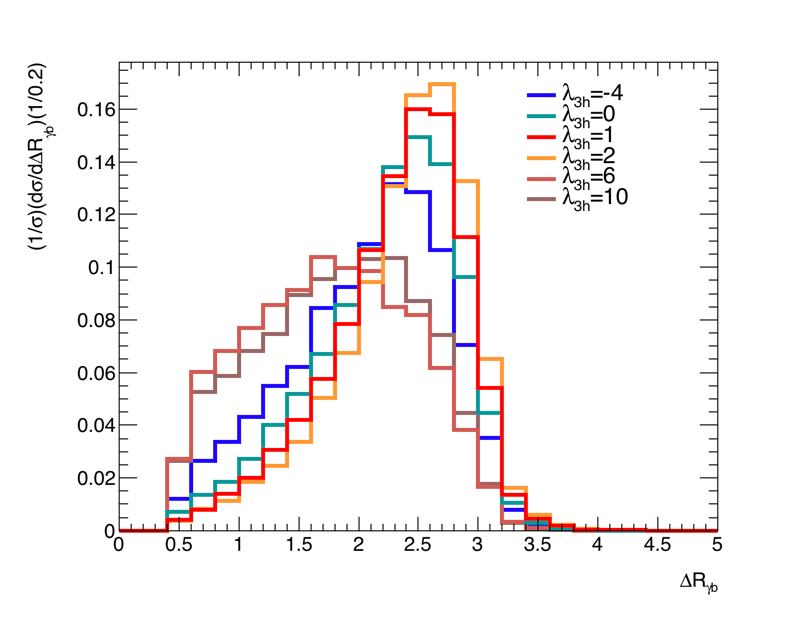}
\includegraphics[width=3.2in]{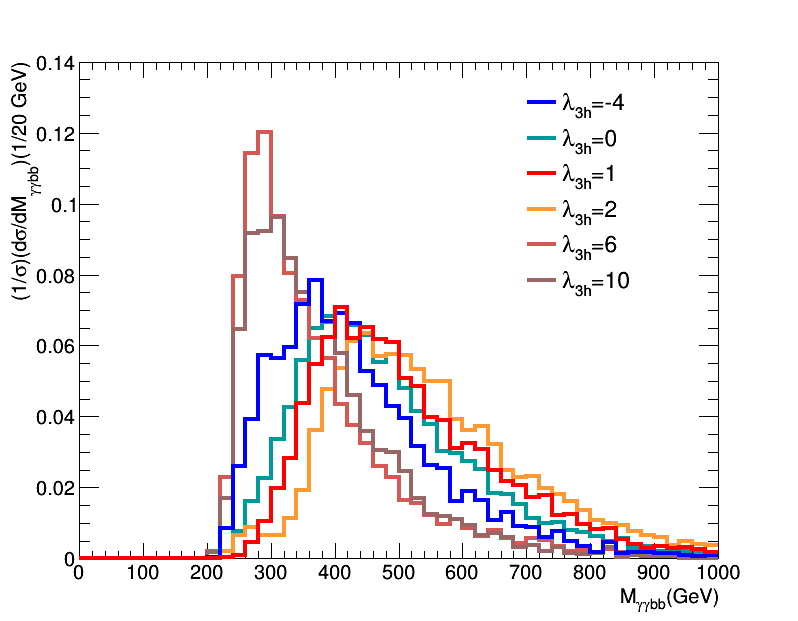}
\caption{\small \label{appfig:distr1}
{\bf HL-LHC}:
The $\Delta R_{\gamma\gamma}$, $P_T^{\gamma\gamma}$,
$\Delta R_{\gamma b}$, and $M_{\gamma\gamma b b}$
distributions for the signal 
taking $\lambda_{3H} = -4,0,1,2,6$, and $10$. 
}
\end{figure}

\begin{figure}[th!]
\centering
\includegraphics[width=3.2in]{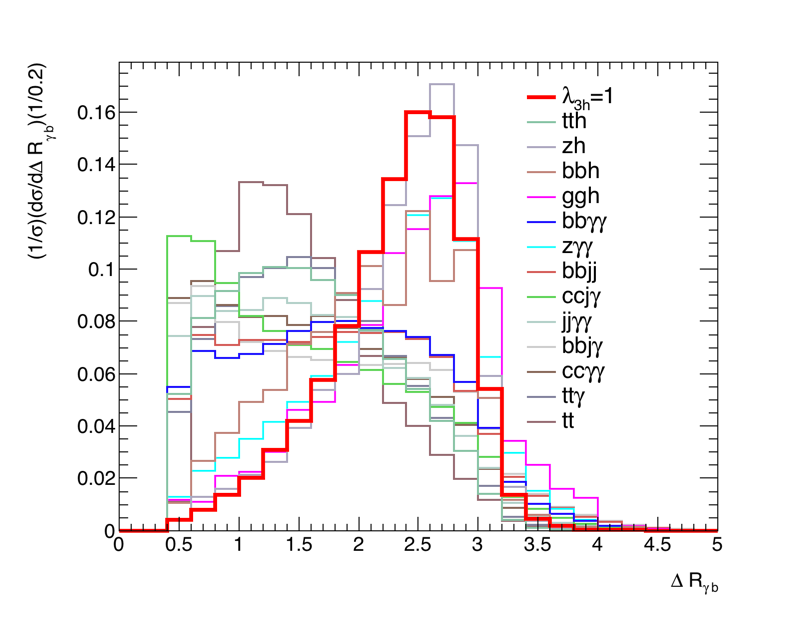}
\includegraphics[width=3.2in]{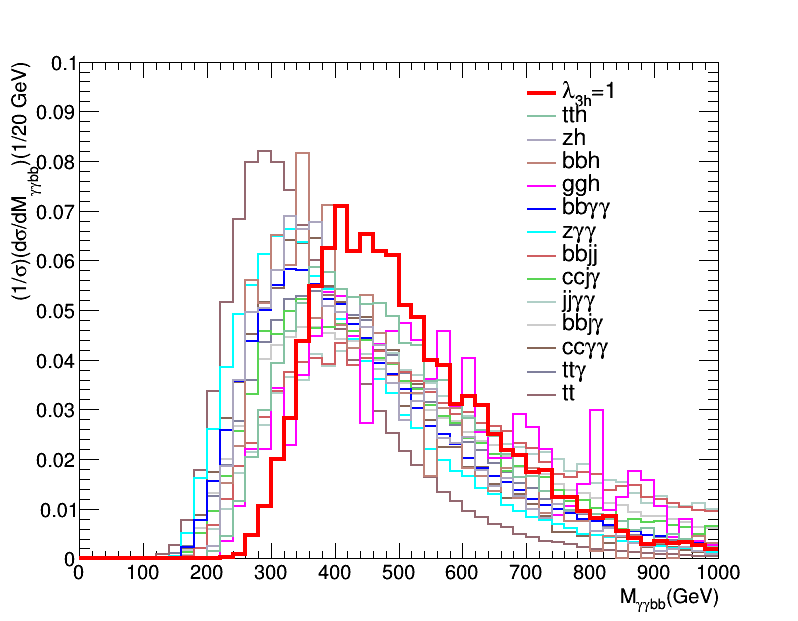}
\caption{\small \label{appfig:distr2}
{\bf HL-LHC}:
The $\Delta R_{\gamma b}$  and $M_{\gamma\gamma bb}$
distributions for the SM signal ($\lambda_{3H}=1$) and all the backgrounds
considered in this work.
}
\end{figure}

Fig.~\ref{appfig:100} is for some distributions
at the HL-100 TeV hadron collider. The most of distributions are very 
similar to those at the HL-LHC.

\begin{figure}[ht]
\centering
\includegraphics[width=3.2in]{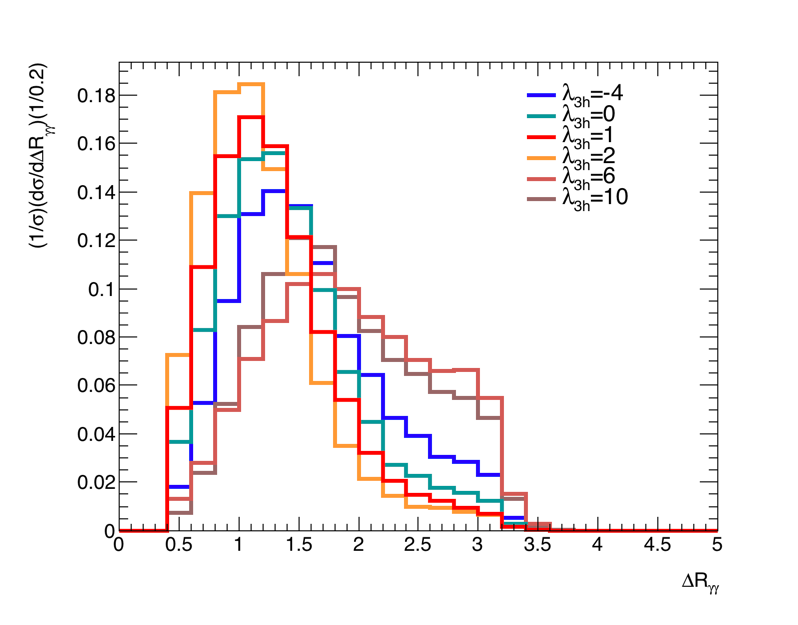}
\includegraphics[width=3.2in]{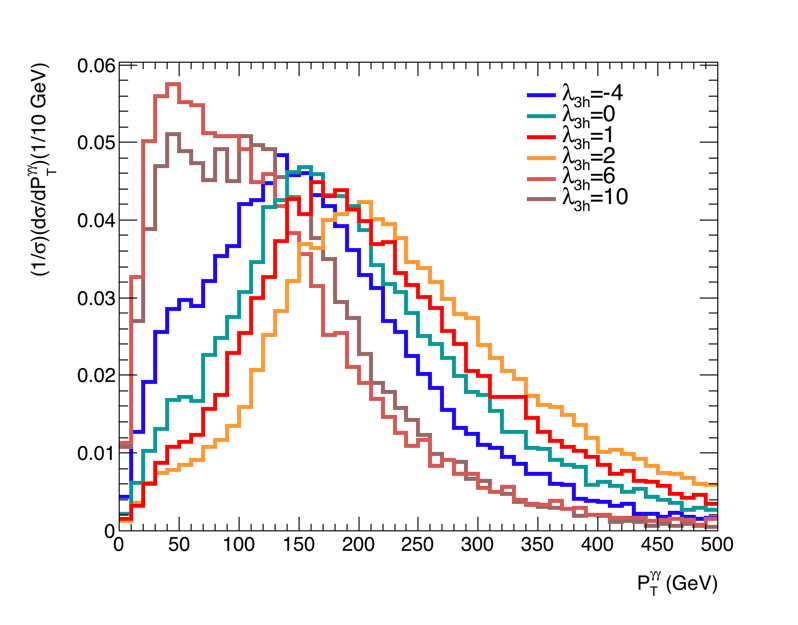}
\includegraphics[width=3.2in]{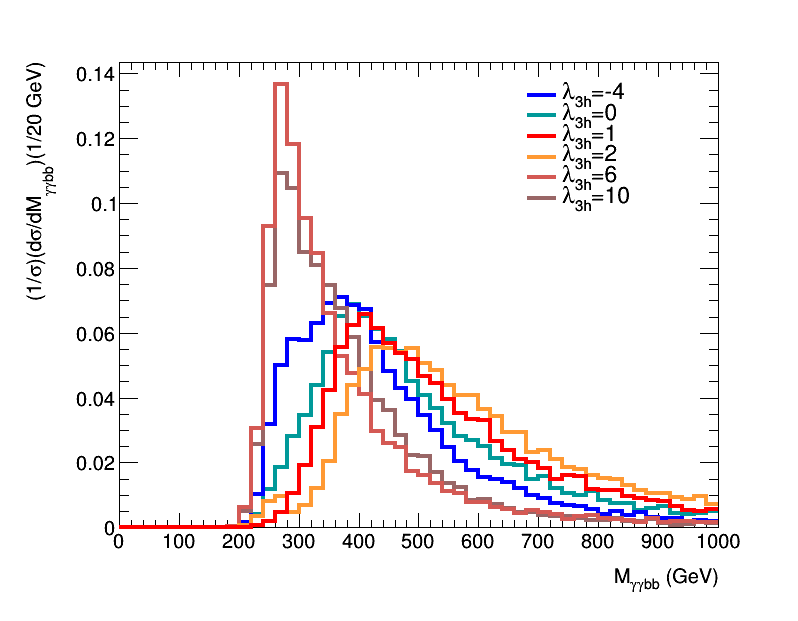}
\includegraphics[width=3.2in]{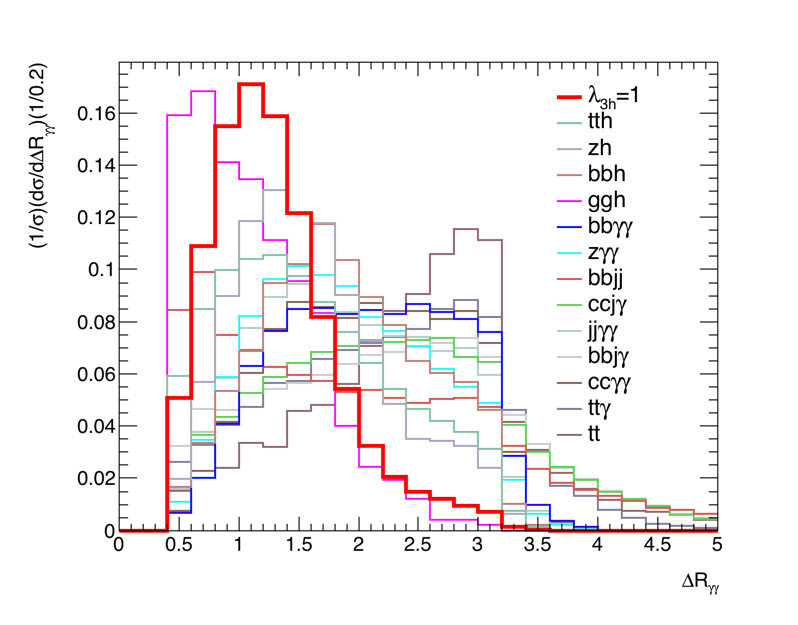}
\caption{\small \label{appfig:100}
{\bf HL-100 TeV}:
The $\Delta R_{\gamma\gamma}$ (upper left), $P_T^{\gamma\gamma}$ (upper right),
and $M_{\gamma\gamma b b}$ (lower left)
distributions for the signal
taking $\lambda_{3H} = -4,0,1,2,6$, and $10$.
In the lower right frame,
the $\Delta R_{\gamma\gamma}$  
distributions for the SM signal ($\lambda_{3H}=1$) and all the backgrounds
are compared. 
}
\end{figure}

\section{Cut flow tables for all the backgrounds at the HL-LHC
and HL-100 TeV hadron collider}
\label{cutflow}

In this appendix, we present the cut flow tables for all the backgrounds at 
the HL-LHC and HL-100 TeV hadron collider, see 
Tables~\ref{apptab:14bkgs} and \ref{apptab:100bkgs}.
We note that the lepton-veto cut does not affect
the $t\bar t$ related BGs in which electrons are faking photons.
\begin{table}[ht]
\scriptsize
\caption {Cut flow table of the backgrounds in terms of 
efficiencies (\%) at the HL-LHC.} \vspace{3mm}
\label{apptab:14bkgs}
\centering
 \begin{tabular}{l | cccc | ccccc }
\hline
  & \multicolumn{4}{c|}{Single-Higgs BG} & 
     \multicolumn{5}{c}{Non-resonant BG}  \\
  \hline
  Cuts & $ggH$ & $t\bar tH$ & $ZH$ & $b\bar b H$
       & $b\bar b\gamma\gamma$ &  $c\bar c\gamma\gamma$ &  $jj\gamma\gamma$ &
       $b\bar b j \gamma$ & $c\bar c j \gamma$ \\
    \hline
1. diphoton trigger & 18.36 & 23.37 & 18.22 & 17.27 &  
    17.86 & 16.81 & 0.22 & $1.43\times 10^{-2}$ & 0.02 \\ \hline
2. $\ge 2$ isolated photons & 7.43 & 21.43 & 11.87 & 2.88 &  
     12.16 & 11.53 & 0.15 & $8.43\times 10^{-3}$ & 0.01   \\ \hline
3-1.  jet candidates & 1.97 & 20.33 & 5.49 & 0.25  &  
    7.33 & 6.82 & 0.09 & $7.75\times 10^{-3}$ & 0.01  \\ \hline
3-2 $\ge 2$ two b-jet & $1.99\times 10^{-2}$ & 6.57 & 0.36 & $6.71\times 10^{-2}$  &  
    2.13 & 0.24 & $2.60\times 10^{-3}$ & $1.33\times 10^{-3}$ & $1.98\times 10^{-4}$   \\ \hline
4.  no. of jets $\leq 5$ & $1.94\times 10^{-2}$ & 5.16 & 0.36 & $6.70\times 10^{-2}$ & 
2.08 & 0.23 & $2.48\times 10^{-3}$ & $1.23\times 10^{-3}$ & $1.75\times 10^{-4}$  \\ \hline
5. lepton veto & $1.91\times 10^{-2}$ & 3.85 & 0.36 & $6.66 \times 10^{-2}$ & 
    2.07 & 0.23 & $2.42\times 10^{-3}$ & $1.23\times 10^{-3}$ & $1.71\times 10^{-4}$ \\ \hline
6. $\Delta R_{\gamma\gamma,bb}$ cut & $1.13 \times 10^{-2}$ & 1.16 & 0.26 & $1.73 \times 10^{-2}$ &  
    0.41 & 0.03 & $7.71\times 10^{-4}$ & $2.93\times 10^{-4}$ & $3.29\times 10^{-5}$ \\ \hline
7-1.  Higgs mass window $M_{\gamma\gamma}$  & $1.08\times 10^{-2}$ & 1.09 & 0.25 & $1.71 \times 10^{-2}$ & 
$1.85 \times 10^{-2}$ & $1.08\times 10^{-3}$ & $3.56 \times 10^{-5}$ & $8.30\times 10^{-6}$ & $9.35 \times 10^{-7}$ \\ \hline
7-2. Higgs mass window $M_{bb}$  & $1.92\times 10^{-3}$ & 0.37 & $5.39\times 10^{-2}$ & $4.20\times 10^{-3}$ & 
$4.85\times 10^{-3}$ & $2.20\times 10^{-4}$ & $1.14\times 10^{-5}$ & $2.33\times 10^{-6}$ & $2.65\times 10^{-7}$ \\ \hline
8. $p_{T_{\gamma\gamma}}$, $p_{T_{bb}}$ & $1.83\times 10^{-3}$ & 0.32 & $5.38 \times 10^{-2}$ & $3.90 \times 10^{-3}$ & 
$4.49 \times 10^{-3}$ & $2.10 \times 10^{-4}$ & $6.88\times 10^{-6}$ & $1.71\times 10^{-6}$ & $1.75\times 10^{-7}$   \\ \hline \hline
    other/barrel ratio & 46.6\% & 34.5\% & 48.3\% & 39.6\% & 
    69.1\% & 57.2\% & 110.0\%  & 80.4\% & 40.1\%
\\ \hline
\end{tabular} 

\vspace{0.5in}

{\small TABLE~\ref{apptab:14bkgs} (continued)}

\vspace{3mm}

\scriptsize
\centering
 \begin{tabular}{l | cc | cc }
\hline
 &  \multicolumn{2}{c|}{Non-resonant BG} &
    \multicolumn{2}{c}{$t\bar t$ related BG} \\
  \hline
Cuts & $b\bar b j j$ & 
       $Z(b\bar b) \gamma\gamma$ & $t\bar t$ & $t\bar t \gamma$ \\
    \hline
1. diphoton trigger & $7.33\times 10^{-6}$ & 18.70 &
 21.25 & 6.00  \\ \hline
2. $\ge 2$ isolated photons & $3.90\times 10^{-7}$ & 13.01 &
9.97 & 4.77 \\ \hline
3-1.  jet candidates & $3.90\times 10^{-7}$ & 6.11  &
8.86 & 4.18   \\ \hline
3-2 $\ge 2$ two b-jet  & $4.01\times 10^{-7}$ & 1.24 &
    2.23 & 1.21  \\ \hline
4.  no. of jets $\leq 5$ &  $2.85\times 10^{-7}$ & 1.22 &  
2.07 & 1.09  \\ \hline
5. lepton veto & $2.80\times 10^{-7}$ & 1.21 &
$2.07$ & $1.09$  \\ \hline
6. $\Delta R_{\gamma\gamma,bb}$ cut & $8.76\times 10^{-8}$ & 0.58 &  
0.37 & 0.18 \\ \hline
7-1.  Higgs mass window $M_{\gamma\gamma}$  &  $2.77 \times 10^{-9}$ & $2.64 \times 10^{-2}$ & 0.01 & $5.86 \times 10^{-3}$ \\ \hline
7-2. Higgs mass window $M_{bb}$  & $6.98\times 10^{-10}$ & $5.89 \times 10^{-3}$ & $3.79 \times 10^{-3}$ & $1.98 \times 10^{-3}$ \\ \hline
8. $p_{T_{\gamma\gamma}}$, $p_{T_{bb}}$ & $4.25 \times 10^{-10}$ & $5.80 \times 10^{-3}$ & $2.40 \times 10^{-3}$ & $1.74 \times 10^{-3}$ \\ \hline \hline
other/barrel ratio & 45.4\% & 66.6\% & 63.8\% & 57.6\%    
\\ \hline
\end{tabular} 
\end{table}

\begin{table}[ht]
\scriptsize
\caption { Cut flow table of the backgrounds
in terms of efficiencies (\%) at the HL-100 TeV hadron collider. }
\vspace{0.3cm}
\label{apptab:100bkgs}
\centering
 \begin{tabular}{l | cccc | ccccc }
\hline
  & \multicolumn{4}{c|}{Single-Higgs BG} &
     \multicolumn{5}{c}{Non-resonant BG}  \\
  \hline
  Cuts & $ggH$ & $t\bar tH$ & $ZH$ & $b\bar b H$
       & $b\bar b\gamma\gamma$ &  $c\bar c\gamma\gamma$ &  $jj\gamma\gamma$ &
       $b\bar b j \gamma$ & $c\bar c j \gamma$ \\
    \hline
1. diphoton trigger & 60.04 & 45.79 & 54.04 & 64.18 &
    44.55 & 44.13 & 0.33 & 0.08 & $7.58\times 10^{-2}$ \\ \hline
2. $\ge 2$ isolated photons & 22.87 & 31.53 & 22.91 & 11.97 &
     15.44 & 16.85 & 0.09 & 0.03 & $2.73 \times 10^{-2}$  \\ \hline
3-1.  jet candidates & 8.85 & 30.71 & 11.31 & 1.22  &
    10.52 & 12.02 & 0.06 & 0.03 & $2.56 \times 10^{-2}$  \\ \hline
3-2 $\ge 2$ two b-jet & 0.14 & 11.59 & 0.81 & 0.36  &
    3.14 & 0.19 & $1.52 \times 10^{-3}$ & 0.01 & $4.19\times 10^{-4}$   \\ \hline
4.  no. of jets $\le 5$ & 0.11 & 7.10 & 0.78 & 0.35 &
2.78 & 0.14 & $1.13\times 10^{-3}$ & $4.35\times 10^{-3}$ & $2.18\times 10^{-4}$
\\ \hline
5. lepton veto & 0.11 & 5.20 & 0.78 & 0.35 &
    2.78 & 0.14 & $1.13\times 10^{-3}$ & $4.35\times 10^{-3}$ & $2.18\times
10^{-4}$ \\ \hline
6. $\Delta R_{\gamma\gamma,bb}$ cut & 0.10 & 3.79 & 0.71 & 0.19 &
    1.62 & 0.08 & $7.78\times 10^{-4}$ & $2.30\times 10^{-3}$ & $1.03 \times
10^{-4}$ \\ \hline
7-1.  Higgs mass window $M_{\gamma\gamma}$  & 0.09 & 3.45 & 0.67 & 0.18 &
0.07 & $3.35\times 10^{-3}$ & $3.23 \times 10^{-5}$ & $6.38\times 10^{-5}$ & $3.29
\times 10^{-6}$ \\ \hline
7-2. Higgs mass window $M_{bb}$  & 0.02 & 0.97 & 0.33 & 0.04 &
0.02 & $9.45\times 10^{-4}$ & $8.20\times 10^{-6}$ & $2.07\times 10^{-5}$ &
$1.08\times 10^{-6}$ \\ \hline
8. $p_{T_{\gamma\gamma}}$, $p_{T_{bb}}$ & 0.02 & 0.40 & 0.22 & 0.02 &
$5.21 \times 10^{-3}$ & $1.64 \times 10^{-4}$ & $2.00\times 10^{-6}$ & $4.23\times
10^{-6}$ & $2.33\times 10^{-7}$   \\ \hline \hline
    other/barrel ratio & 19.9\% & 31.8\% & 37.4\% & 40.3\% &
    49.6\% & 100.0\% & 53.8\%  & 42.0\% & 35.7\%  \\
\hline
\end{tabular}

\vspace{0.5in}
{\small TABLE~\ref{apptab:100bkgs} (continued)}

\vspace{3mm}

\scriptsize
 \begin{tabular}{l | cc | cc }
\hline
 &  \multicolumn{2}{c|}{Non-resonant BG} &
    \multicolumn{2}{c}{$t\bar t$ related BG} \\
  \hline
Cuts & $b\bar b j j$ &
       $Z(b\bar b) \gamma\gamma$ & $t\bar t$ & $t\bar t \gamma$ \\
    \hline
1. diphoton trigger & $1.33\times 10^{-4}$ & 45.38 &
 14.61 & 10.49  \\ \hline
2. $\ge 2$ isolated photons & $5.77\times 10^{-5}$ & 14.85 &
5.98 & 5.62 \\ \hline
3-1.  jet candidates & $5.77\times 10^{-5}$ & 9.28  &
5.85 & 5.39   \\ \hline
3-2 $\ge 2$ two b-jet  & $1.01\times 10^{-5}$ & 2.06 &
    1.81 & 1.88  \\ \hline
4.  no. of jets $\le 5$ &  $5.41\times 10^{-6}$ & 1.92 &
1.28 & 1.32  \\ \hline
5. lepton veto & $5.41 \times 10^{-6}$ & 1.92 &
1.28 & 1.32  \\ \hline
6. $\Delta R_{\gamma\gamma,bb}$ cut & $3.17\times 10^{-6}$ & 1.68 &
0.75 & 0.75 \\ \hline
7-1.  Higgs mass window $M_{\gamma\gamma}$  &  $8.44\times 10^{-8}$ & 0.07 & 0.02
& 0.02 \\ \hline
7-2. Higgs mass window $M_{bb}$  & $2.79\times 10^{-8}$ & 0.04 & 0.01 & 0.01 \\
\hline
8. $p_{T_{\gamma\gamma}}$, $p_{T_{bb}}$ & $7.44 \times 10^{-9}$ & 0.02 & $1.31
\times 10^{-3}$ & $1.95 \times 10^{-3}$ \\ \hline \hline
other/barrel ratio & 55.6\% & 53.6\% & 54.8\% & 69.0\%     \\
\hline
\end{tabular}
\end{table}

$$ $$
\newpage

\section{On the cross sections of non-resonant backgrounds}
\label{app:C}

For the non-resonant continuum backgrounds of
$b\bar{b} \gamma\gamma$, $c\bar{c} \gamma\gamma$,
$jj\gamma\gamma$, $b\bar{b}j\gamma$, $c\bar{c}j\gamma$, $b\bar{b}jj$
and $Z(b\bar{b})\gamma\gamma$,
we have estimated the cross sections by applying the 
generator-level pre-selection cuts listed in Eq.~(\ref{eq:presel}). 
As explained in the main text, in each background, we consider a process with
an additional hard parton\footnote{In this appendix, we use the term of
`parton' instead of
`jet' to make distinction from a clustered jet obtained by collecting several hard and
soft partons.}
at the matrix-element level
to capture the bulk of the NLO corrections.

In our estimation, there might be a worry of double counting 
between the leading process and the sub-leading one with an additional hard parton
when generated
background event samples are interfaced with $\mathtt{PYTHIA8}$
for showering and hadronization.
To study the double counting issue, taking the PDF set of {\tt CT14LO},
we consider the following three types of cross sections:
\begin{itemize}
\item  $\sigma_{\rm Eq.(6)}$ without matching: the cross section obtained 
by applying the 
generator-level pre-selection cuts listed in Eq.~(\ref{eq:presel})
\item  $\sigma_{\rm\bf xqcut}$ without matching: the cross section obtained 
by varying {\bf xqcut}. The variation of {\bf xqcut} affects
the pre-selection cuts on $P_{T_j}$, $M_{jj}$, and $\Delta R_{jj}$.
Otherwise, the other pre-selection cuts remain the same as
in Eq.~(\ref{eq:presel}).
\item  $\sigma_{\rm merged}$ with MLM matching: the cross section obtained
after implementing MLM matching. The merged cross section depends on  
the parameters of {\bf xqcut} and $Q_{\rm cut}$.
In the default $\mathtt{MG5\_aMC@NLO}$ setting, 
when a value of {\bf xqcut} is given,
three merged cross sections are provided for the three values of
$Q_{\rm cut}/{\rm\bf xqcut}$: $1.5$, $2.25$, and $3$. 
For the representative value, the merged cross section
with $Q_{\rm cut}/{\rm\bf xqcut}=1.5$ is taken.
\end{itemize}

For further discussion, it is helpful to introduce
the distance between the two objects ($d_{ij}$) and
that between an object and the beam direction ($d_{iB}$).
Here an object could stand for a hard parton at the matrix-element level,
a showering soft parton, or a clustered jet.
Precisely,
\begin{equation}
d_{ij} = {\mathrm{min} \left( P_{T_{i}}^{2p}\, , P_{T_{j}}^{2p} \right) 
\frac{\Delta R_{ij}^2}{R^2}~}\,, \ \ \
d_{iB} = {P_{T_{i}}^{2p}}\,,
\end{equation}
where the parameter $R$ defines the jet size and the parameter
$p$ the jet algorithm used. In MLM matching, the $k_T$ algorithm with $p=1$
is used.
We note that $\sqrt{d_{iB}}$ in the $k_T$ algorithm  is nothing but $P_{T_i}$ or
the transverse momentum of an object.

Roughly speaking, the calculation of the merged cross section proceeds
as the following steps:
\begin{itemize}
\item[$(i)$] generation of hard partons with 
$\sqrt{d_{ij}}\,,\sqrt{d_{iB}}>{\rm\bf xqcut}$ at the matrix-element level
\item[$(ii)$] showering soft partons with $\sqrt{d_{ij}}\,,\sqrt{d_{iB}}<\mu_F$
with $\mu_F$ being the factorization scale
\item[$(iii)$] clustering partons and pseudo-partons
into jets according to a certain jet algorithm
until all the distances among clustered jets and the beam direction
are smaller than $Q_{\rm cut}^2$
\item[$(iv)$] matching by requiring that the number of jets obtained at the step
$(iii)$ should be equal to the number of hard partons at the step $(i)$ 
\footnote{Sometimes, for the highest multiplicity sample, the number of jets 
is required to be equal to or larger than the number of hard partons.}
and the distance between a jet and its nearest hard parton is smaller than
${\rm max}\{Q_{\rm cut}^2, P_{T}^2\}$ with $P_{T}$ being 
the transverse momentum of the nearest hard parton
\item[$(v)$] calculating the merged cross section by exploiting  the weight factors 
and other information obtained in the matching step $(iv)$
\end{itemize}

\begin{table}[t]
\caption{ {\bf HL-LHC}:
The cross sections for the non-resonant backgrounds
taking the PDF set of {\tt CT14LO}.
For the three merged cross sections,
$Q_{\rm cut}/{\rm GeV}=30$ (upper), 45 (middle), 60 (low) are taken
with the parameter {\bf xqcut} set to 20 GeV. 
%
} 
\vspace{3mm}
\label{table:updatedXsection2}
\begin{tabular}{|c||c|c|c|c|c|c|c|}
\hline
 Cross Section & $b\bar{b}\gamma\gamma$ & $c \bar{c} \gamma \gamma$ &
$jj \gamma\gamma$ & $b \bar{b} j\gamma$ & $c \bar{c} j \gamma$ &
$b \bar{b} j j$ & $Z(b\bar{b}) \gamma\gamma$ \\
\hline
\hline
 $\sigma_{\mathrm{\,Eq.(6)}}~\mathrm{[fb]}$ & 112 & 1081 & $1.40\times 10^4$
& $2.72\times 10^5$ & $0.91\times 10^6$ & $3.00\times 10^8$ & $5.03$  \\
\hline
\multirow{3}{*}{}      & 82.5 & 647 & $0.59 \times 10^4$ &
$1.22 \times 10^5$ & $0.35 \times 10^6$ & $0.67\times 10^8$ & 3.65 \\
   $\sigma_{\rm merged}~\mathrm{[fb]}$ & 82.3 & 662 & $0.44 \times 10^4$ &
$0.96 \times 10^5$ & $0.25 \times 10^6$ & $0.28 \times 10^8$ & 3.68 \\
                         & 81.5 & 662 & $0.34 \times 10^4$ &
$0.78 \times 10^5$ & $0.18 \times 10^6$ & $0.13\times 10^8$ & 3.68 \\
\hline
$\delta\sigma/\sigma$ [\%] & $1.2$ & $2.3$ & $42$ & $36$ & $49$ & $81$ & $0.8$ \\
\hline
\end{tabular}
\end{table}
%
%
In Table~\ref{table:updatedXsection2}, we present the cross sections of
$\sigma_{\rm\,Eq.(6)}$ and $\sigma_{\rm merged}$.
For the three merged cross sections,
$Q_{\rm cut}/{\rm GeV}=30$ (upper), 45 (middle), 60 (low) are taken
with the parameter {\bf xqcut} set to 20 GeV. 
Note that the smaller value of $Q_{\rm cut}$ usually results in the larger 
$\sigma_{\rm merged}$.
First of all, we observe that $\sigma_{\rm\,Eq.(6)}$'s are smaller than
those presented in Table~\ref{tab:ParticleList}. 
This is because the PDF set of {\tt CT14LO} is taken for this table
while, in Table~\ref{tab:ParticleList}, the PDF set of {\tt CTEQ6L1} is taken.
The difference between $\sigma_{\rm Eq.(6)}$ and $\sigma_{\rm merged}$
could be interpreted as the degree of double counting.
Further, the variation of the merged cross sections depending
on the choice of $Q_{\rm cut}$  may
provide a measure of the quality of the matching.
%
For quantitative estimation of the matching quality, we introduce the following
quantity:
$$\frac{\delta\sigma}{\sigma} \equiv
\frac{\left|\sigma_{\rm merged}^{Q_{\rm cut}/{\bf\rm xqcut}=1.5}
-\sigma_{\rm merged}^{Q_{\rm cut}/{\bf\rm xqcut}=3}\right|}{
\sigma_{\rm merged}^{Q_{\rm cut}/{\bf\rm xqcut}=1.5}}\,.$$
We observe $\delta\sigma/\sigma$ is less than about 2\% for
$b\bar b \gamma\gamma$, $c\bar c \gamma\gamma$, and $Z(b\bar b)\gamma\gamma$
and it is about 40\% for $b\bar b j\gamma$, $c\bar c j\gamma$, and $jj\gamma\gamma$.
For $b\bar b jj$, on the other hand, it amounts to more than $80$\%.

\begin{figure}
\centering
\includegraphics[width=3in]{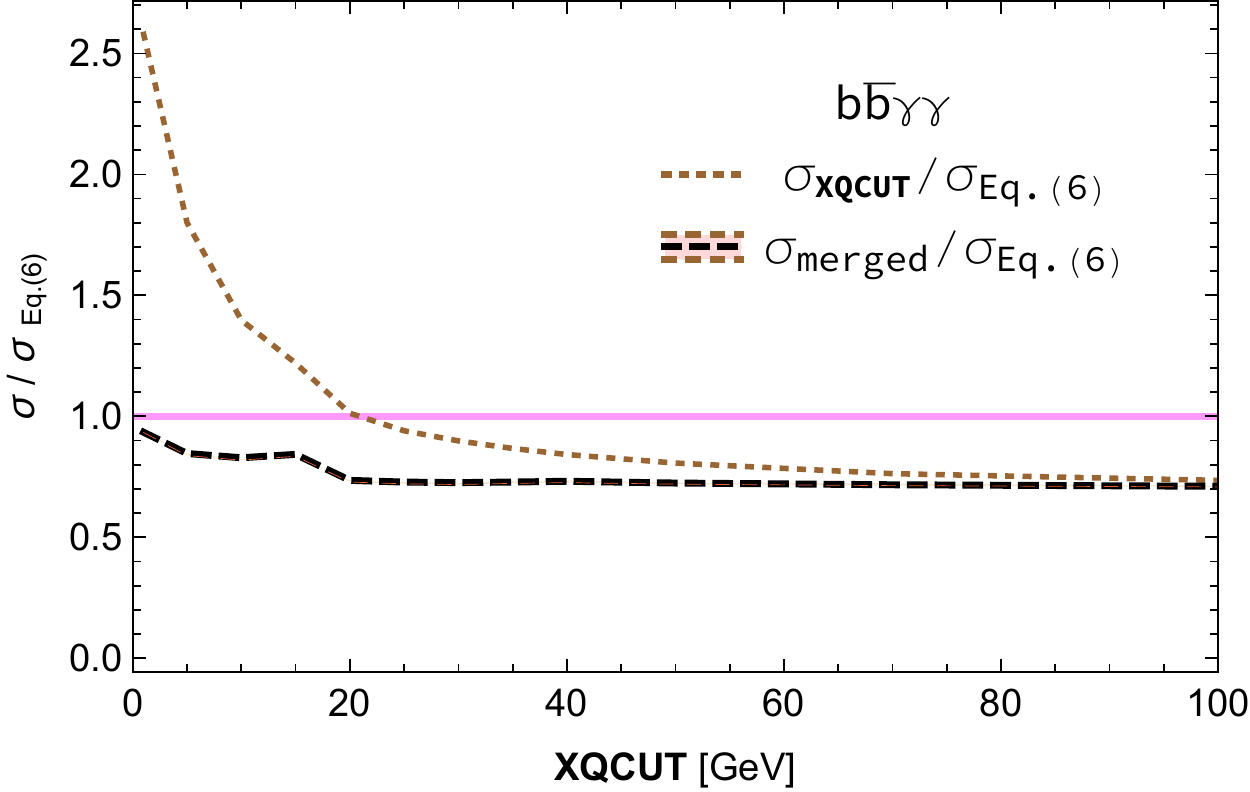}
\includegraphics[width=3in]{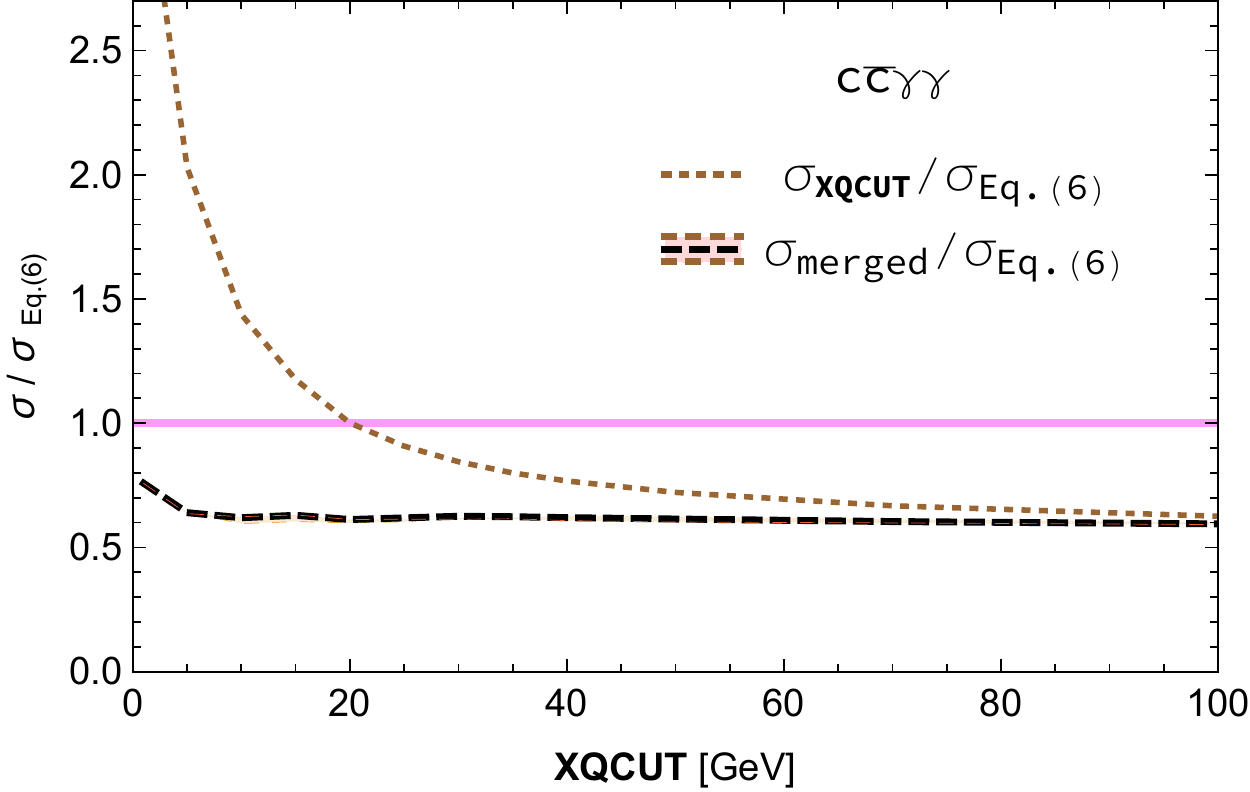}
\includegraphics[width=3in]{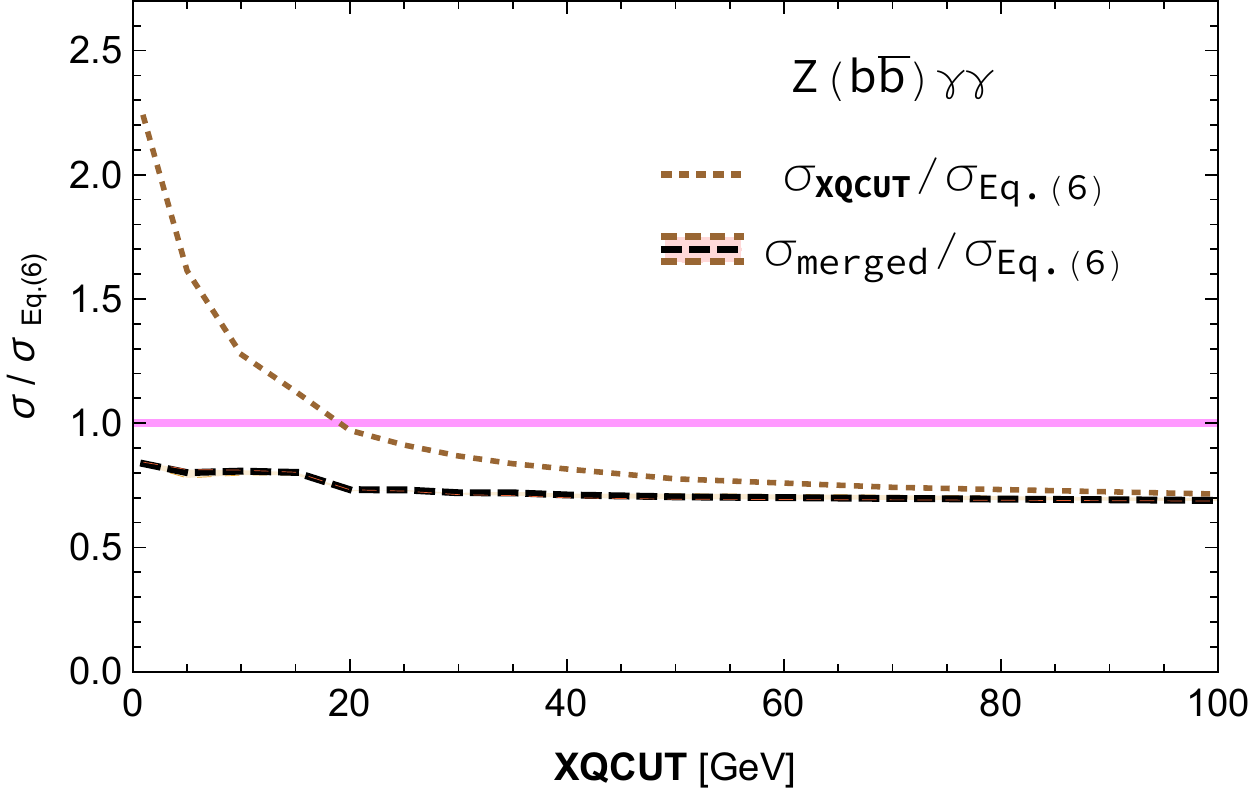}
\caption{\label{fig:XXaa}
The dependence of the ratios of $\sigma_{\rm\bf xqcut}/\sigma_{\rm Eq.(6)}$ 
(dotted lines) and $\sigma_{\rm merged}/\sigma_{\rm Eq.(6)}$ 
(bands) on {\bf xqcut} for the non-resonant
backgrounds of $b\bar b \gamma\gamma$ (upper left), 
$c\bar c \gamma\gamma$ (upper right),
and $Z(b\bar b) \gamma\gamma$ (lower) . 
The horizontal magenta lines locate the positions where
$\sigma_{\rm\bf xqcut}=\sigma_{\rm Eq.(6)}$. 
The bands show the variation of the merged cross sections depending 
on the choice of $Q_{\rm cut}/{\rm\bf xqcut}$: $1.5$, $3$ 
(upper and lower boundaries) and $2.25$ (middle dashed line).
The band width for all these 3 processes is negligible.
}
\end{figure}
Fig.~\ref{fig:XXaa} shows the ratios of $\sigma_{\rm\bf xqcut}/\sigma_{\rm Eq.(6)}$ 
and $\sigma_{\rm merged}/\sigma_{\rm Eq.(6)}$
as functions of {\bf xqcut} for the non-resonant backgrounds 
of $b\bar b \gamma\gamma$ (upper left), $c\bar c \gamma\gamma$ (upper right),
and $Z(b\bar b) \gamma\gamma$ (lower).
In each frame,
the dotted curve is for $\sigma_{\rm\bf xqcut}/\sigma_{\rm Eq.(6)}$ and
the band with a dashed line at its center for 
$\sigma_{\rm merged}/\sigma_{\rm Eq.(6)}$. A band is delimited by 
the choices of $Q_{\rm cut}/{\rm\bf xqcut}=1.5$ and $3$
while the center line is obtained by taking $Q_{\rm cut}/{\rm\bf xqcut}=2.25$.
For a given value of {\bf xqcut},
the larger value of $Q_{\rm cut}$ usually leads to
the smaller merged cross section.
First of all, we observe that 
$\sigma_{\rm\bf xqcut}=\sigma_{\rm Eq.(6)}$ around ${\rm\bf xqcut}\simeq 20$ GeV
which is nothing but the value of $P_{T_j}$ cut, see Eq.~\ref{eq:presel}.
And $\sigma_{\rm merged}$ is always smaller than $\sigma_{\rm\bf xqcut}$ and
the difference between them could be interpreted as the degree of double counting.
We note that the difference becomes smaller when {\bf xqcut} grows. This is 
because the leading process without an additional hard parton dominates more and more
as the value of {\bf xqcut} becomes large.
For the choice of $Q_{\rm cut}/{\rm\bf xqcut}=1.5$ and ${\rm\bf xqcut}=20$ GeV,
compared to $\sigma_{\rm\bf xqcut}$,
the merged cross sections for 
$b\bar b \gamma\gamma$, $c\bar c \gamma\gamma$,
and $Z(b\bar b) \gamma\gamma$ 
decrease by about 30\%.
Incidentally, we note the band widths are negligible 
for $b\bar b \gamma\gamma$, $c\bar c \gamma\gamma$,
and $Z(b\bar b) \gamma\gamma$.

\begin{figure}
\centering
\includegraphics[width=3in]{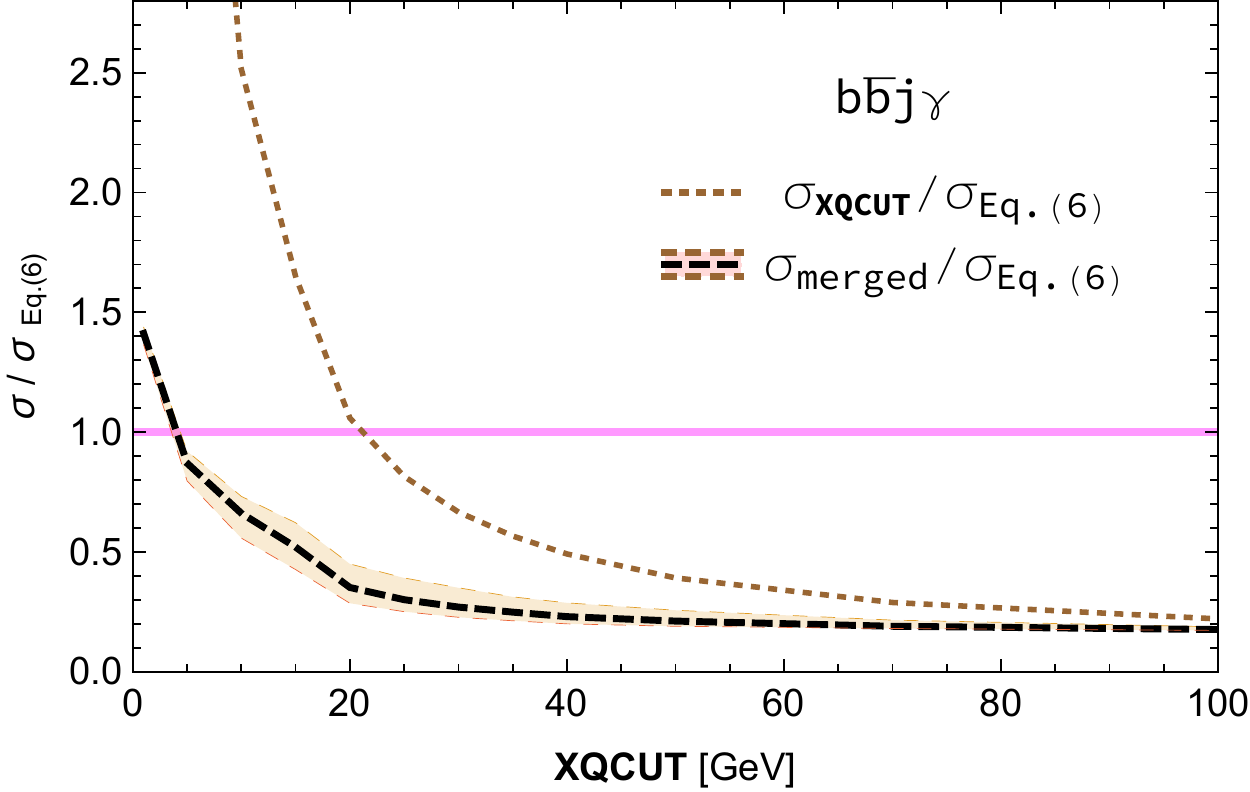}
\includegraphics[width=3in]{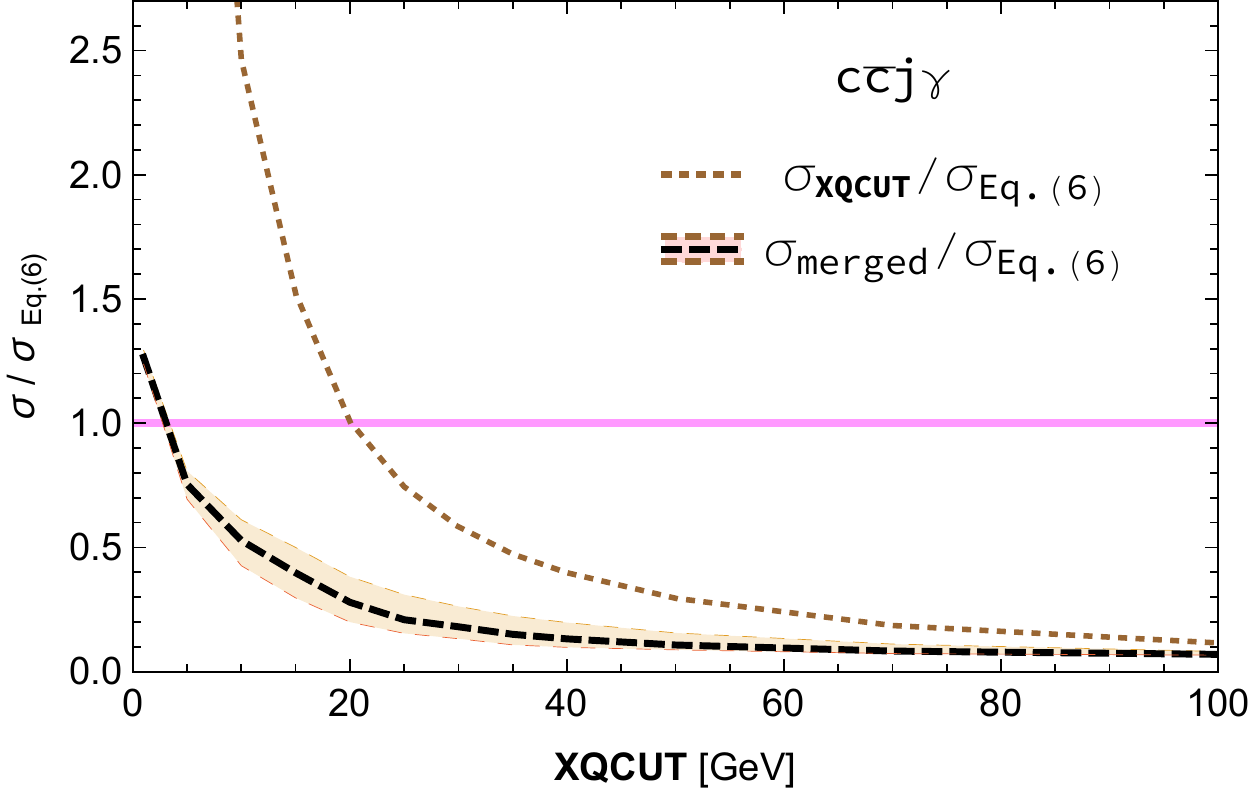}
\includegraphics[width=3in]{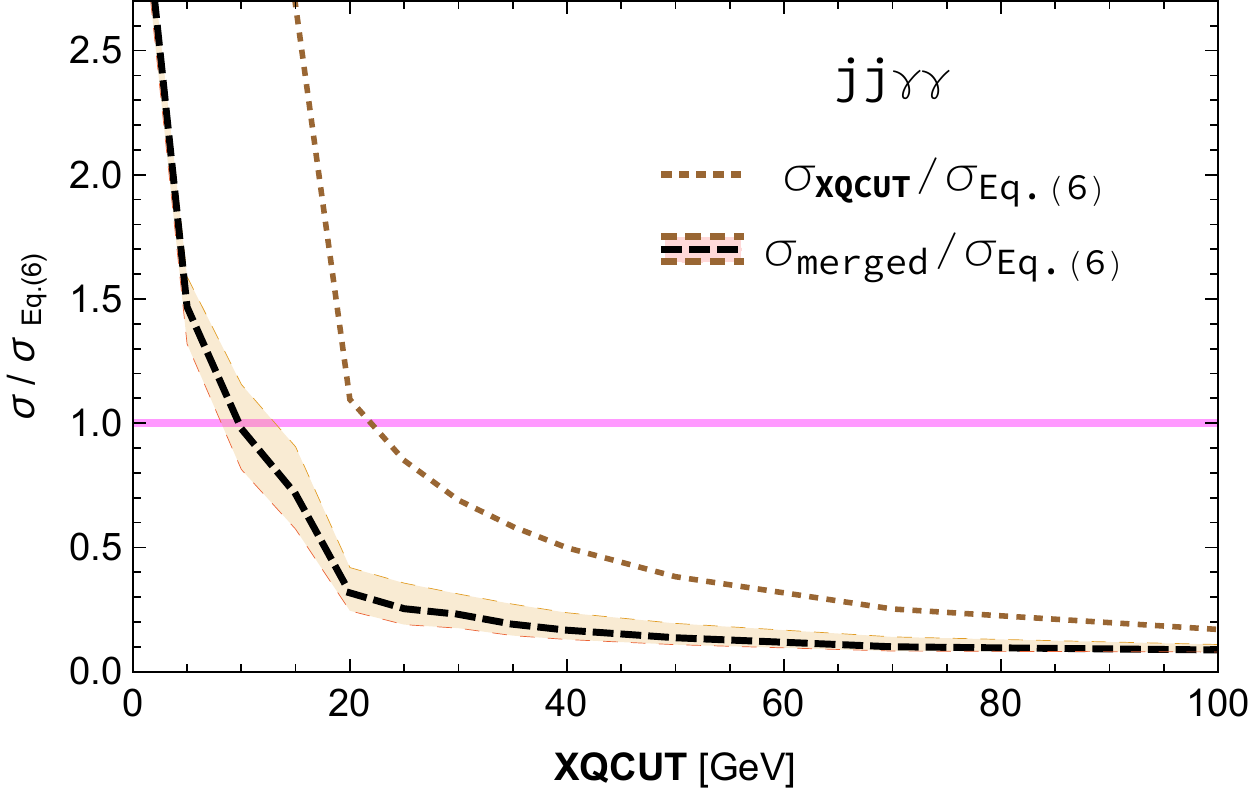}
\includegraphics[width=3in]{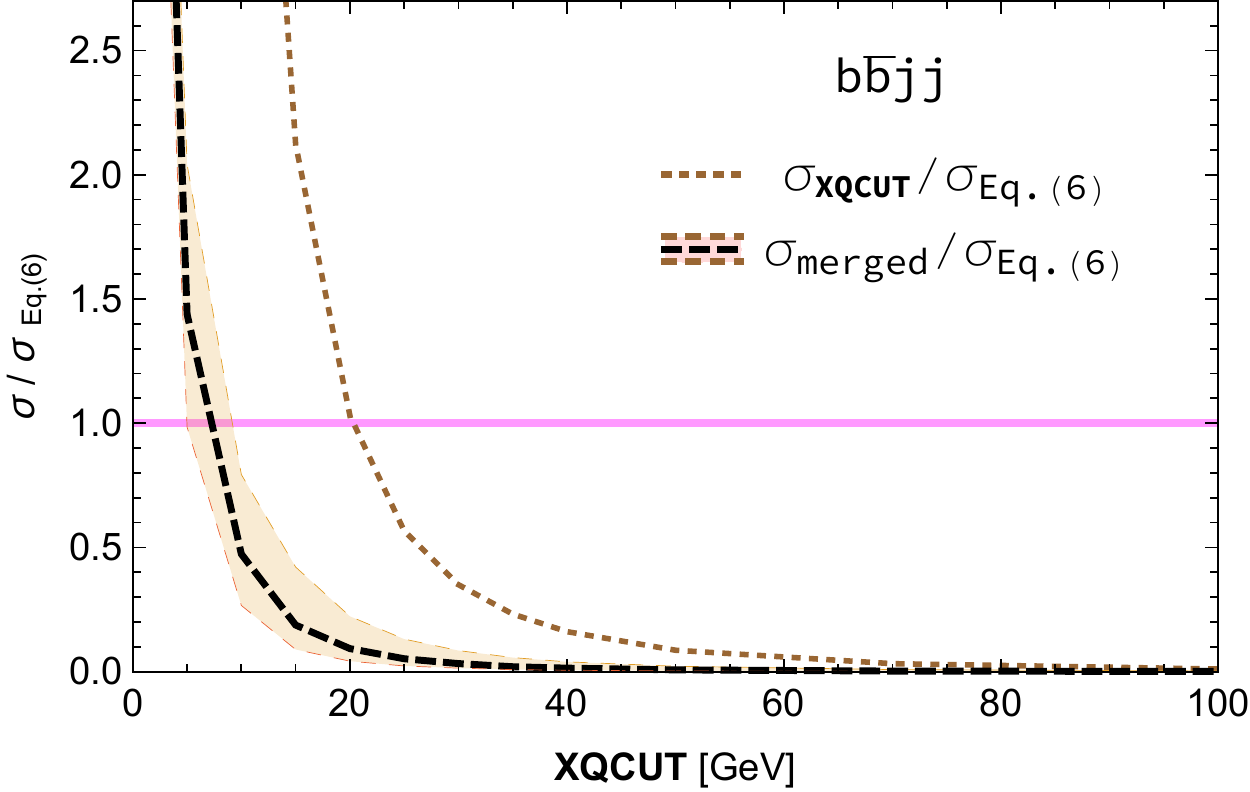}
\caption{\label{fig:XXjY}
The same as in Fig.~\ref{fig:XXaa} but for the the non-resonant backgrounds 
of $b\bar b j\gamma$ (upper left), $c\bar c j\gamma$ (upper right), 
$jj\gamma\gamma$ (lower left) and $b\bar b jj$ (lower right).
}
\end{figure}
Fig.~\ref{fig:XXjY} shows the ratios of $\sigma_{\rm\bf xqcut}/\sigma_{\rm Eq.(6)}$ 
and $\sigma_{\rm merged}/\sigma_{\rm Eq.(6)}$
as functions of {\bf xqcut} for the non-resonant backgrounds 
of $b\bar b j\gamma$ (upper left), $c\bar c j\gamma$ (upper right), 
$jj\gamma\gamma$ (lower left) and $b\bar b jj$ (lower right).
%
%
Compared to $b\bar b \gamma\gamma$, $c\bar c \gamma\gamma$,
and $Z(b\bar b) \gamma\gamma$ in Fig.~\ref{fig:XXaa},
the reduction of the merged cross sections is larger 
and the band width is sizeable.

\begin{figure}
\centering
\includegraphics[width=2.0in]{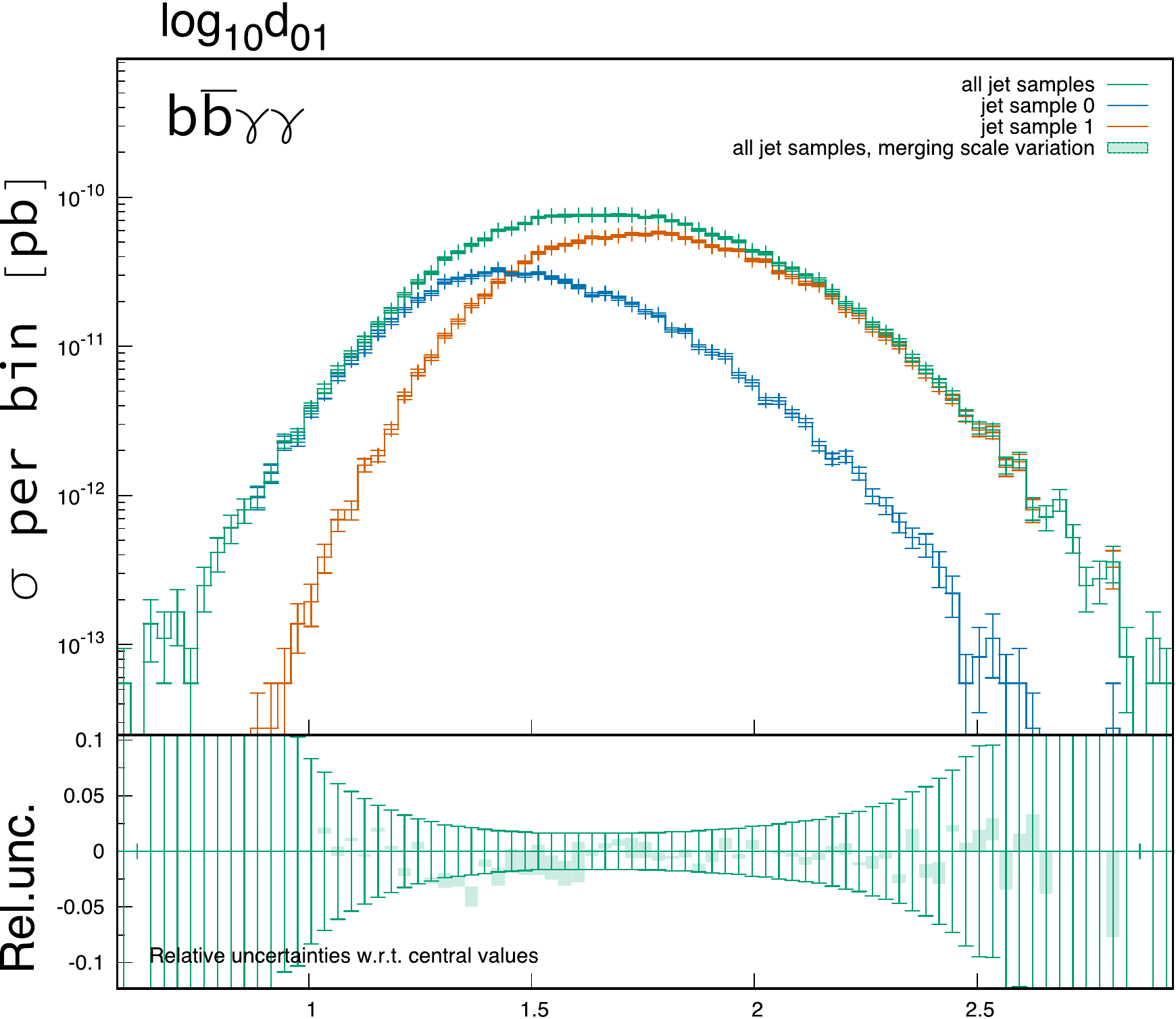}
\includegraphics[width=2.0in]{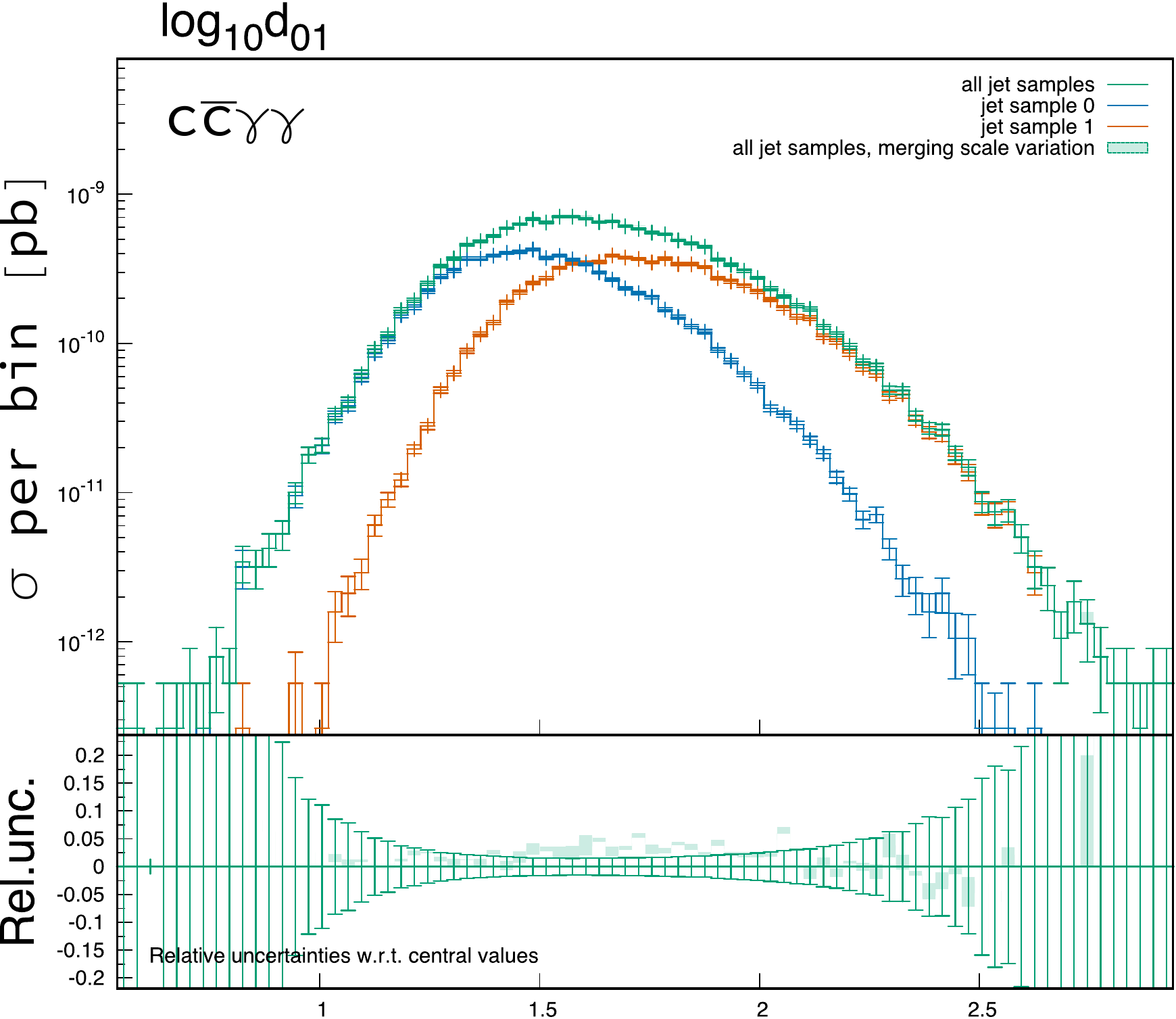}
\includegraphics[width=2.0in]{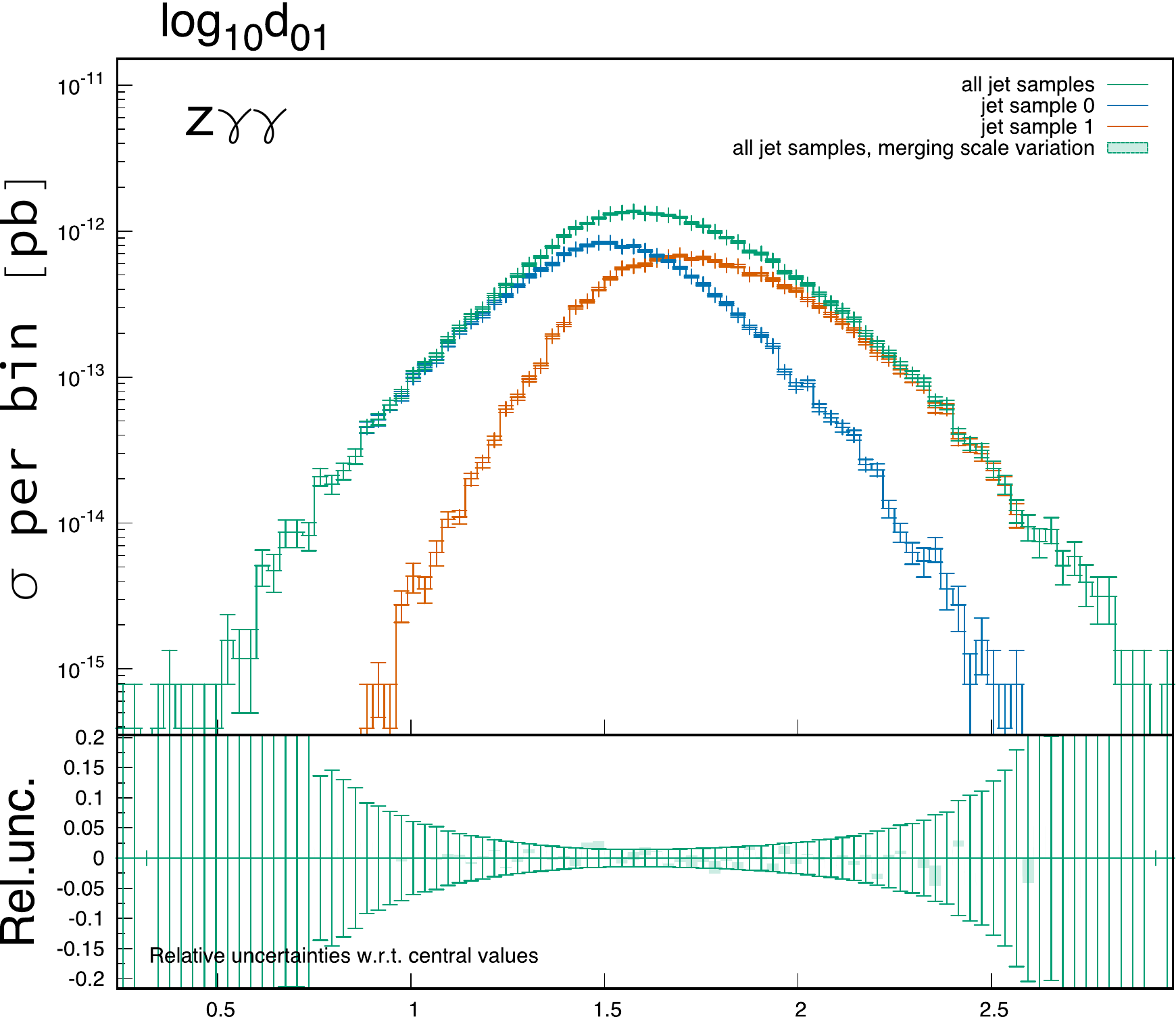}
\caption{\label{fig:djr_1}
{\bf HL-LHC}:
The Differential Jet Rate (DJR) distributions for the non-resonant
backgrounds of
$b\bar b \gamma\gamma$ (left), $c\bar c \gamma\gamma$ (middle), 
and $Z(b\bar b)\gamma\gamma$ (right)
taking {\bf xqcut}$=20$ GeV and $Q_{\rm cut}=30$ GeV.
Here, ``jet sample 0" and ``jet sample 1" 
refer to the samples containing
$0$ and $1$ hard parton, respectively, with 
$\sqrt{d_{ij}}\,,\sqrt{d_{iB}}>{\rm\bf xqcut}$ at the matrix-element level.
}
\end{figure}
\begin{figure}
\centering
\includegraphics[width=2.0in]{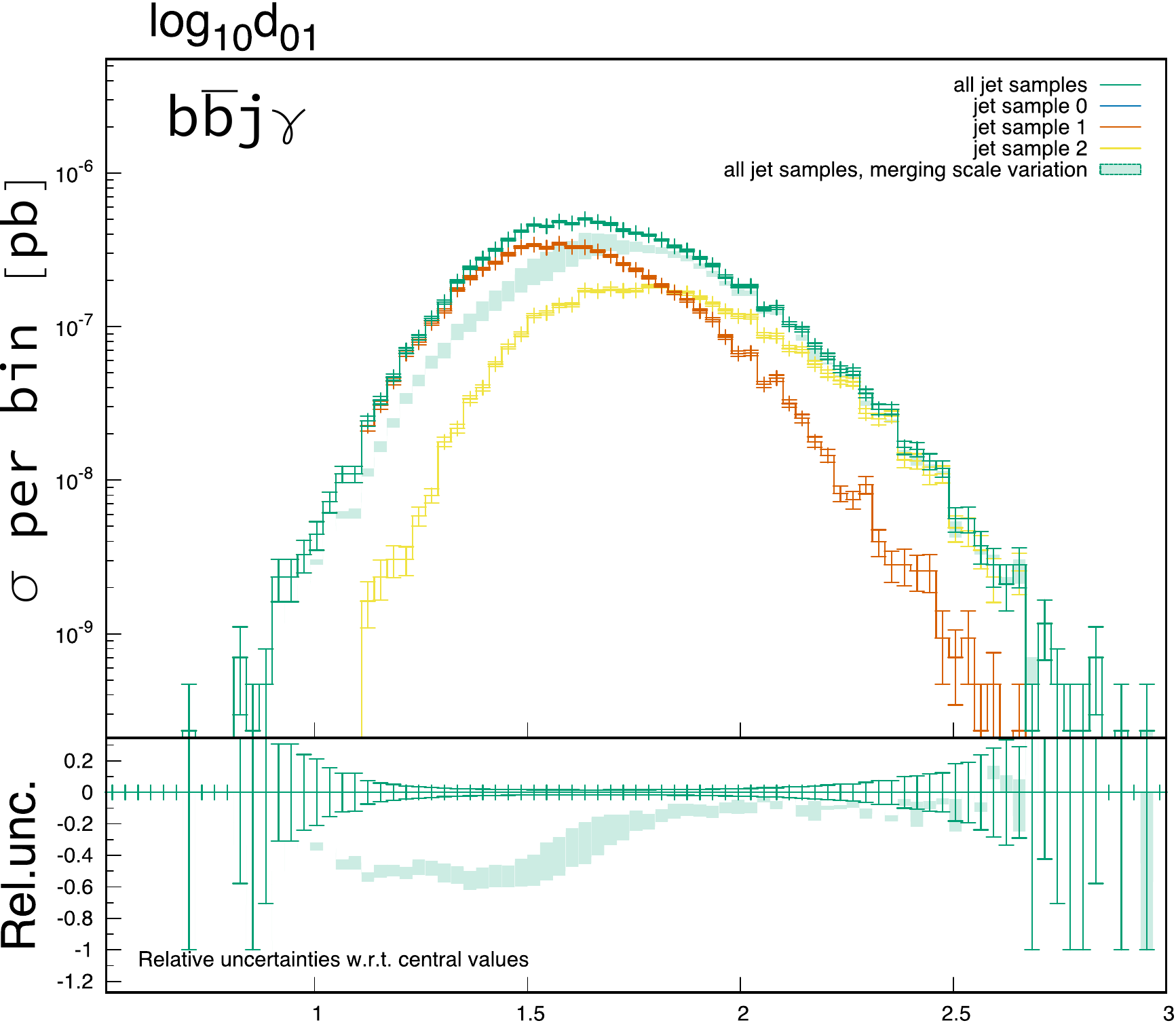}
\includegraphics[width=2.0in]{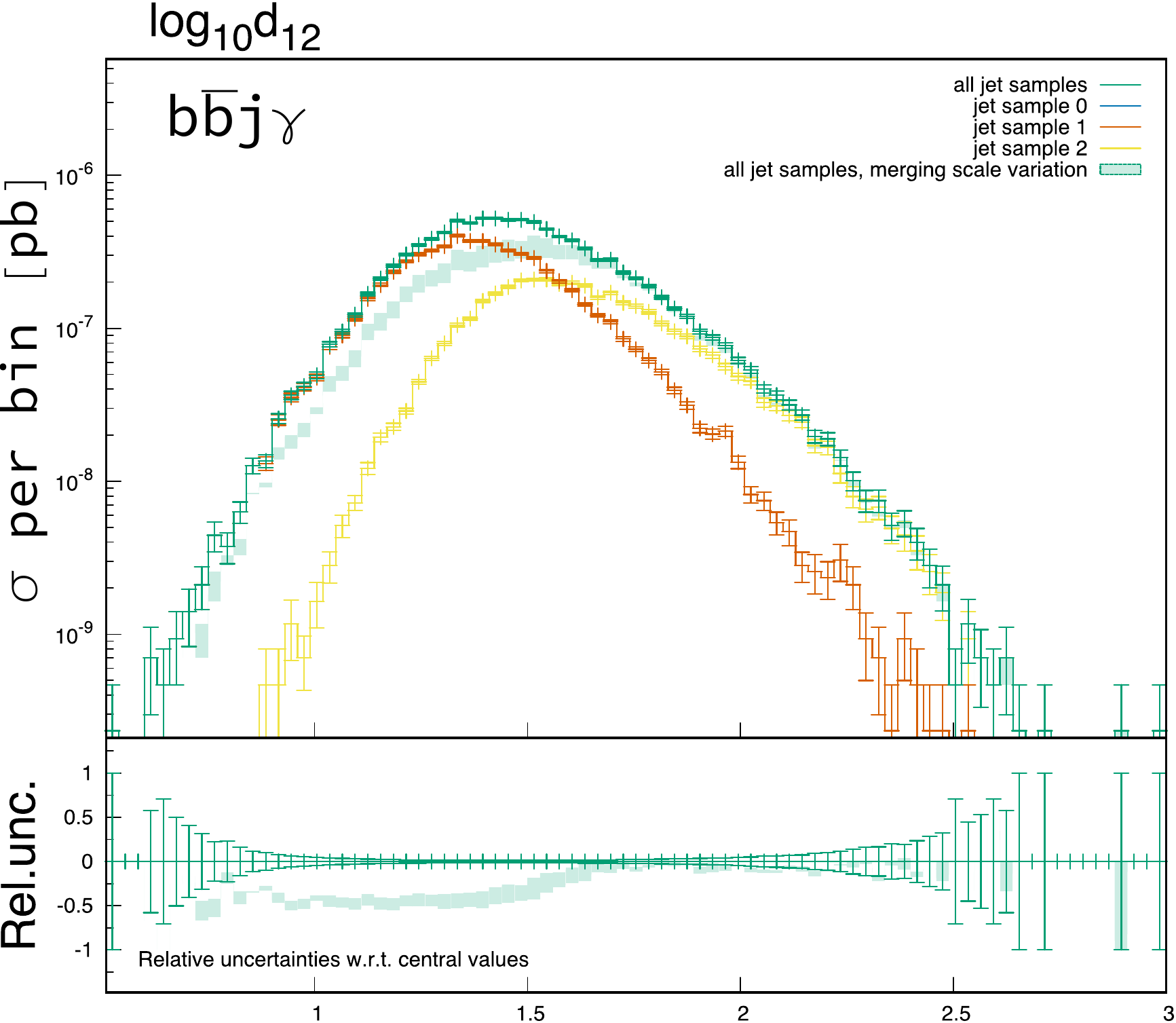} \\
\includegraphics[width=2.0in]{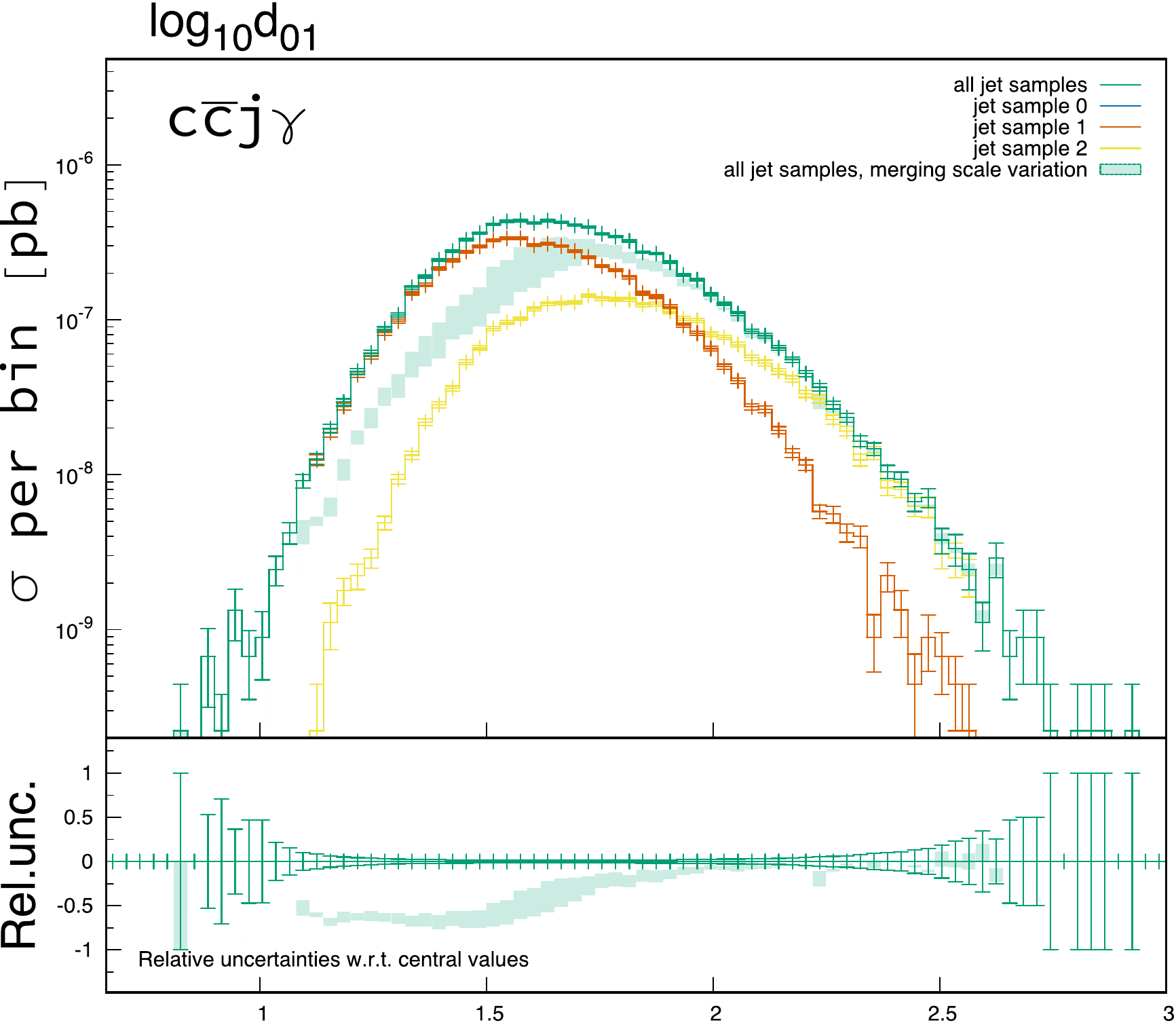}
\includegraphics[width=2.0in]{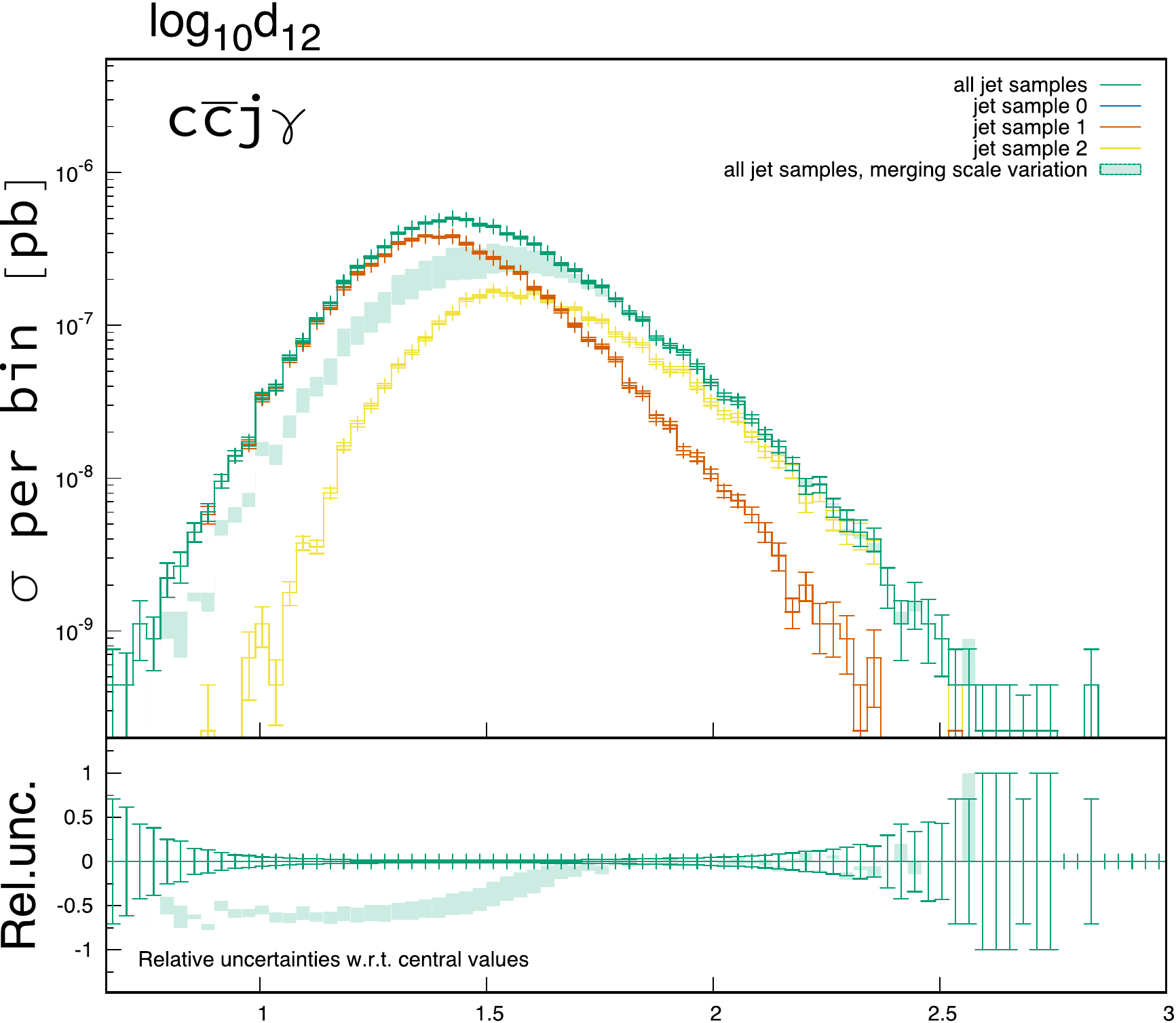}
\caption{\label{fig:djr_2}
{\bf HL-LHC}:
The DJR distributions for the non-resonant
backgrounds of
$b\bar b j\gamma$ (upper) and $c\bar c j\gamma$ (lower)
taking {\bf xqcut}$=20$ GeV and $Q_{\rm cut}=30$ GeV.
Here, ``jet sample $n$" refers to the sample containing $n$ hard partons
at the matrix-element level.
}
\end{figure}
\begin{figure}
\centering
\includegraphics[width=2.0in]{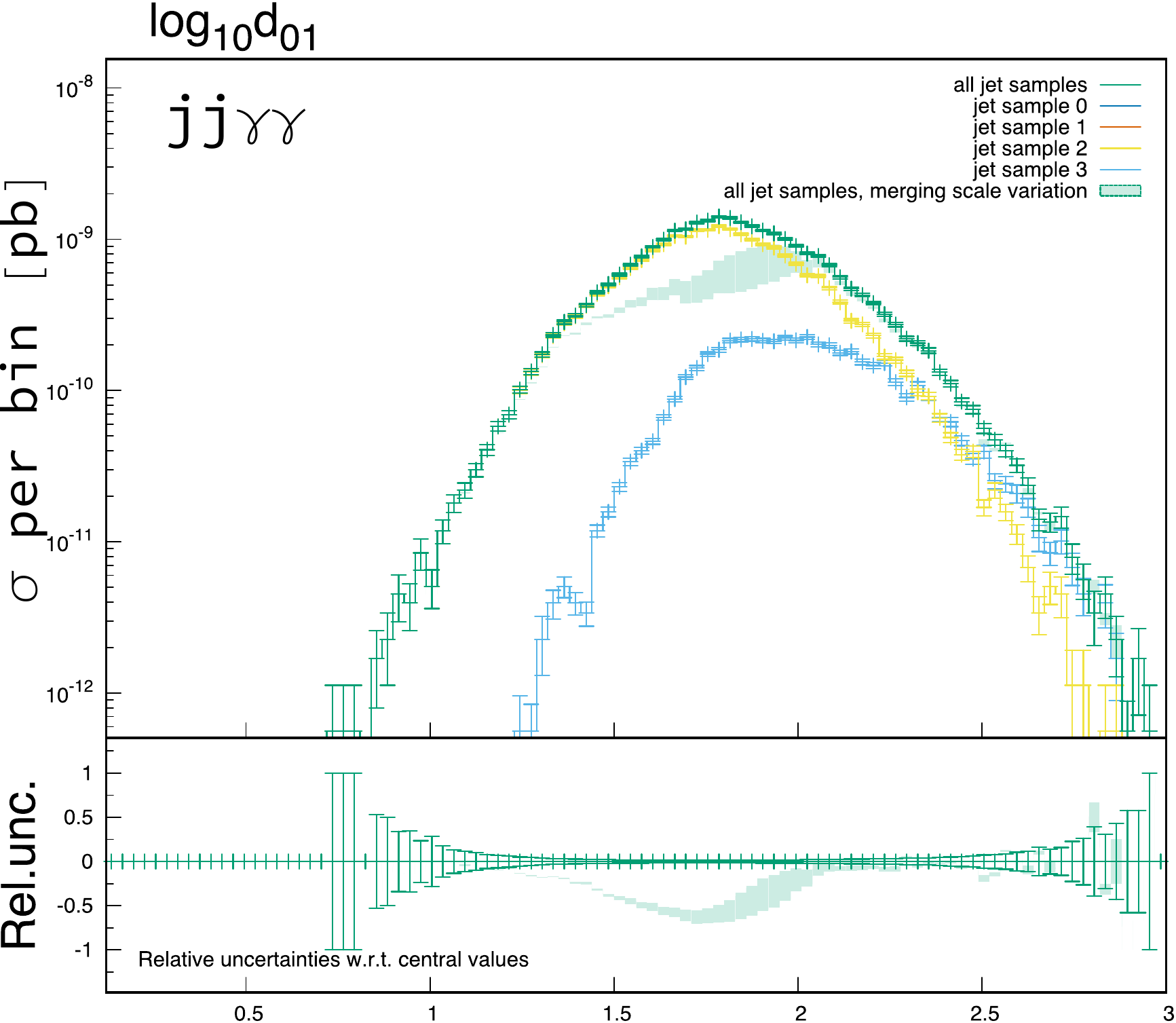}
\includegraphics[width=2.0in]{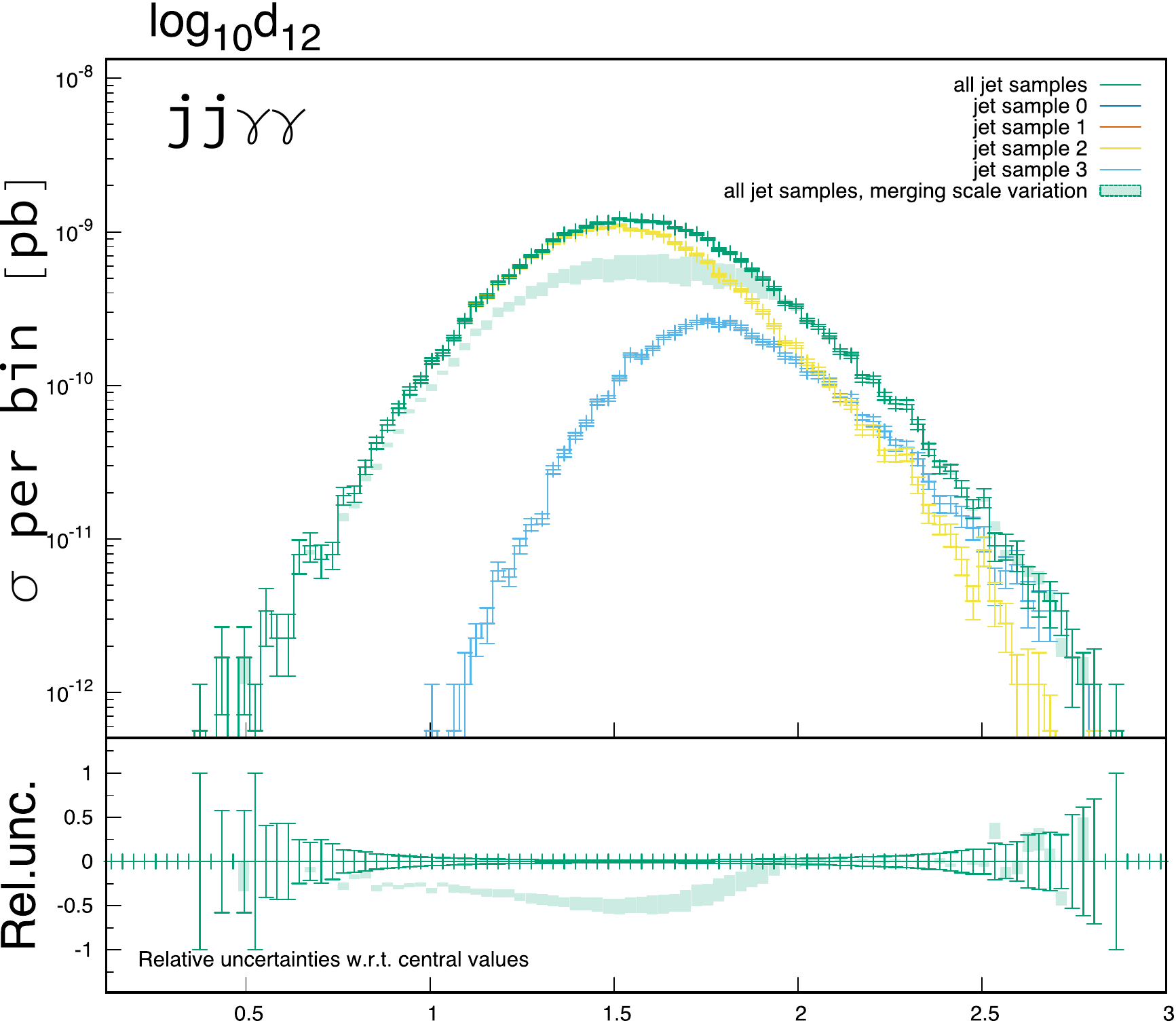}
\includegraphics[width=2.0in]{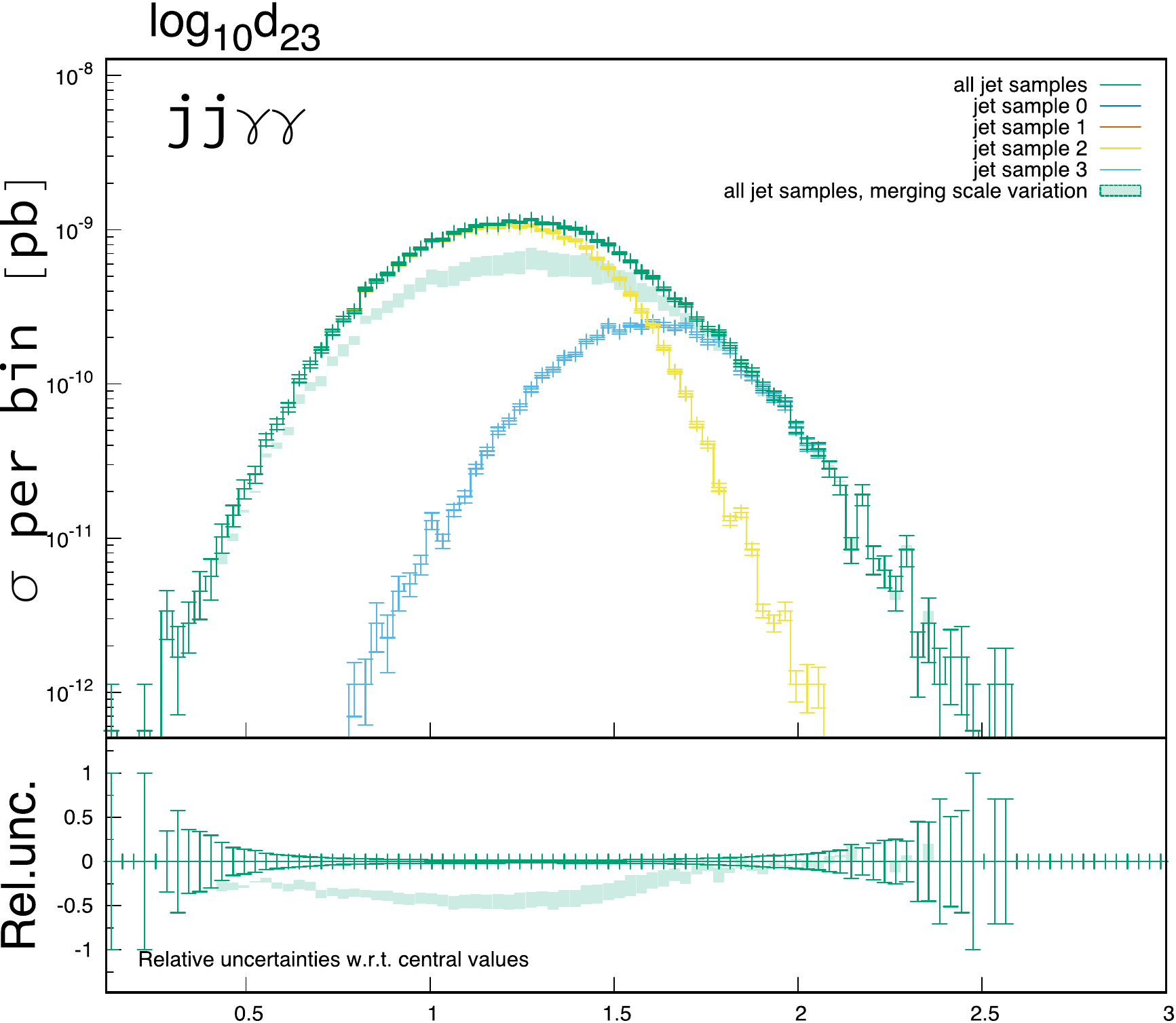}
\includegraphics[width=2.0in]{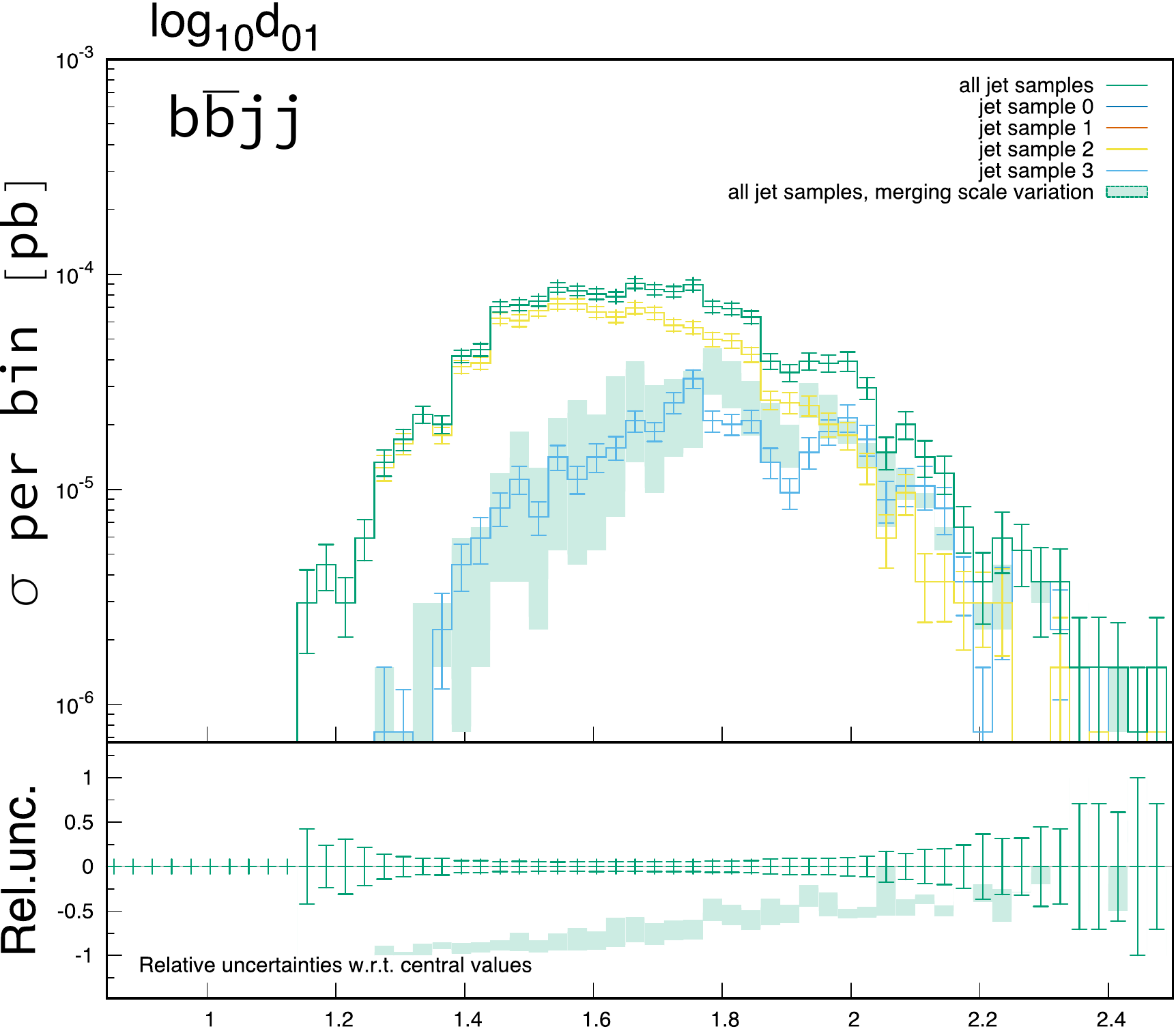}
\includegraphics[width=2.0in]{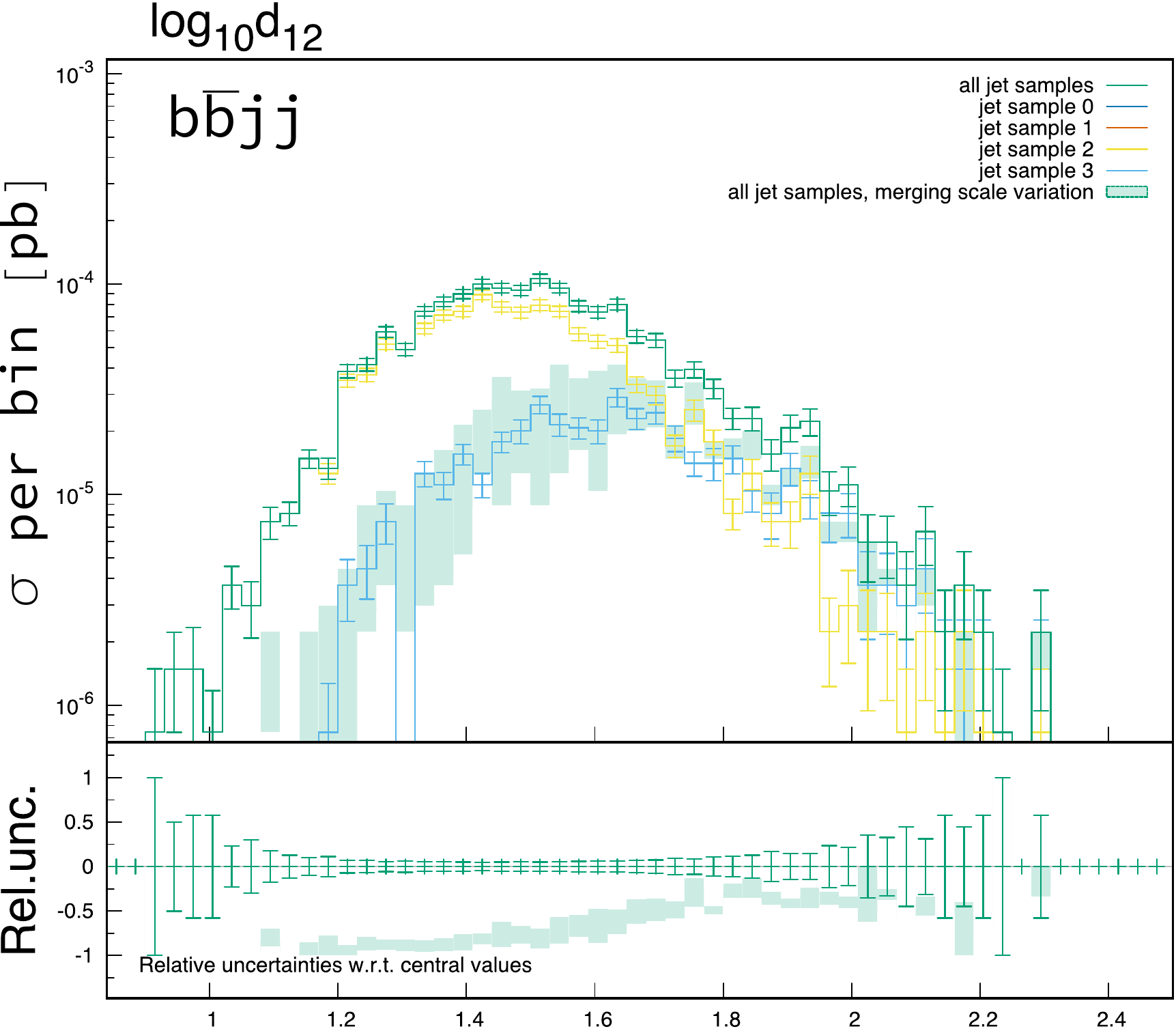}
\includegraphics[width=2.0in]{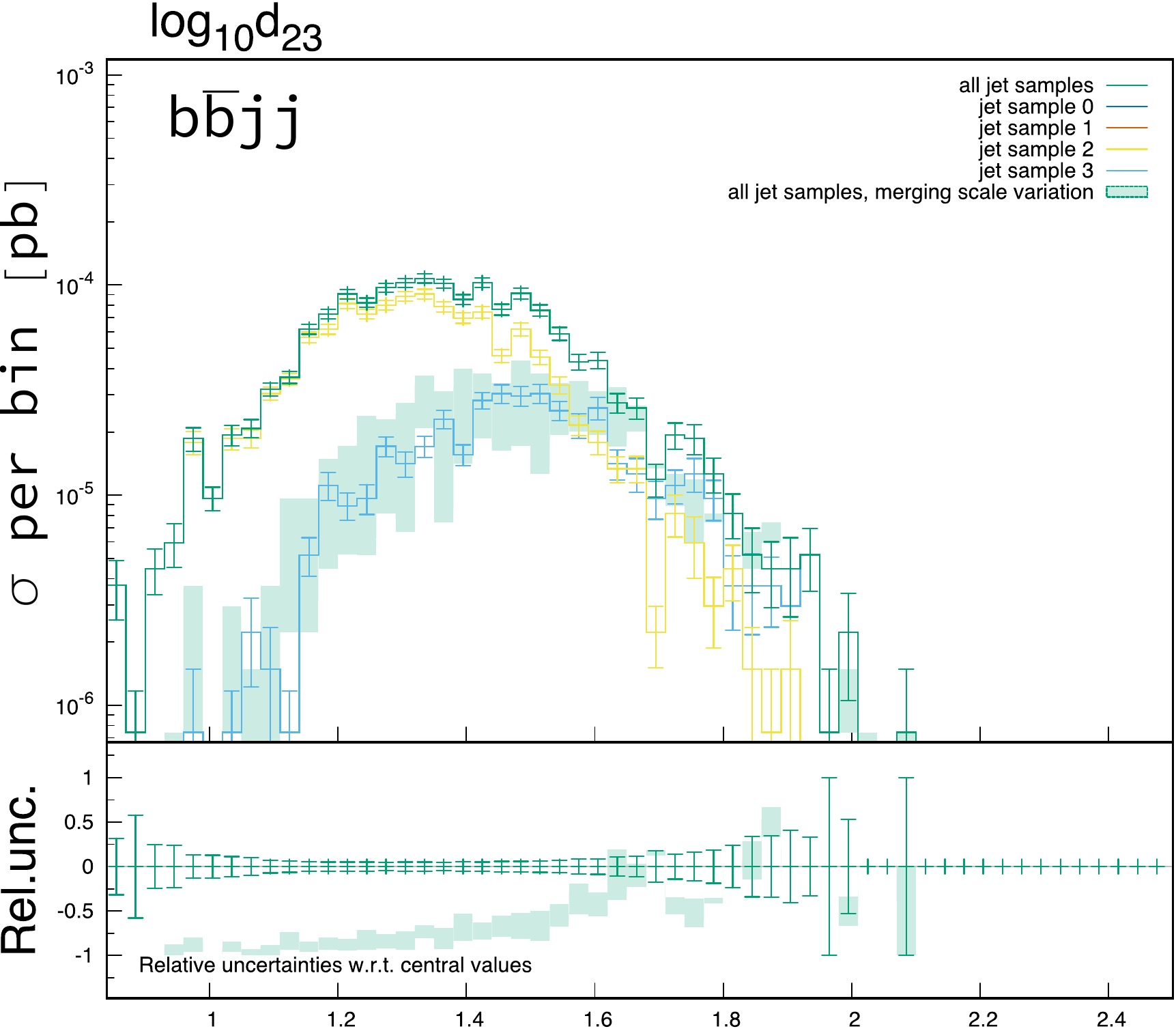}
\caption{\label{fig:djr_3}
{\bf HL-LHC}:
The DJR distributions for the non-resonant backgrounds of
$jj\gamma\gamma$ (upper) and $b\bar b jj$ (lower)
taking {\bf xqcut}$=20$ GeV and $Q_{\rm cut}=30$ GeV.
Here, ``jet sample $n$" refers to the sample containing $n$ hard partons
at the matrix-element level.
}
\end{figure}
Figs.~\ref{fig:djr_1}, \ref{fig:djr_2}, and \ref{fig:djr_3}
show the Differential Jet Rate (DJR) distributions 
after hadronization, multi-parton interactions (MPI), and decays
for  all the non-resonant backgrounds taking
{\bf xqcut}$=20$ GeV and $Q_{\rm cut}=30$ GeV.
%
%
%
%
We observe the DJR distributions for $b\bar b \gamma\gamma$,
$c\bar c \gamma\gamma$, and $Z(b\bar b)\gamma\gamma$ are very smooth and
the variation of the merged cross sections depending
on the choice of $Q_{\rm cut}$ is negligible.
For $b\bar b j\gamma$, $c\bar c j\gamma$, and $jj\gamma\gamma$
the distributions are smooth and the variation is small.
For $b\bar b jj$, the DJR distributions are coarse and the
variation of the merged cross section is sizeable. 

To conclude, the matching has been excellently implemented
for $b\bar b \gamma\gamma$, $c\bar c \gamma\gamma$, and
$Z(b\bar b)\gamma\gamma$ backgrounds and it is less successful
for $jj \gamma\gamma$, $b\bar b j \gamma$, and  $c\bar c j \gamma$.
On the other hand, for $b\bar b jj$,
it is doubtful whether the merged cross section is trustworthy.
Therefore, for $b\bar b \gamma\gamma$,
$c\bar c \gamma\gamma$, and $Z(b\bar b)\gamma\gamma$,
one may safely use the merged cross sections obtained by matching
the leading and sub-leading processes.
For $jj\gamma\gamma$, $b\bar b j\gamma$, and $c\bar c j\gamma$, they are less
reliable.
And, for $b\bar b jj$, it might be 
recommended to use $\sigma_{\rm Eq.(6)}$ 
for conservative estimation of the background,

\begin{figure}
\centering
\includegraphics[width=4.5in]{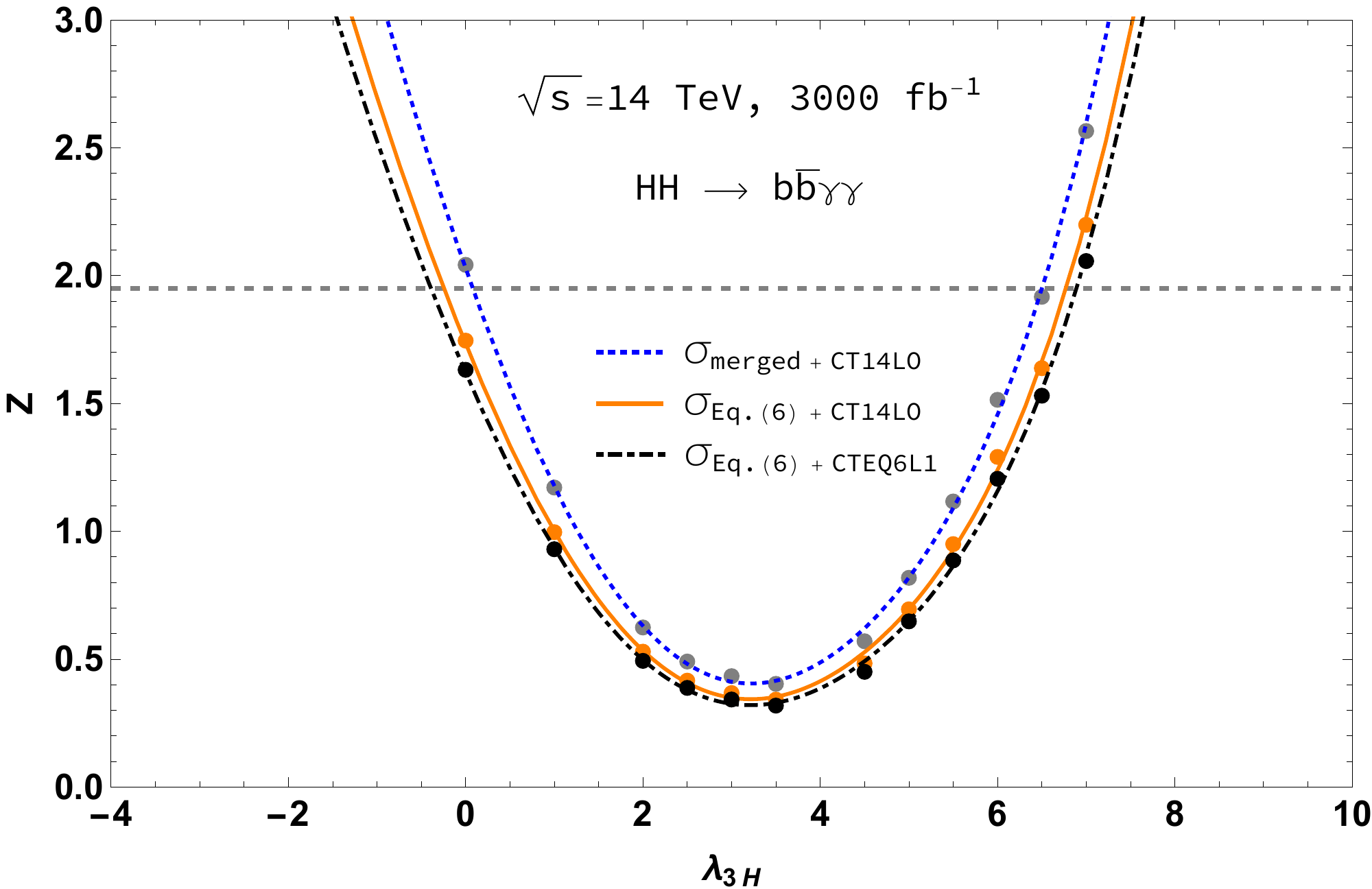}
\caption{\label{fig:Zcomp}
{\bf HL-LHC}:
Significance of the signal over the background versus $\lambda_{3H}$ 
taking $\sigma_{\rm Eq.(6)}$ (red solid) and
$\sigma_{\rm merged}$ (blue dashed) for the non-resonant backgrounds.
The PDF set of {\tt CT14LO} is taken. For comparison, also shown is
the case with the PDF set of {\tt CTEQ6L1}  (black dash-dotted).
Note that the NNLO cross section
$\sigma(gg\to HH)=36.69$ fb in the FT approximation is taken
and the $\lambda_{3H}$-dependent QCD corrections have been included,
see Fig.~\ref{fig:kfactor}.}
\end{figure}
To see the impact of matching for the non-resonant backgrounds, we show the
significance of the signal over the background versus $\lambda_{3H}$
in Fig.~\ref{fig:Zcomp}.
We find that the $ 95\% $ CL region is reduced by the amount of about 15\%
taking the merged cross sections for the non-resonant backgrounds
with {\tt CT14LO}. 

$$ $$
\newpage


\end{document}